\documentclass[prd,aps,longbibliography,twocolumn,nofootinbib,superscriptaddress,showkeys,reprint]{revtex4-1}

\usepackage{extramarks} 
\usepackage{graphicx} 
\usepackage{amsmath}
\usepackage[usenames,dvipsnames]{color} 
\usepackage{listings} 
\usepackage{courier} 

\usepackage{float}
\usepackage{enumerate}
\usepackage[linktocpage=true]{hyperref}
\graphicspath{{./fig/}}
\hypersetup{colorlinks=true, linkcolor=blue, filecolor=magenta,  urlcolor=blue, citecolor=red,}

\makeatletter
\def\@bibdataout@aps{%
\immediate\write\@bibdataout{%
@CONTROL{%
apsrev41Control%
\longbibliography@sw{%
    ,author="08",editor="1",pages="1",title="0",year="1"%
    }{%
    ,author="08",editor="1",pages="1",title="",year="1"%
    }%
  }%
}%
\if@filesw \immediate \write \@auxout {\string \citation {apsrev41Control}}\fi 
}
\makeatother

\newcommand{\BoldVec}[1]{\mathchoice%
  {\mbox{\boldmath $\displaystyle     #1$}}%
  {\mbox{\boldmath $\textstyle        #1$}}%
  {\mbox{\boldmath $\scriptstyle      #1$}}%
  {\mbox{\boldmath $\scriptscriptstyle#1$}}%
}

\def\half{{\textstyle{1\over2}}}
\def\onethird{{\textstyle{1\over3}}}
\def\threehalf{{\textstyle{3\over2}}}
\def\threefourth{{\textstyle{3\over4}}}

\def\twothird{{\textstyle{2\over3}}}

\newcommand{\EEM}{{\cal E}_{\rm M}}
\newcommand{\HH}{{\cal H}}

\newcommand{\TeV}{{\rm \, TeV}}
\newcommand{\GeV}{{\rm \, GeV}}
\newcommand{\MeV}{{\rm \, MeV}}
\newcommand{\Mpc}{{\rm \, Mpc}}
\newcommand{\kpc}{{\rm \, kpc}}
\newcommand{\pc}{{\rm \, pc}}
\newcommand{\J}{{\rm \, J}}
\newcommand{\km}{{\rm \, km}}
\newcommand{\m}{{\rm \, m}}
\newcommand{\cm}{{\rm \, cm}}
\newcommand{\s}{{\rm \, s}}

\newcommand{\Hz}{{\rm \, Hz}}

\newcommand{\OmM}{\Omega_{\rm M}}
\newcommand{\OmMmax}{\Omega_{\rm M}^*}
\newcommand{\OmMmaxsq}{{\Omega_{\rm M}^{*}}^2}
\newcommand{\OmMmaxcub}{{\Omega_{\rm M}^{*}}^3}
\newcommand{\OmGW}{\Omega_{\rm GW}}

\newcommand{\OmK}{\Omega_{\rm K}}
\newcommand{\EM}{E_{\rm M}}
\newcommand{\EMmax}{E_{\rm M}^*}

\newcommand{\EGW}{E_{\rm GW}}

\newcommand{\kGW}{k_{\rm GW}}

\newcommand{\tGW}{t_{\rm GW}}
\newcommand{\fGW}{f_{\rm GW}}
\newcommand{\kf}{k_\ast}
\newcommand{\kbr}{k_{\rm br}}
\newcommand{\vA}{v_{\rm A}}
\newcommand{\EErad}{{\cal E}_{\rm rad}}

\newcommand{\xx}{\BoldVec{x}{}}
\newcommand{\kk}{\BoldVec{k}{}}
\newcommand{\pp}{\BoldVec{p}{}}

\newcommand{\BB}{\BoldVec{B}{}}
\newcommand{\uu}{\BoldVec{u}{}}
\newcommand{\JJ}{\BoldVec{J}{}}
\newcommand{\SSS}{\BoldVec{S}{}}

\newcommand{\dd}{{\rm \, d}} 
\newcommand{\hc}{h_{\rm c}}
\newcommand{\hij}{h_{ij}}

\newcommand{\nab}{\BoldVec{\nabla} {}}
\newcommand{\Ci}{{\rm Ci}}
\newcommand{\Si}{{\rm Si}}

\newcommand{\tini}{t_*}
\newcommand{\tfin}{t_{\rm fin}}
\newcommand{\te}{t_{\rm e}}
\newcommand{\tend}{t_{\rm end}}
\newcommand{\fyr}{f_{\rm yr}}
\newcommand{\Omyr}{\Omega_{\rm yr}}

\newcommand{\bra}[1]{\langle #1\rangle}

\newcommand{\Eq}[1]{Eq.~(\ref{#1})}
\newcommand{\EEq}[1]{Equation~(\ref{#1})}
\newcommand{\Eqs}[2]{Eqs.~(\ref{#1}) and (\ref{#2})}
\newcommand{\Eqss}[2]{Eqs.~(\ref{#1})--(\ref{#2})}
\newcommand{\Fig}[1]{Fig.~\ref{#1}}
\newcommand{\FFig}[1]{Figure~\ref{#1}}
\newcommand{\Tab}[1]{Table \ref{#1}}

\newcommand{\Sec}[1]{Sec.~\ref{#1}}
\newcommand{\Secs}[2]{Secs.~\ref{#1} and \ref{#2}}
\newcommand{\Secss}[2]{Secs.~\ref{#1}--\ref{#2}}

\newcommand{\cf}{{\em cf.~}}

\usepackage{amsfonts}

\begin{document}

\title{Gravitational wave signal from primordial magnetic fields in the Pulsar Timing Array frequency band}

\author{Alberto Roper Pol}
\email[corresponding author: ]{roperpol@apc.in2p3.fr}
\affiliation{Universit\'e de Paris, CNRS, Astroparticule et Cosmologie,
F-75006 Paris, France}
\affiliation{Faculty of Natural Sciences and Medicine, Ilia State University, 
GE-0194 Tbilisi, Georgia}

\author{Chiara Caprini}
\email{chiara.caprini@cern.ch}
\affiliation{D\'epartement de Physique Th\'eorique, Universit\'e de Gen\`eve,
CH-1211 Gen\`eve, Switzerland}
\affiliation{Theoretical Physics Department, CERN, CH-1211 Gen\`eve, Switzerland}

\author{Andrii Neronov}
\email{andrii.neronov@apc.in2p3.fr}
\affiliation{Universit\'e de Paris, CNRS, Astroparticule et Cosmologie,  F-75006 Paris, France}
\affiliation{Laboratory of Astrophysics, \'Ecole Polytechnique F\'ed\'erale de Lausanne, CH-1015 Lausanne, Switzerland}

\author{Dmitri Semikoz}
\email{semikoz@apc.in2p3.fr}
\affiliation{Universit\'e de Paris, CNRS, Astroparticule et Cosmologie,  F-75006 Paris, France}
\affiliation{Institute for Nuclear Research, Russian Academy of Sciences,
117312 Moscow, Russia}
\affiliation{National Research Nuclear University, MEPHI, 115409 Moscow, Russia}
\begin{abstract}
The NANOGrav, Parkes, European, and International Pulsar Timing Array (PTA) Collaborations
have reported evidence for a common-spectrum process that can potentially correspond to a
stochastic gravitational wave background (SGWB) in the 1--100 nHz 
frequency range.
We consider the scenario in which this signal is produced by magnetohydrodynamic (MHD)
turbulence in the early Universe, induced by a nonhelical primordial
magnetic field at the energy scale corresponding to the quark confinement phase transition.
We perform MHD simulations to study the dynamical
evolution of the magnetic field and compute the resulting SGWB. 
We show that the SGWB output from the simulations can be very well approximated by assuming that the
magnetic anisotropic stress is constant in time,
over a time interval related to the eddy turnover time.
The analytical spectrum that we derive under this assumption 
features a change of slope at a frequency corresponding to the GW source duration that we confirm with the numerical simulations.
We compare the SGWB signal with 
the PTA data 
to constrain the temperature scale at which the SGWB is sourced, as well as the amplitude and characteristic scale of the initial magnetic field.
We find that the generation temperature is 
constrained to be in the 1--200 MeV range, the
magnetic field amplitude must be $>1$\% of the radiation energy density at that time, and the magnetic field characteristic scale is constrained to be $>10$\% of the horizon scale.
We show that the turbulent decay of this magnetic field will lead to
a field at recombination that can help to alleviate the Hubble tension
and can be tested by measurements in the voids of the Large Scale Structure with gamma-ray telescopes like the Cherenkov Telescope
Array.

\end{abstract}
\maketitle

\tableofcontents

\section{Introduction}

The magnetic fields observed today in the voids of the Large Scale Structure (LSS) \cite{Neronov:2010,Fermi-LAT:2018jdy} could be of primordial origin, generated for example during inflation or primordial phase 
transitions (for a review, see Ref.~\cite{Durrer:2013pga}, and for a recent update, see Ref.~\cite{Vachaspati:2020blt}).
Once produced, the magnetic field  interacts  with  the  primordial
plasma, leading to magnetohydrodynamic (MHD) turbulence due to the high
conductivity of the early Universe \cite{Ahonen:1996nq,Brandenburg:1996fc}.
The energy-momentum tensor of both the magnetic field and the bulk fluid motions feature a tensor component, 
which can source a stochastic gravitational  wave background  (SGWB)
(see Ref.~\cite{Caprini:2018mtu} for a review and
references therein).
The epoch in the early Universe at which the gravitational wave (GW) production occurs sets the typical frequency of the GW signal. 
In particular, anisotropic stresses present at the energy scale of the quantum chromodynamics (QCD) phase transition can lead to a GW signal around the nanohertz frequency, in the frequency band of pulsar timing arrays (PTA) \cite{Sazhin:1978,Detweiler:1979wn,Deryagin:1986qq,Hogan:1986qda,Witten:1984rs,Signore:1988ud,Thorsett:1996dr,Caprini:2010xv}. 

Recently, the North American Nanohertz Observatory for Gravitational Waves 
(NANOGrav) \cite{NANOGrav:2020bcs}, followed by the Parkes Pulsar Timing Array (PPTA) \cite{Goncharov:2021oub}, the European Pulsar Timing Array (EPTA) \cite{Chen:2021rqp}, and the International Pulsar Timing Array (IPTA) \cite{Antoniadis:2022pcn} Collaborations,
have reported the detection of a signal common to all the analyzed pulsars, 
with very marginal evidence for a quadrupole correlation (following the Hellings-Downs curve), characteristic
of a SGWB according to general relativity \cite{Hellings:1983fr}.

Several sources could produce such a signal, in particular a 
population of merging supermassive black hole binaries \cite{1994MNRAS.269..199H,Sesana:2004sp,2008MNRAS.390..192S}.
Cosmological sources have also been proposed, including inflationary GWs \cite{Vagnozzi:2020gtf,Li:2020cjj,Kuroyanagi:2020sfw,Sharma:2021rot,Lazarides:2021uxv,Yi:2021lxc,He:2021kah},
cosmic strings and domain walls \cite{Ellis:2020ena,Blasi:2020mfx,Buchmuller:2020lbh,Samanta:2020cdk,Bian:2020urb,Liu:2020mru,Chiang:2020aui,Gorghetto:2021fsn,Blanco-Pillado:2021ygr,Chang:2021afa,Lazarides:2021uxv,Buchmuller:2021mbb},
primordial black holes \cite{Vaskonen:2020lbd,DeLuca:2020agl,Kohri:2020qqd,Domenech:2020ers,Wu:2021zta}, supercooled and dark phase transitions \cite{Nakai:2020oit,Addazi:2020zcj,Ratzinger:2020koh,Moore:2021ibq}, and the 
QCD phase transition \cite{Abe:2020sqb,NANOGrav:2021flc,Li:2021qer}, as well as primordial magnetic fields \cite{Neronov:2020qrl, Brandenburg:2021tmp,Khodadi:2021ees}.

In this work, we study 
the SGWB generated by
MHD turbulence due to the presence
of a nonhelical magnetic field at the QCD energy scale
and compare it with the common noise measured by NANOGrav, PPTA, EPTA,
and IPTA. 
We refine and extend the analysis 
of Ref.~\cite{Neronov:2020qrl}, where we used
an approximate analytical estimate of the SGWB signal and compared it qualitatively with the NANOGrav measurement.
Here, we have conducted numerical MHD simulations to accurately
predict the SGWB.
We set as initial condition of the simulations a fully developed MHD spectrum for the magnetic field, and zero initial bulk velocity.
For the numerical simulations we use the {\sc Pencil Code} \cite{PencilCode:2020eyn}, as in
similar numerical works on the SGWB produced by MHD turbulence
during the radiation-dominated era
\cite{RoperPol:2018sap,RoperPol:2019wvy}.

We derive a simple analytical formula for the SGWB spectrum that fits 
the simulation results and widely improves the one given in Ref.~\cite{Neronov:2020qrl}.
We then use this formula to compare the MHD-generated SGWB spectrum to the PTA data, including NANOGrav, PPTA, EPTA, and IPTA.

The results we obtain are broadly consistent with those of Ref.~\cite{Neronov:2020qrl}, namely: the PTA observations can be accounted for by GW production from MHD turbulence at the QCD scale, provided \emph{i)} the
temperature is in the range $1 \MeV < T_* < 200 \MeV$,
\emph{ii)} the magnetic field 
characteristic scale is close to the horizon, and it can correspond to the scale of the largest processed eddies if $2 \MeV < T_* < 50 \MeV$, and
\emph{iii)} the
magnetic field energy density is larger than a few percent of the radiation energy density at $T_*$.
In particular, we show here that the spectral slope of a SGWB from MHD turbulence at the QCD scale is fully compatible with the PTA constraints. 

As already pointed out in Ref.~\cite{Neronov:2020qrl}, a primordial magnetic field with these characteristics has a particularly interesting phenomenology since, besides accounting for the PTA common noise, it could also change the sound horizon at the cosmic microwave background (CMB) epoch, easing the Hubble tension, and explain the magnetic fields observed today in matter structures \cite{Jedamzik:2020krr, Jedamzik:2020zmd,Galli:2021mxk}. Such a field is also within the sensitivity range of the next-generation $\gamma$-ray observatory Cherenkov Telescope Array (CTA) \cite{Korochkin:2020pvg}.

We stress that our results depend on the particular initial conditions that we have chosen, namely a fully developed, nonhelical MHD spectrum for the magnetic field, with no initial bulk velocity. 
In Refs.~\cite{RoperPol:2019wvy,Kahniashvili:2020jgm,Brandenburg:2021tmp,Brandenburg:2021bvg,RoperPol:2021uol},
similar simulations have been performed using the {\sc Pencil Code} by inserting an electromotive force to model the initial magnetic field
obtaining SGWB spectra that differ from ours, especially at large
frequencies, which are of less observational relevance.
We have chosen the aforementioned initial conditions for mainly two reasons. 
First of all, they are conservative: whatever the initial generation mechanism, the magnetic field is expected to enter a phase of fully developed and freely decaying turbulence \cite{Ahonen:1996nq, Brandenburg:1996fc}. 
Any initial phase of magnetic field growth would increase the GW production
\cite{RoperPol:2019wvy,Kahniashvili:2020jgm,RoperPol:2021uol}. 
Secondly, simple initial conditions make it easier to build an analytical description of the simulation outcome.  
Since we want to model magnetically driven turbulence, we also neglect the presence of initial bulk velocity for simplicity.

The paper is organised as follows. In Sec.~\ref{sec:GW}, we analyze GW generation from MHD turbulence.
First, we present the equations governing the dynamics of the source and the
model for the initial condition (Secs.~\ref{sec:MHD} and \ref{sec:mag}),
followed by a derivation of the general expression for the SGWB spectrum, \cf Sec.~\ref{sec:OmGW}.
We then derive the SGWB spectrum obtained under the assumptions of a constant
source (Sec.~\ref{analytical_spec}), and from the MHD simulations 
(Sec.~\ref{sec:numerical_spec}), and we compare the two results (Sec.~\ref{sec:comparison_approx}). 
In Sec.~\ref{sec:comparison_data}, we adopt the analytical form of the SGWB spectrum, validated with the simulations, and compare it with the PTA measurements, introduced in Sec.~\ref{PTA_res}, to constrain
the magnetic field parameters (Sec.~\ref{sec:constraints}).
Furthermore, we consider the effect the magnetic field could have on the CMB at recombination and in the cosmic
voids of the LSS (Sec.~\ref{comp_Bl}), and compare our results to those of previous publications, to show the implications of
our initial conditions (Sec.~\ref{init}).
In Sec.~\ref{sec:MBHB}, we compare the SGWB produced by MHD turbulence
with the one produced by supermassive black hole binaries.
Finally, we conclude in \Sec{conclusions}.

We use the $(-\!+\!\!+ +)$ metric signature and
set $c = k_B = 1$.
Magnetic fields are expressed in Lorentz-Heaviside units by
setting the vacuum permeability to unity.
The Kronecker delta is indicated by $\delta_{ij}$, the
$n$-dimensional Dirac delta function by $\delta^n (x)$,
the Gamma function by $\Gamma(x)$, and the cosine and sine
integral functions by $\Ci(x)$ and $\Si(x)$, respectively.
Fields in Fourier space are indicated by a tilde and we use the
Fourier convention $\tilde B (\kk) = \int B(\xx) \exp{(i \xx\cdot \kk)} \dd^3 \xx$ and
$B(\xx) = (2\pi)^{-3} \int \tilde B(\kk) \exp{(-i \xx\cdot \kk)} \dd^3 \kk$.
In \Secss{sec:MHD}{sec:OmGW}, space coordinates, time coordinate and (inverse) wave vectors are normalized to the comoving Hubble scale at initial time $\mathcal{H}_*^{-1}=t_*$, and a subscript $*$ in general denotes quantities at initial time. 
In \Sec{analytical_spec} and following, we restore their dimensions for convenience. 
Energy densities are normalized by the
radiation energy density.

\section{SGWB spectrum: constant source model and MHD simulations}
\label{sec:GW}

\subsection{Magnetohydrodynamics}
\label{sec:MHD}

In this work, we
perform simulations solving the MHD equations and the subsequent GW production with the {\sc Pencil Code} \cite{PencilCode:2020eyn}.
A fully developed stochastic magnetic field $\BB$
is inserted as initial condition of the simulations, while
the initial velocity field $\uu$ is zero (though it can be driven by
the primordial magnetic field at later times via Lorentz forcing).
The MHD equations for a relativistic fluid with $p = \rho/3$ in the
Friedmann-Lema\^itre-Robertson-Walker (FLRW) background metric [$a(t)$ is the scale factor],
\begin{equation}
\dd s^2 = a^2 (t) \left[-\dd t^2 + \delta_{ij} \dd x^i \dd x^j\right],
\label{FLRW}
\end{equation}
are \cite{Brandenburg:1996fc,
Brandenburg:2017neh,Durrer:2013pga}
\begin{align}
    \frac{\partial \ln \rho}{\partial t} = &\, 
    - \frac{4}{3} \left(\nab \cdot \uu + \uu \cdot 
    \ln \rho \right)  \nonumber \\
    &\, + \frac{1}{\rho} \left[\uu \cdot \left(\JJ \times 
    \BB \right) + \eta \JJ^2\right], \label{dlnrhodt} \\
    \frac{\partial \uu }{\partial t} = &\, - \uu \cdot 
    \nab \uu + \frac{\uu}{3} \left(\nab \cdot \uu + \uu \cdot 
    \nab \ln \rho \right) \nonumber \\
    &\, - \frac{\uu}{\rho} \left[ \uu \cdot \left(\JJ \times \BB 
    \right) + \eta \JJ^2 \right] - \frac{1}{4} \nab \ln \rho 
    \nonumber \\
    &\, + \frac{3}{4\rho} \JJ \times \BB + \frac{2}{\rho} 
    \nab \cdot \left(\rho \nu \SSS\right), \label{dudt} \\
    \frac{\partial \BB}{\partial t} = &\, \nab \times
    \left(\uu \times \BB - \eta \JJ \right), \quad
    \JJ =  \nab \times \BB, \label{dBdt}
\end{align}
where $\rho$ is the energy density, $\JJ$ the current density,
$S_{ij} = \half (u_{i,j} + u_{j,i}) - \onethird \nab \cdot
\uu$ the rate-of-strain tensor, $\nu$ the kinematic
viscosity, and $\eta$ the magnetic diffusivity.
The space coordinates are comoving with the expansion of the Universe and
normalized by the comoving Hubble radius $\HH_*^{-1}$;
$t$ denotes conformal time, also normalized by $\HH_*^{-1}$.
All MHD fields are comoving and normalized by the radiation energy density, $\EErad=3\HH^2/8\pi G a^2$
(where $G$ is the gravitation constant) \cite{Brandenburg:1996fc,Brandenburg:2017neh,RoperPol:2019wvy,RoperPol:2018sap}.

\subsection{Magnetic field}
\label{sec:mag}
    
We model the (normalized) magnetic field as a stochastic nonhelical field,
statistically homogeneous, isotropic and Gaussian.
Hence, the two-point autocorrelation function is sufficient
to describe it statistically due to the Isserlis theorem \cite{Isserlis:1916, MY75}.
At unequal times, the autocorrelation takes the form
\begin{align}
\bra{\tilde B_i^*(\kk, t_1)\, \tilde B_j(\kk', t_2)}=&\nonumber \\
(2\pi)^6 \delta^3 (\kk - &\, \kk')  P_{ij} (\kk) \frac{\EM (k, t_1,t_2)}{4 \pi k^2},
\label{equaltime}
\end{align}
where $\EM(k,t_1,t_2)$ is the unequal-time correlator (UETC), reducing, at equal time $t_2 = t_1$, to the normalized magnetic field energy density spectrum $\EM(k,t_1)$. Furthermore, $P_{ij} = \delta_{ij} - 
\hat k_i \hat k_j$ is the projection tensor, with $\hat \kk$ indicating
the unit wave vector $\hat \kk = \kk/|\kk|$.
The angle brackets indicate ensemble average over stochastic realizations,
which can be approximated by a volume average in homogeneous
turbulence \cite{MY75}.

At the initial time $\tini$,
we assume that the magnetic energy spectrum $\EM(k,t_*)$ is characterized by a Batchelor spectrum at large
scales and a Kolmogorov spectrum at small scales,
peaking at the characteristic scale $k_*=2\pi/l_*$.
Magnetic fields produced by causal processes,
e.g., during cosmological phase transitions, feature a finite
correlation length, which leads to a Batchelor magnetic spectrum
in the limit $k\rightarrow 0$ \cite{MY75,Durrer:2003ja}.
The Kolmogorov-type $k^{-5/3}$ spectrum is found and 
well established in purely hydrodynamic turbulence \cite{1941DoSSR..30..301K}.
In general MHD, different models have been proposed, e.g.,
Iroshnikov-Kraichnan $k^{-3/2}$ \cite{1965PhFl....8.1385K, 1964SvA.....7..566I}, Goldreich-Sridhar $k^{-5/3}$ \cite{Goldreich:1994zz}, weak turbulence $k^{-2}$ \cite{1997PhPl....4..605N}, and Boldyrev $k^{-3/2}$ \cite{Boldyrev:2005ms}.
Simulations of MHD turbulence in this context seem to indicate
the development of a turbulent spectrum with a $k^{-5/3}$ scaling \cite{Christensson:2000sp, 2000PhRvL..84..475M,Kahniashvili:2012uj,Brandenburg:2014mwa,Brandenburg:2016odr,Brandenburg:2017neh}.
We therefore adopt the following spectral shape: 
\begin{align}
    \EM (k,\tini)=&\,(1+\mathcal{D})^{1/\alpha} \EMmax \nonumber \\
    \times &\,\frac{(k/\kf)^4}{\left[1+\mathcal{D} (k/\kf)^{\alpha(4+5/3)}
    \right]^{1/\alpha}},
\label{eq:Bspec2}
\end{align}
where $\EMmax = \EM(\kf,\tini)$ denotes the magnetic amplitude at the peak $k_*$ and at initial time,  the parameter $\mathcal{D}=12/5$ is tuned\footnote{The parameter ${\cal D}$ depends on the slopes in the subinertial and inertial ranges. 
Here we set $a=4$
in the subinertial range, and $b=5/3$ in the inertial range, which gives ${\cal D}=a/b=12/5$.} so that the spectrum peaks at $k=\kf$, and
the parameter $\alpha$ indicates the smoothness of the
transition between the Batchelor ($\sim\!k^4$) and Kolmogorov ($\sim\!k^{-5/3}$)
scalings around the spectral peak: we set it to $\alpha=2$ based on previous fits of simulated spectra \cite{Brandenburg:2017neh}.
Note that $\EMmax$ is the maximal value of the magnetic amplitude, since we consider decaying MHD turbulence.

The simulations are initialized with the magnetic
field \cite{Brandenburg:2017neh,RoperPol:2019wvy,RoperPol:2021uol}
\begin{equation}
    \tilde B_i (\kk, t_*) = P_{ij} (\kk) g_j (\kk) g_0 (k),
    \label{B_stoch}
\end{equation}
where
$g_j(\kk)$ is the Fourier transform of a $\delta$-correlated
vector field in three dimensions with Gaussian fluctuations,
i.e., $g_i(\xx) g_j(\xx') = \delta_{ij} \delta^3(\xx - \xx')$,
and $g_0 (k) = \sqrt{\EM(k, t_*)}/k$ corresponds to the magnetic
spectral shape defined in \Eq{eq:Bspec2}.
We therefore assume that the MHD-processed magnetic spectrum is already  established 
when the GW generation starts. 
The simulations output the GW generation by the subsequent magnetic
turbulent decay, by solving the full MHD system of \Eqss{dlnrhodt}{dBdt}.
The initial amplitude $\EMmax$
and the position of the spectral peak $\kf$ are the input parameters of the
numerical simulations in \Sec{sec:numerical_spec}.

The total normalized magnetic energy density is
$\OmM(t) = \EEM(t)/\EErad (t) = \half \langle {\bf B}^2({\xx},t)
\rangle$. 
Using \Eqs{equaltime}{eq:Bspec2}, at initial time, this becomes
\begin{align}
    \OmMmax = \OmM (t_*) = & \, \int_0^{\infty} \EM(k, t_*) \dd k \nonumber \\
    = &\, \kf \EMmax {\cal A} (\alpha),
    \label{integ_OmM}
\end{align}
with
\begin{align}
    {\cal A} (\alpha=2)=&\frac{3^{19\over 34}\, \Gamma\bigl[\textstyle{1\over 17}\bigr]\,\Gamma\bigl[\textstyle{15\over 34}\bigr] }{5^{1\over 17}\, 2^{32\over 17} \sqrt{17 \pi}}
    \approx 2.064.
    \label{A_alp}
\end{align}

The (normalized) magnetic stress tensor components are $T_{ij} (\xx, t)= - B_i  (\xx, t) B_j  (\xx, t) + \half 
\delta_{ij} \BB^2  (\xx, t)$, and the traceless and transverse (TT) projection
of the stress tensor in Fourier space is $\tilde \Pi_{ij}  (\kk, t) =\tilde T_{ij}^{\rm TT}  (\kk, t) =
\Lambda_{ijlm} (\hat \kk) \tilde T_{lm}  (\kk, t)$, with
$\Lambda_{ijlm} =P_{il} P_{jm} - \half P_{ij} P_{lm}$.
The TT-projected stress $\Pi_{ij}$ sources the GWs. As $B_i  (\xx, t)$, $\Pi_{ij}$ is also a random variable: the GW production can therefore be described statistically using the UETC of the tensor stress $E_\Pi(k,t_1,t_2)$, defined in analogy\footnote{%
The UETC $E_\Pi(k, t_1, t_2)$ satisfies
\begin{equation}
\bra{\tilde \Pi_{ij}^*(\kk, t_1) \, \tilde \Pi_{ij}(\kk', t_2)}=
(2\pi)^6 \delta^3 (\kk - \kk')  \frac{E_\Pi (k, t_1,t_2)}{4 \pi k^2}.\nonumber
\end{equation}} with \Eq{equaltime}, as
\begin{align}
  \langle \Pi_{ij}({\xx},t_1) &\, \Pi_{ij}({\xx},t_2)\rangle= 
  \nonumber \\
  &\, \int_{0}^\infty
  E_\Pi(k,t_1,t_2) \dd k.
  \label{Pi_gaussian}
\end{align}
For a Gaussian magnetic field (as the one with which we initialize the simulations),
the UETC is \cite{Caprini:2009yp}
\begin{align}
E_\Pi (k,t_1,t_2) & \, = {k^2\over 4\pi} \int \frac{\dd^3 \pp}{p^2 |{\kk-\pp}|^2}
\EM(p,t_1,t_2) \nonumber \\ &\, \times
\EM(|{\kk-\pp}|,t_1,t_2) \Bigl(1+(\hat \kk \cdot \hat \pp)^2\Bigr)
\nonumber \\ &\, \times
\Bigl(1+(\hat \kk \cdot \widehat{\kk-\pp})^2\Bigr).
\label{Pi_conv}
\end{align}

\subsection{Gravitational wave production}
\label{sec:OmGW}

GWs are defined as the metric tensor perturbations $\bar h_{ij}$
over the FLRW metric, defined in \Eq{FLRW},
\begin{equation}
    \label{GWconf}
\dd s^2 = a^2(t) \, \left[ - \dd t^2 + (\delta_{ij} + \bar h_{ij}) \, \dd x^i \dd x^j \right] \,.
\end{equation}
We restrict to the radiation era and, as previously, space and time are normalized with $\HH_*^{-1}$.
We assume that the scale factor evolves linearly with conformal time during the GW sourcing and normalize it such that
$a=t$ with  $a_*=1$ \cite{RoperPol:2018sap}. 
The wave equation in the radiation era for the scaled variable
$h_{ij}=a \bar h_{ij}$  \cite{Gris74},
following the normalization of Refs.~\cite{RoperPol:2018sap, RoperPol:2019wvy},
becomes
\begin{equation}
\Bigl(\partial_t^2 + \kk^2\Bigr)\, \tilde h_{ij} (\kk, t) = 
\frac{6\, \tilde \Pi_{ij}(\kk, t)}{t}\,,
\label{GW1_Fourier}
\end{equation}
where $k$ is the wave number normalized with $\HH_*$ and the tensor stress sourcing the GWs is defined above \Eq{Pi_gaussian}.
Assuming that the source is acting until a finite time\footnote{Note that the parameter $\tfin$ is a mathematical artifact, inserted to separate the sourced phase from the phase of free propagation. 
In the simulations, the source evolves via
turbulent MHD decay and, as in previous numerical works \cite{RoperPol:2019wvy, Kahniashvili:2020jgm,
RoperPol:2021uol, Brandenburg:2021bvg, Brandenburg:2021tmp}, we run them until
all GW wave numbers in the simulation box have reached a stationary state, i.e., are oscillating around a fixed amplitude.
Hence, the simulations indeed output spectra in the free propagation phase.} $\tfin$,
the solution to \Eq{GW1_Fourier} with initial conditions $\tilde h_{ij}(\kk, t_*) = \partial_t \tilde h_{ij}(\kk, t_*) = 0$,
matched with the homogeneous solution at $\tfin$,
describes a freely propagating wave at times $t > \tfin$:
\begin{equation}
\tilde h_{ij}(\kk, t) = 
\frac{6}{k} \int_{t_*}^{t_{\rm fin}} \frac{\tilde \Pi_{ij}(\kk,t_1)}{t_1}
\sin k(t - t_1) \dd t_1. 
\label{sol2}
\end{equation}

The GW energy density, normalized to the radiation energy density
$\EErad$, is
\begin{align}
\OmGW (t)  = &
\,\frac{1}{12}  \, \big\langle\big(\partial_t \hij({\bf x},t) -
\hij({\bf x},t)/t \big)^2\big\rangle \nonumber\\
= &\,
\int_{-\infty}^\infty \OmGW(k,t) \dd \ln k,
\label{eq:OmGWdef}
\end{align}
where $\OmGW(k,t)$ denotes the normalized logarithmic GW energy density spectrum.
Fourier transforming Eq.~\eqref{eq:OmGWdef} and inserting solution \eqref{sol2}, it becomes, at times $t > \tfin$,
\begin{align}
\OmGW  (k,t) &\, = 3 \, k \int_{t_*}^{\tfin}
\frac{dt_1}{t_1} \int_{t_*}^{\tfin} 
 \frac{dt_2}{t_2} \nonumber \\
&\, \times  E_\Pi(k,t_1,t_2)\,S(k,t,t_1,t_2),
\label{eq:OmGWt} 
\end{align}
with
\begin{align}
S(k,t,t_1,t_2)
\approx &\, \cos k(t-t_1)\cos k(t-t_2).
\label{eq:S}
\end{align}
In the above equation, we have omitted the terms proportional to $1/(kt)$ that appear due to
the $h_{ij}/t$ term in \Eq{eq:OmGWdef} since,
at present time, all relevant wave numbers of signals produced in the early Universe are inside the horizon $kt\gg 1$.

\subsection{Analytical GW spectrum for constant-in-time stress}
\label{analytical_spec}

The characteristic time of the magnetic field decay in a turbulent MHD cascade is the eddy turnover time
$\delta \te = \te - \tini = (\vA \kf)^{-1}$,
where $\vA = \sqrt{\threehalf \OmMmax}$ is
the Alfv\'en speed at initial time\footnote{The Alfv\'en speed is $\vA^2 = \langle\BB^2\rangle/\langle p + \rho\rangle$, such that
in the radiation-dominated era, with $p = \onethird \rho$, it becomes 
    $\vA^2 = \threefourth  \langle\BB^2\rangle/ \langle \rho\rangle = 
    \threehalf \OmM$.}
and $k_*$
the initial characteristic wave number.

The GW sourcing, on the other hand, occurs on a characteristic time interval $\delta t = t - \tini \sim 1/k$ for a given GW wave number 
$k$, as can be inferred from the Green function of the wave 
equation \eqref{sol2}, which has period $2\pi/k$. 
Indeed, recent simulations of GW production by MHD turbulence have shown that each mode $k$ of the GW spectrum in the simulation box reached a stationary amplitude after an initial growth period lasting
$\delta t \sim 1/k$
\cite{RoperPol:2019wvy, RoperPol:2021uol}. 
Since the GW spectrum is expected to peak around $k_*$, 
the GW sourcing is faster than
the magnetic field decay for all wave numbers satisfying $k>\vA k_*$, i.e., around and above
the GW spectrum peak. 
Correspondingly, the simulations have also shown that the total GW energy density, which is dominated by the spectral amplitude at the peak, enters a stationary regime shortly after the time interval
$ \delta \tGW \approx 1/\kf$ \cite{RoperPol:2019wvy, RoperPol:2021uol}.
Comparing the latter to the eddy turnover time gives, by causality, $\delta \tGW/\delta \te = \vA \lesssim 1 $.
More precisely, one has $\vA \lesssim 0.4$, since big bang nucleosynthesis (BBN) limits\footnote{%
The upper bound $\OmMmax \lesssim 0.1$ is reported in Refs.~\cite{Shvartsman:1969mm, Grasso:1996kk, Kahniashvili:2009qi}.
However, the constraint from nucleosynthesis has been recently revisited in Ref.~\cite{Kahniashvili:2021gym}, taking into account the MHD turbulent
decay from the time when the magnetic field is generated to BBN, allowing larger values of $\OmMmax$.}
$\OmM^* \lesssim 0.1$ 
\cite{Shvartsman:1969mm, Grasso:1996kk, Kahniashvili:2009qi}.

Since the dynamical evolution of GW production is faster than
that of the magnetic field for all relevant wave numbers, we can assume, as a first approximation, that the magnetic stresses
in \Eq{eq:OmGWt} are constant in time. 
The role of $\tfin$ in
\Eq{eq:OmGWt} becomes then to cut off the constant source, inserting an effective source duration $\delta \tfin = \tfin - \tini$. 
The latter is expected to be related to the characteristic time of the magnetic field decay, 
$\delta \tfin \gtrsim \delta \te$. 
As we shall see, the effective source duration $\delta \tfin$ introduces a feature in the SGWB spectrum, i.e., a change of spectral slope between the wave numbers for which the GW production is faster than the source decay $k\gtrsim 1/\delta \te\gtrsim 1/\delta \tfin$ and those for which the GW production is slower
$k<1/\delta \tfin$.

In the following,
we first present the analytic calculation of the GW
spectrum under the assumption that the magnetic stress is constant
in time up to $\tfin$. 
We then revise the qualitative analysis of Ref.~\cite{Neronov:2020qrl} using the more accurate prediction for the GW spectrum derived previously. 
In the next subsection (\Sec{sec:numerical_spec}), we discuss the results of the simulations. We then show, in \Sec{sec:comparison_approx}, that the constant source model approximates very well the  GW spectra output from the simulations and use the latter to compute the specific value 
of the parameter $\delta \tfin$ and infer empirically its relation to $\delta \te$.

\subsubsection{Analytical GW spectrum}
\label{sec:analytical_detail}

If we assume that the stress is constant in time, \Eq{eq:OmGWt} can be easily integrated to find (note that from this section on, dimensions in wave numbers and time are restored for clarity)
\begin{align}
\OmGW  (k, t) = &\, 3\,
k \, E_\Pi^* (k)  \Bigl\{\cos kt \bigl[ \Ci(k\tfin) - \Ci(kt_*) \bigr]
\nonumber \\
&\ \, + \sin kt \bigl[ \Si(k\tfin) - \Si(kt_*) \bigr] \Bigr\}^2,
\label{OmGW_constant_tfin}
\end{align}
where $t \geq \tfin$ and $E_\Pi^* (k) = E_\Pi(k, t_1=\tini, t_2=\tini)$ is the autocorrelation
function of the magnetic stresses at time $t_* = \HH_*^{-1}$, 
defined in \Eq{Pi_gaussian}. 

The GW spectrum oscillates in time and wave number.
Fixing the time to\footnote{We choose $t=\tfin$ here since, as shown in \Sec{sec:comparison_approx}, $\OmGW(k, \tfin)$ approximates very well the envelope of the SGWB that we obtain via the simulations.
The evolution after $\tfin$ in the constant-in-time model
yields an enhancement of the SGWB at high frequency as a consequence
of the abrupt switching off of the source.
We investigate this feature in a separate publication \cite{paper2}.
} $t = \tfin$, we can approximate the envelope of the 
oscillations over $k$
as \cite{paper2}
\begin{align}
     \OmGW  & (k, \tfin)  \approx 3 \,
     k \,E_\Pi^* (k) \nonumber \\
     \times &  \left\{ \begin{array}{ll}
        \ln^2 [1+\HH_*\delta \tfin]
        & \text{ if } k\,\delta \tfin < 1,\\
        \ln^2[1 + (k/\HH_*)^{-1}]
        & \text{ if } k\, \delta \tfin
        \geq 1.
        \end{array} \right.
        \label{OmGW0_kt}
\end{align}

Let us first investigate the  proportionality to $k\, E_\Pi^*(k)$.
The UETC of the anisotropic stress energy tensor is given in \Eq{Pi_conv} for a Gaussian
magnetic field. 
In general, the MHD evolution can yield non-Gaussianities
in the statistical distribution of the magnetic field, such that 
$E_\Pi(k, t_1, t_2)$ might deviate from the expression  given in \Eq{Pi_conv}.
However, here we assume that the initial magnetic field
is Gaussian and that the magnetic anisotropic stress is constant in time: \Eq{Pi_conv} is therefore appropriate. 
In terms of the anisotropic stress power spectral density $P_\Pi^*(k)=2\pi^2 E_\Pi^* (k) /k^2$,
we can write
\begin{equation}
    k \,E_\Pi^* (k) = 
    \frac{k^3}{2\pi^2} \,P_\Pi^* (0) \,p_\Pi\biggl({k\over\kf}\biggr),
    \label{kEPi}
\end{equation} 
where $P_\Pi^* (0)$ can be computed using Eqs.~(\ref{eq:Bspec2}), (\ref{integ_OmM}), and
(\ref{Pi_conv}) (see Ref.~\cite{paper2} for a detailed derivation):
\begin{equation}
     P_\Pi^* (0)= 2 \pi^2 \OmMmaxsq\,{\cal C} (\alpha) {\cal A}^{-2}(\alpha)\, \kf^{-3},
     \label{P_Pi0}
\end{equation}
with
\begin{align}
    {\cal C} (\alpha=2) =  \frac{7 \,\Gamma\bigl[\textstyle{21\over 34}\bigr] 
    \Gamma\bigl[\textstyle{13\over 34}\bigr]}{2^{4\over 17} 3^{21\over34}
    5^{47\over34}} \approx 1.0987,
\end{align}
and
\begin{equation}
p_\Pi \biggl(\frac{k}{k_*}\biggr)\equiv \frac{P_\Pi^*(k)}{P_\Pi^*(0)}\in (0, 1)
\end{equation}
is a monotonically decreasing
function that is computed numerically using \Eq{Pi_conv}.
At small wave numbers, $p_\Pi$ is constant by causality \cite{Caprini:2009fx} and equal to one, while at large wave numbers $k \gtrsim 2 \kf$, $p_\Pi (k/\kf) \propto k^{-2} \EM(k)\sim k^{-11/3}$
for a Kolmogorov magnetic spectrum $\EM(k)\sim k^{-5/3}$ \cite{Caprini:2009yp}.

We can now proceed to investigate the residual $k$ dependence of the SGWB spectrum. 
In terms of $p_\Pi(k/k_*)$, combining \Eqss{OmGW0_kt}{P_Pi0}, the SGWB takes the form
\begin{align}
    \OmGW (k,  \tfin) \approx  &\, 3 \, \biggl(\frac{k}{k_*}\biggr)^3 \, \OmMmaxsq \frac{{\cal C} (\alpha)}{{\cal A}^{2}(\alpha)} 
     \,\,p_\Pi \biggl(\frac{k}{k_*}\biggr)
    \nonumber \\ &\,  \hspace{-18mm}
   \times   \left\{
    \begin{array}{lr}
       \ln^2 [1 + \HH_* \delta \tfin]  &
       \text{ if } k\, \delta \tfin < 1, \\
        \ln^2 [1 + (k/\HH_*)^{-1}]
       & \text{ if }  k\, \delta \tfin \geq 1.
    \end{array}
    \right.
    \label{OmGW_envelope}
\end{align}
The hierarchy of scales appearing in the above equation is as follows.
The source duration is $\delta \tfin\gtrsim \delta \te \geq 1/\kf$, since $\vA \leq 1$. 
Furthermore, while $\delta \tfin$ can be longer (long source) or shorter (short source) than one Hubble time $\HH_*^{-1}$, one always has 
$k_* \HH_*^{-1} \geq 2 \pi$, since the Hubble scale
corresponds to $k_{\rm H} \HH_*^{-1} = 2 \pi$ and for causally generated turbulent sources $k_*\geq k_{\rm H}$.

At wave numbers below $1/\delta
\tfin$, the spectrum is decorrelated from the source, leading to the usual cubic increase with wave number: $\OmGW \propto  (k/k_*)^3\,\OmMmaxsq \ln^2 [1 + \HH_* \delta \tfin]$. 
For a short source satisfying $\delta
\tfin< \HH_*^{-1}$, the prefactor becomes $\ln^2 [1 + \HH_* \delta \tfin]\simeq (\HH_* \delta \tfin)^2$, further suppressing the spectrum amplitude, as indicated in Ref.~\cite{Caprini:2009yp}.
The quadratic increase with time of $\OmGW (k, t) \propto \delta t^2$  at early times has also been observed in the 
simulations of Ref.~\cite{RoperPol:2019wvy}.

The spectral $k$ dependence changes at wave numbers $k>1/\delta
\tfin$,  turning into $\OmGW \propto (k/\kf)^3 \, \OmMmaxsq \ln^2 [1 + (k/\HH_*)^{-1}]$
in the range $k \in (1/\delta \tfin,  k_*)$. 
For a short source satisfying $\delta
\tfin< \HH_*^{-1}$, one can 
approximate
$\ln^2 \, [1 + (k/\HH_*)^{-1}] \simeq (\HH_*/k)^2$, such that the spectrum is nearly linear in $k\in (1/\delta \tfin,  k_*)$: $\OmGW \propto (k/k_*) \, \OmMmaxsq \,  (\HH_*/k_*)^2$.  
If, on the other hand, the source is long, $\delta
\tfin> \HH_*^{-1}$, the transition to the nearly linear spectrum is smoother,  preceded by the logarithmic dependence  $\OmGW \propto (k/k_*)^3\, \OmMmaxsq \ln^2 [\HH_*/k]$ in the region  
$k \in (1/\delta \tfin,  \HH_*)$. 

This spectral shape is in accordance with what was previously found in
Ref.~\cite{Caprini:2009fx} in the context of a coherent and instantaneous GW source, of which our approximation can be seen as a special case:
we assume in fact that the source is constant in time, but that it turns on and off instantaneously, with a discontinuity in time.

Note that, when $\vA \sim 1$ and the eddy turnover time $\delta \te$ is very short, 
$\delta \tfin\sim \delta \te\sim 1/k_*$ and the quick magnetic field evolution does
not allow the linear regime $\OmGW \sim k$ to form.
Since we set $\vA \lesssim 0.4$ because of nucleosynthesis constraints, the linear increase regime is present in all the spectra output by the simulations (\cf \Sec{sec:numerical_spec}),
in agreement with earlier numerical results \cite{RoperPol:2019wvy, Kahniashvili:2020jgm,RoperPol:2021uol,Brandenburg:2021bvg, Brandenburg:2021tmp}.

Finally, at wave numbers $k\gtrsim k_* $, the function $p_\Pi (k/k_*)$ changes slope, slowly transitioning from constant to $k^{-11/3}$ at large wave numbers. 
As pointed out in Ref.~\cite{Caprini:2009fx}, for a coherent and  discontinuous source the peak of the GW spectrum is determined by the 
behavior of $p_\Pi (k/k_*)$: in the present case, the $p_\Pi (k/\kf)$ slope combines with the previously derived linear increase to give
$\OmGW \sim k^{-8/3}$.

The peak $\kGW$ of the logarithmic GW energy density is located where $(k/\kf)^3  \ln^2 [1 + (k/\HH_*)^{-1}] \, p_\Pi (k/\kf)  \simeq (k/\kf) \, (\HH_*/\kf)^{2} \, p_\Pi (k/\kf)$ is maximum (the approximation is justified since, as previously mentioned, $\kf\HH_*^{-1}\geq 2\pi$). 
This occurs at $\kGW \simeq 1.6\, \kf$, where $p_\Pi (\kGW/\kf) \simeq 0.5$, as can be computed by solving \Eq{Pi_conv} numerically with $\alpha = 2$ \cite{paper2}.
Note that previous numerical works reported a peak at $\kGW \approx 2\, \kf$ \cite{RoperPol:2018sap,RoperPol:2019wvy,RoperPol:2021uol,Brandenburg:2021bvg}.
This can be explained by the fact that they investigated the linear GW energy density $\EGW(k) = \OmGW(k)/k$,
which becomes flat: since the transition from constant to $k^{-11/3}$
of $p_\Pi (k/\kf)$ is slow, different characteristic GW wave numbers can be picked up by different functions.  

We find that the value of the GW spectrum at the peak always scales as $\OmMmaxsq (\HH_*/k_*)^{2}$.
Approximating $\ln^2 \, [1 + (k/\HH_*)^{-1}] \simeq (\HH_*/k)^2$, one gets, in fact, from \Eq{OmGW_envelope},
\begin{equation}
    \OmGW(\kGW, \tfin) \simeq
    A_\Omega\,
    \OmMmaxsq \,\biggl(\frac{\HH_*} {k_*}\biggr)^{2},
    \label{GW_peak}
\end{equation}
where the amplitude $A_\Omega$ is
\begin{align}
    A_\Omega (\alpha) = 
    3\, {\kGW\over \kf} \,{{\cal C} (\alpha) \over{\cal A}^{2}(\alpha)}
    p_\Pi \biggl({\kGW\over \kf}\biggr),
\end{align}
which gives $A_\Omega (\alpha = 2) \simeq 0.6$ \cite{paper2}.

To summarize, the main properties of the GW spectrum $\OmGW (k)$ derived in the approximation of a constant source operating over a time interval $\delta\tfin$ are the following.
\begin{itemize}
    \item There is a nearly linear increase in the region ${\rm max}(\HH_*,1/\delta\tfin)< k \lesssim \kGW$, which sharply transitions to the causal $k^3$ slope at $k< \kbr \equiv 1/\delta\tfin$ if the source is short ($\delta\tfin<\HH_*^{-1}$). If the source is long, the transition is smoother, logarithmic in the region $\kbr < k < \HH_*$.
    As we shall see, the transition toward the linear regime in the spectrum characterizing the GW signal from MHD can occur in the PTA frequency range; notably, it can occur at
    $\HH_{\rm QCD}$.
    \item The GW signal peaks at $\kGW\simeq 1.6 \,k_*$, and the scaling of the GW spectrum amplitude at the peak is $\OmGW\propto \OmMmaxsq (\HH_*/k_*)^{2}$, regardless of whether the source is long or short.
\end{itemize}
The first property is in agreement with what was theoretically derived in Ref.~\cite{Caprini:2009fx} for a coherent source with instantaneous turn on, and both properties were observed in  the simulations of Refs.~\cite{RoperPol:2019wvy,RoperPol:2021uol, Brandenburg:2021bvg}. 
In \Sec{sec:numerical_spec}, we validate the spectral shape of \Eq{OmGW_constant_tfin} and its envelope [\cf \Eq{OmGW_envelope}] with a set of dedicated simulations.
We study the dynamics of the proposed model in further detail
in a separate publication \cite{paper2}.

Previous semianalytical analyses of (M)HD turbulence predicted different 
spectral shapes and scaling with the parameters $\OmM^*$ (or $\OmK^*$ in the
case of kinetic turbulence) and $k_*$. 
The main difference with what is proposed here resides in the fact that 
we assume a constant-in-time magnetic stress, while previous analyses accounted for some form of time decorrelation of the source.
To give some examples, the UETC were modeled with the top hat ansatz in Ref.~\cite{Caprini:2009yp}, providing a scaling as $\OmGW\propto {\OmMmax}^{3/2} (\HH_*/k_*)$, and with the Kraichnan random sweeping model in Ref.~\cite{Niksa:2018ofa}, providing a scaling as $\OmGW\propto {\OmK^*}^{3/2} (\HH_*/k_*)^2$.
The typical scaling of GW production by sound waves, $\OmGW\propto K^2 (\HH_*/k_*)^2/(\sqrt{K}+\HH_*/k_*)$, where $K$ denotes the normalized kinetic energy, cannot be reproduced by our model either, as it is typical of a stationary, decorrelating source \cite{Hindmarsh:2019phv,Hindmarsh:2020hop}. 
An heuristic model inspired by this scaling was adopted in Ref.~\cite{Caprini:2019egz}, providing for turbulence the scaling $\OmGW\propto {\OmK^*}^{3/2} (\HH_*/k_*)$ for a long source and $\OmGW\propto {\OmK^*} (\HH_*/k_*)^2$ for a short source.

At $t>\tfin$, the source stops operating and the GW energy density only decreases due to the expansion of the Universe.
The GW energy density today, normalized to the critical energy density today, becomes
\begin{align}
    h^2\OmGW^0 &\, (k) = \biggl(\frac{a_{\rm fin}}{a_0}\biggr)^4
    \biggl(h \frac{H_{\rm fin}}{H_0}\biggr)^2
    \OmGW (k,\tfin) \nonumber \\
    \simeq &\, 3.5 \times 10^{-5}\,  \OmGW (k,\tfin) \biggl(\frac{10}{g_{\rm fin}}\biggr)^{1\over3},
    \label{eq:evolGW}
\end{align}
where the factor $(a_{\rm fin}/a_0)^4$ gives the ratio between
the GW energy
density at $\tfin$ and today
and $(H_{\rm fin}/H_0)^2$ is the
ratio between the critical energy densities,
being $H_0 = 100\,h \km\s^{-1}\Mpc^{-1}$ the Hubble
rate today with $h \simeq 0.68$ \cite{Planck:2018vyg}.
The prefactor in \Eq{eq:evolGW} is computed
using \cite{Kolb:1990vq}
\begin{align}
    \frac{a_{\rm fin}}{a_0} = &\, \frac{T_0}{T_{\rm fin}}\biggl(\frac{g_0}{g_{\rm fin}}\biggr)^{1\over3},
    \label{arat}
    \\
    H_{\rm fin} = &\, \sqrt{\frac{4 \pi^3 G}{45 \hbar^3}} g_{\rm fin}^{1\over2} T_{\rm fin}^2,
    \label{Hstar}
\end{align}
where we take the entropic degrees of freedom and temperature today
to be $g_0=3.91$ and $T_0= 2.755$\,K, respectively.
The gravitational and reduced Planck constants are
$G = 2.76 \times 10^{-53} \J^{-1} \s$ and $ \hbar = 1.05 \times 10^{-34} \J \s$, respectively.
At the QCD phase transition $T_* \sim 100 \MeV$, we take the entropic and relativistic degrees of freedom to be equal with value $g_{\rm fin} \approx g_* \sim 10$ \cite{Kolb:1990vq}.

\subsubsection{Analysis of NANOGrav results with the analytical GW spectrum model}

A thorough comparison of the MHD GW signal with the common noise reported 
by NANOGrav \cite{NANOGrav:2020bcs}, PPTA \cite{Goncharov:2021oub},
EPTA \cite{Chen:2021rqp}, and IPTA \cite{Antoniadis:2022pcn} is performed in \Sec{sec:comparison_data}, after
we present the results of simulations in \Sec{sec:numerical_spec}.
In this subsection, we redo the analysis of Ref.~\cite{Neronov:2020qrl}, in order to show how it changes with the more 
accurate model of the SGWB spectrum given in \Eq{OmGW_envelope}.

Note that \Eq{OmGW_envelope} corresponds to the envelope of the SGWB spectrum: 
indeed, the SGWB in \Eq{OmGW_constant_tfin} is rapidly oscillating. 
In principle, time averages appropriate to the specific GW observatory and its detection strategy should be performed, but here and in \Sec{sec:comparison_data}, for simplicity, we are comparing the PTA data with the SGWB envelope. 
This procedure is conservative.

In Ref.~\cite{Neronov:2020qrl}, the analysis was simplified
setting a reference amplitude for the NANOGrav observation
of $h^2 \OmGW^{\rm ref}=10^{-9}$ at $f_{\rm yr}=3\times 10^{-8} \Hz$ and comparing it with an order-of-magnitude estimate of the GW signal from MHD turbulence, obtained by taking $\OmGW \sim \OmMmaxsq (\HH_* l_*)^2$ for the GW energy density at the peak frequency $f_{\rm GW}\sim 2/l_*$.
In \Sec{sec:analytical_detail}, we have demonstrated that a more
careful analysis of the GW signal, which we validate with the simulations
in \Sec{sec:numerical_spec}, leads instead to 
$\OmGW \simeq A_{\Omega'}\, \OmMmaxsq (\HH_* l_*)^2$ with 
$A_{\Omega'} = A_\Omega/(2\pi)^2 \simeq 1.5 \times 10^{-2}$ [\cf \Eq{GW_peak}],
and $f_{\rm GW}= 2B/l_*$ with $B\simeq 0.8$ (corresponding to $\kGW\simeq 1.6\,k_*$).

Using \Eqs{GW_peak}{eq:evolGW}, one finds the amplitude of the GW signal at the peak at present time, for a signal produced at the QCD phase transition epoch:
\begin{align}
    h^2 \OmGW^0 (\kGW) \simeq 
    &\, 
    \, 3.5 \times 10^{-5}
    \nonumber \\
    \times & \,  A_{\Omega'} \, \OmMmaxsq (\HH_* l_*)^2 
    \biggl({10\over g_*}\biggr)^{1\over 3}.
\end{align}
If we take the characteristic scale to be the largest processed eddies of
the magnetic field $\HH_*\left.l_*\right|_{\rm LPE} = \sqrt{\threehalf \OmMmax}$, as we assumed in Ref.~\cite{Neronov:2020qrl}, we get
\begin{align}
   \left. h^2 \OmGW^0 (\kGW)\right|_{\rm LPE} \simeq &\,\, 5.3 \times 10^{-5}  \nonumber \\
    \times &\, A_{\Omega'} \, \OmMmaxcub \biggl({10\over g_*}\biggr)^{1\over 3}.
    \label{OmGW_Aomega}
\end{align}
Furthermore, the peak of the GW spectrum at $\kGW=1.6\,k_* = 2B\,\kf$ translates at present time into the peak frequency
\begin{align}
    \fGW= 
    \frac{2B}{l_*} \simeq &\,\, 2.24 \times 10^{-8}
    \nonumber \\
    \times &\,     
    \frac{B}{\HH_* l_*} \frac{T_*}{100 \MeV}
    \biggl(\frac{g_*}{10}\biggr)^{1\over 6}\, \Hz ,
    \label{fGW}
\end{align}
where we have used the characteristic Hubble frequency
\begin{equation}
    \HH_* \simeq  1.12 \times 10^{-8} \,
    \frac{T_*}{100 \MeV}
    \biggl(\frac{g_*}{10}\biggr)^{1\over 6}\, \Hz.
    \label{HHstar}
\end{equation}
For the largest processed eddies,
\begin{align}
	\left.\fGW \right|_{\rm LPE} \simeq &\,\, 1.8 \times 10^{-8} \nonumber \\
	\times &\, \, \frac{B}{\sqrt{\Omega_{\rm M}^{*}}} \, \frac{T_*}{100 \MeV} \biggl(\frac{g_*}{10}\biggr)^{1\over 6}
	\, \Hz.
\end{align}

According to the model of \Sec{sec:analytical_detail}, $h^2\OmGW^0 \propto f$ at frequencies $f\lesssim \fGW$.
To reproduce the estimate of Ref.~\cite{Neronov:2020qrl}, we fix the magnetic characteristic scale to the largest processed eddies and 
adopt the aforementioned reference amplitude and frequency of the NANOGrav observation to compute the
amplitude of the magnetic field that could account for it: 
\begin{align}
     \left. h^2 \OmGW^0 (\kGW)\right|_{\rm LPE}  \simeq  & \,\, 10^{-9} \frac{\left.\fGW \right|_{\rm LPE}}{3 \times 10^{-8}\,\Hz}, \nonumber \\
    5.3 \times 10^{-5} A_{\Omega'}\, \OmMmaxcub \simeq &\,\, 6.1 \times 10^{-10}\, B\,\, {\Omega_{\rm M}^{*}}^{-1/2},
    \nonumber \\
    \OmMmax \simeq \, 0.04\, (B/ & A_{\Omega'})^{2/7} \simeq \, 0.12.
    \label{eq:OmMNANO}
\end{align}
In
Ref.~\cite{Neronov:2020qrl}, this was  estimated  to be $\OmMmax = 0.03$, setting $B = A_{\Omega'} = 1$ and using $\vA = \sqrt{2\OmMmax}$ .
Using the most accurate GW signal model developed in \Sec{sec:analytical_detail}, it turns out that one needs a higher magnetic field amplitude to explain the NANOGrav observation at the largest processed eddies scale and for $T_*=100$ MeV.
The value in \Eq{eq:OmMNANO} exceeds the nucleosynthesis bound \cite{Shvartsman:1969mm, Grasso:1996kk, Kahniashvili:2009qi}:
we confirm this finding in \Sec{sec:comparison_data}; \cf
\Fig{OmGW_nanograv}. 
We will find that one needs to consider smaller $T_*$, or characteristic scales $l_*< \left.l_*\right|_{\rm LPE}$, to have a signal compatible with PTA observations.

\subsection{GW spectrum from MHD simulations}
\label{sec:numerical_spec}

In this section, we present the MHD simulations we have performed,
which are listed in \Tab{summary} with their characteristics.
We  use the
{\sc Pencil Code} \cite{PencilCode:2020eyn} to evolve the magnetic field via \Eqss{dlnrhodt}{dBdt} and compute the SGWB spectrum sourced by the magnetic field via \Eq{GW1_Fourier},
following the methodology\footnote {In particular, we use the methodology
described in Sec.~2.6 of Ref.~\cite{RoperPol:2018sap}, which is denoted 
there as approach II.} of 
Refs.~\cite{RoperPol:2018sap, RoperPol:2019wvy}. 
We do so for a range of parameters 
$\OmMmax$ and $\kf$, to accurately study the resulting GW spectra
and compare them with the prediction of the model derived in \Sec{analytical_spec} under the assumption
of constant magnetic stresses.
The simulations are initiated with a fully developed stochastic and
nonhelical magnetic field according to \Eqss{equaltime}{B_stoch}, and zero initial velocity field. The magnetic field later decays following the turbulent MHD description.

\begin{table*}[t!]
\renewcommand{\arraystretch}{1.5}
{\scriptsize
\vspace{12pt}\centerline{\begin{tabular}{|l|cccccccccc|}
\hline
Run & $\OmMmax$ & $\kf\HH_*^{-1}$ & $\HH_* \delta t_e$ & $\HH_* \delta \tfin$ & $\Omega^{\rm num}_{\rm GW} (k_{\rm GW})$
& $[\Omega^{\rm env}_{\rm GW}/\Omega^{\rm num}_{\rm GW}] (\kGW)$ &
$n$ & $\HH_*L$ & $\HH_*t_{\rm end}$ & $\HH_* \eta$\\
\hline
 A1 & $9.6 \times 10^{-2}$ & 15 & 0.176 & 0.60 & $2.1 \times 10^{-9}$ & 1.357 & 768 & $6\pi$ & 9 & $10^{-7}$ \\
 A2 & -- & -- & -- & -- & -- & -- & 768 & $12\pi$ & 9 & $10^{-6}$ \\
\hline
 B & $1.0 \times 10^{-1}$ & 11 & 0.233 & 0.60 & $4.0 \times 10^{-9}$ & 1.250 & 768 & $6\pi$ & 8 & $10^{-6}$ \\
\hline
 C1 & $9.9 \times 10^{-2}$ & 8.3 & 0.311 & 0.75 & $5.6 \times 10^{-9}$ & 1.249 & 768 & $6\pi$ & 8 & $10^{-6}$ \\
 C2 & -- & -- & -- & -- & -- & -- & 768 & $12\pi$ & 10 & $10^{-7}$ \\
\hline
 D1 & $1.1 \times 10^{-1}$ & 7 & 0.354 & 0.86 & $1.1 \times 10^{-8}$ & 1.304 & 768 & $6\pi$ & 5 & $10^{-7}$ \\
 D2 & -- & -- & -- & -- & -- & -- & 768 & $12\pi$ & 9 & $10^{-7}$ \\
\hline
 E1 & $8.1 \times 10^{-3}$ & 6.5 & 1.398 & 2.90 & $5.5 \times 10^{-11}$ & 1.184 & 512 & $4\pi$ & 8 & $10^{-7}$ \\
 E2 & -- & -- & -- & -- & -- & -- & 512 & $10\pi$ & 18 & $10^{-7}$ \\
 E3 & -- & -- & -- & -- & -- & -- & 512 & $20\pi$ & 61 & $10^{-7}$ \\
 E4 & -- & -- & -- & -- & -- & -- & 512 & $30\pi$ & 114 & $10^{-7}$ \\
 E5 & -- & -- & -- & -- & -- & -- & 512 & $60\pi$ & 234 & $10^{-7}$ \\
\hline
\end{tabular}}}
\caption{
Summary of runs.}
\label{summary}\end{table*}

Guided by the findings of 
\Sec{analytical_spec} and of Ref.~\cite{Neronov:2020qrl}, in runs~A--D we have chosen for initial conditions a characteristic scale
$k_* \HH_*^{-1}$
near the Hubble horizon $2\pi$, and total magnetic energy density
around 10\% the total radiation energy density at the time of generation $\OmMmax \simeq 0.1$. 
These runs have eddy turnover times in the range
$\HH_* \delta t_e \in (0.17, 0.4) $. The evolution of the magnetic field
is expected to play a role at wave numbers 
below the
peak of the GW spectrum, since $\vA \simeq 0.4$; \cf \Sec{analytical_spec}.
To check the validity of the model developed in \Sec{analytical_spec}
also in the limit of large $\delta \te$,
we have included runs~E, which feature a smaller value of $\OmMmax \simeq 10^{-2}$ and a
characteristic wave number $\kf$, again close to the Hubble
horizon $2\pi \HH_*$, corresponding to an eddy turnover time of $1.4\, \HH_*^{-1}$.

The simulations in the present work use
a periodic cubic domain of comoving size $\HH_* L$ with a
discretization of $n^3$ mesh points
(see \Tab{summary}), such that the smallest wave number computed
is $k_0 = 2\pi/L$.
We have chosen $L$ and $n$ such that the resulting dynamical range
above the spectral peak (at $k > \kGW$) allows an accurate prediction of
the dynamical evolution of the velocity and magnetic fields.
At the same time, since we are particularly interested in the GW spectral region around $\HH_*$ and below, we need to use domains of size
$\HH_* L > 2\pi$
(see \Tab{summary}).
Following Ref.~\cite{RoperPol:2019wvy}, we fix the viscosities $\nu = \eta$
and choose them to be as small as possible (see \Tab{summary}), in order to appropriately resolve the inertial range \cite{Brandenburg:2017neh}.
The numerical values of the viscosities are still much larger than their physical values at the QCD epoch,\footnote{At the QCD scale $T_* \sim 100\MeV$, we
can use Eq.~(1.11) of Ref.~\cite{Arnold:2000dr}, adapted in Eq.~(19) 
of Ref.~\cite{Brandenburg:2017rcb}, to get $\eta \sim 4\times 10^{-6}
(T_*/100 \MeV)^{-1} \cm^2 \s^{-1}$, which corresponds to $\HH_* \eta \sim 2.91
\times 10^{-23} (T_*/100\MeV)(g_*/10)^{1/2}$ in our normalized
units \cite{RoperPol:2019wvy}.} which would require a much larger resolution.
We are anyway able to properly resolve the interesting part of the inertial range, which is closest to the peak (very high frequencies are of little observational interest since
the GW amplitude at those frequencies is several orders of magnitude smaller).

\begin{figure*}[t]
    \centering
    \includegraphics[width=.49\textwidth]{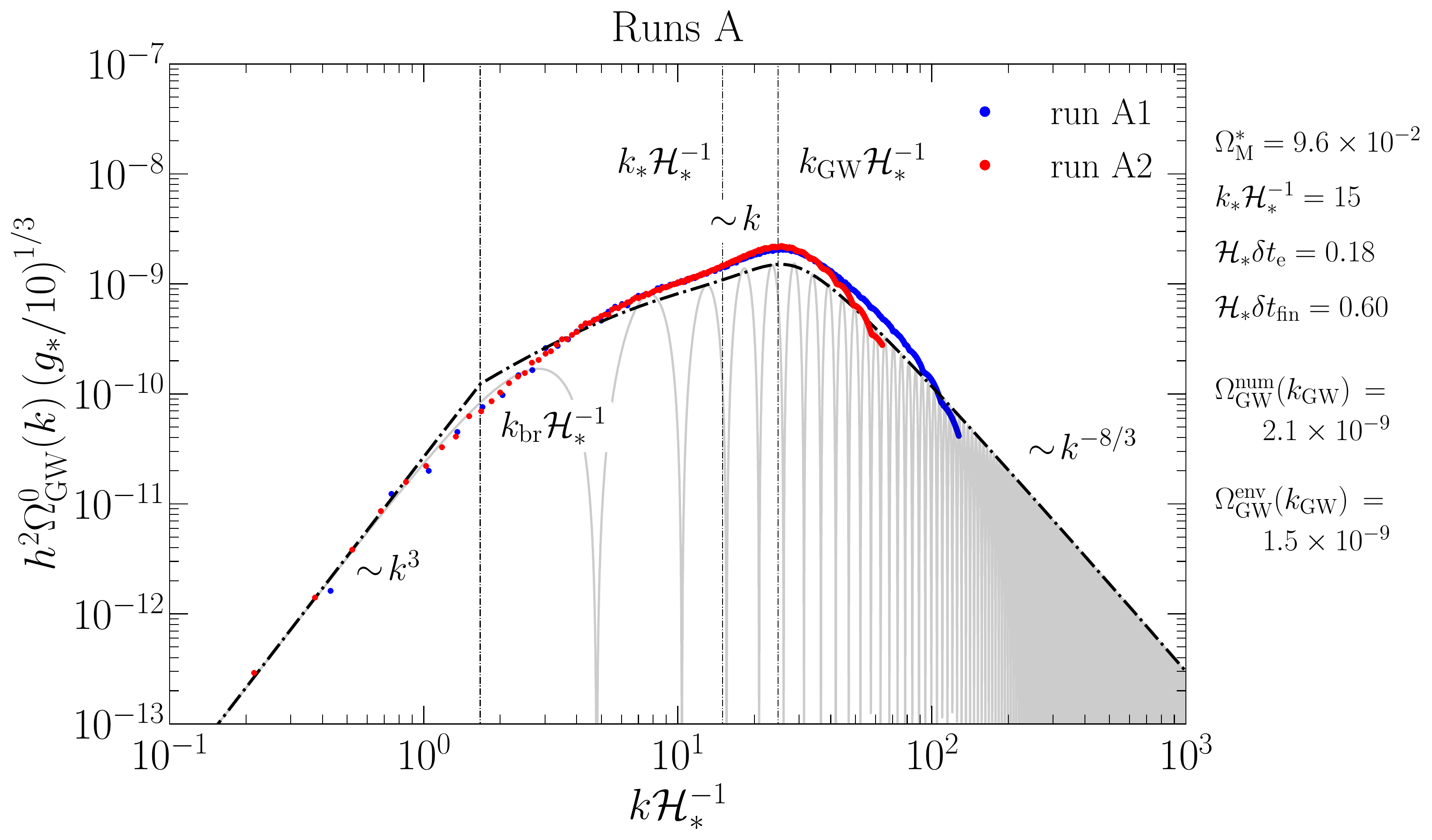}
    \includegraphics[width=.49\textwidth]{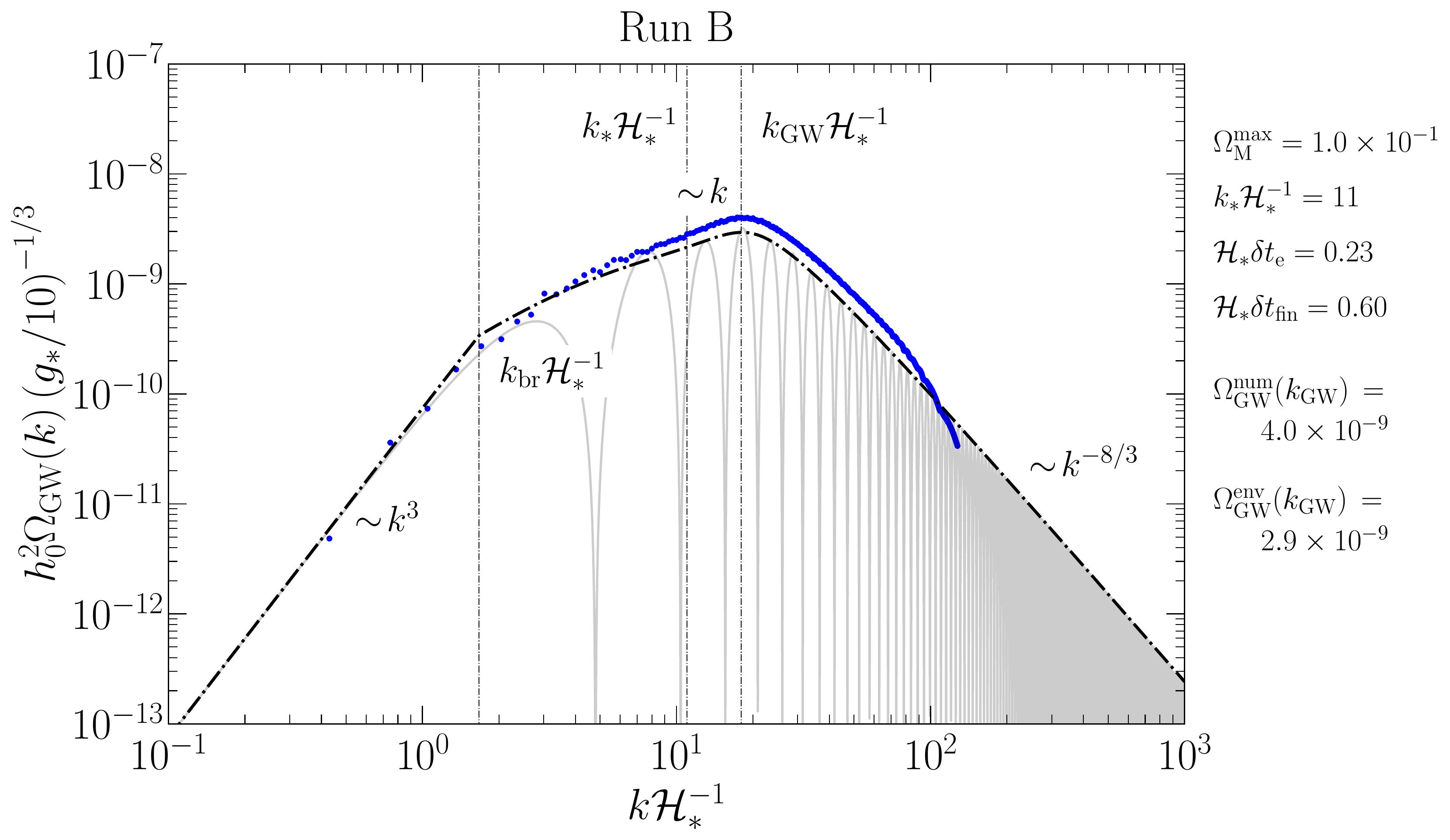}
    \includegraphics[width=.49\textwidth]{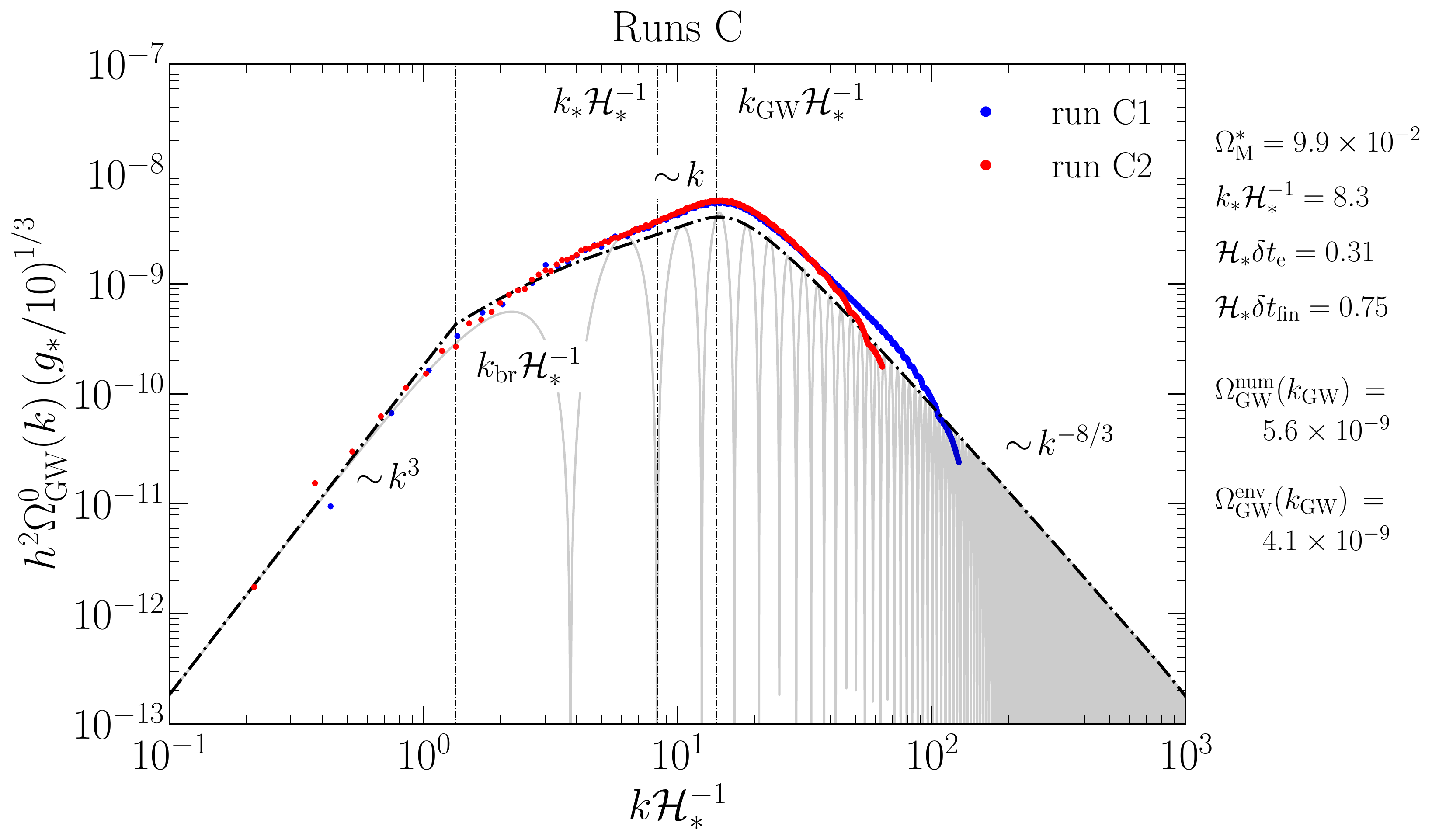}
    \includegraphics[width=.49\textwidth]{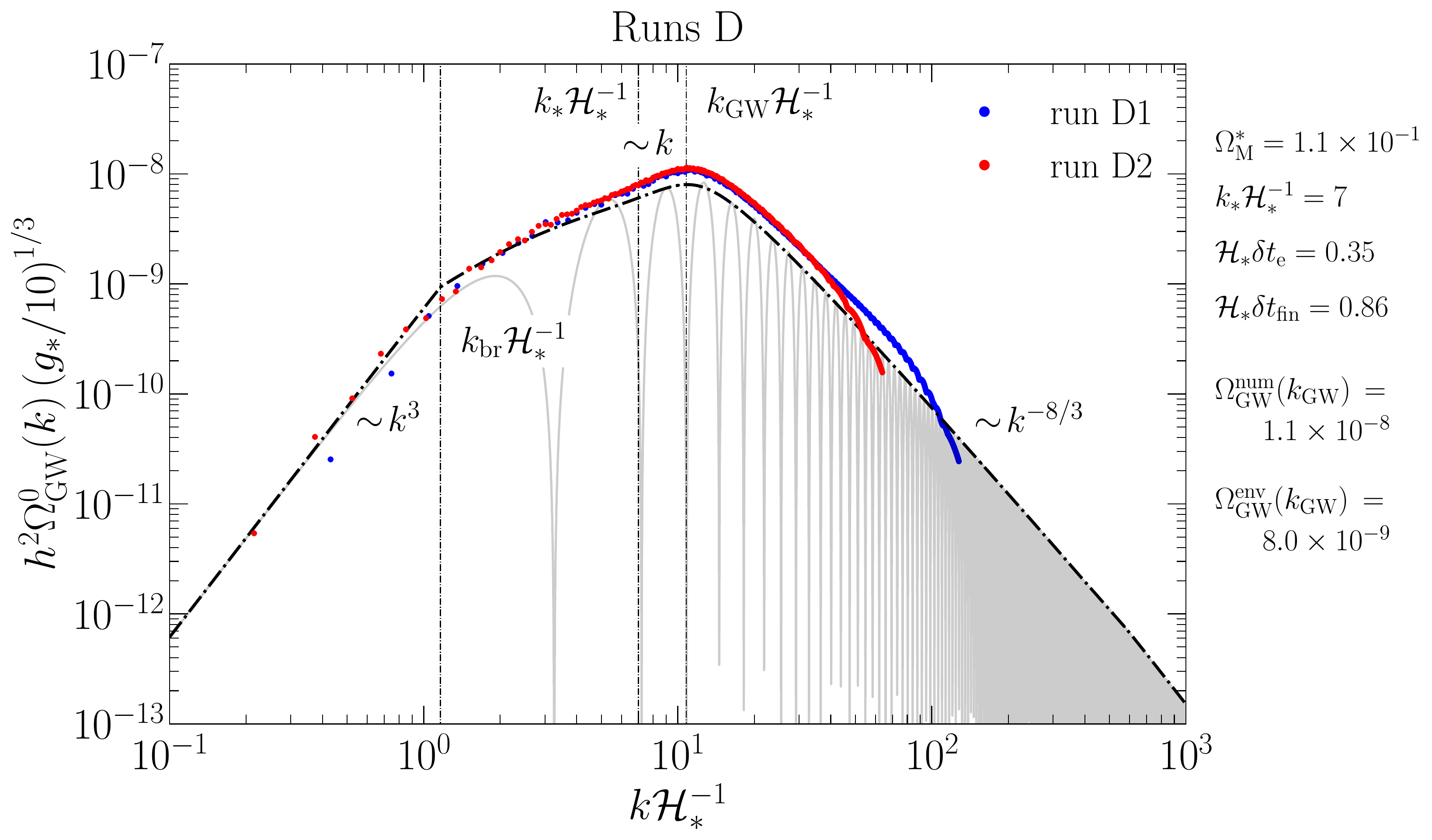}
    \includegraphics[width=.6\textwidth]{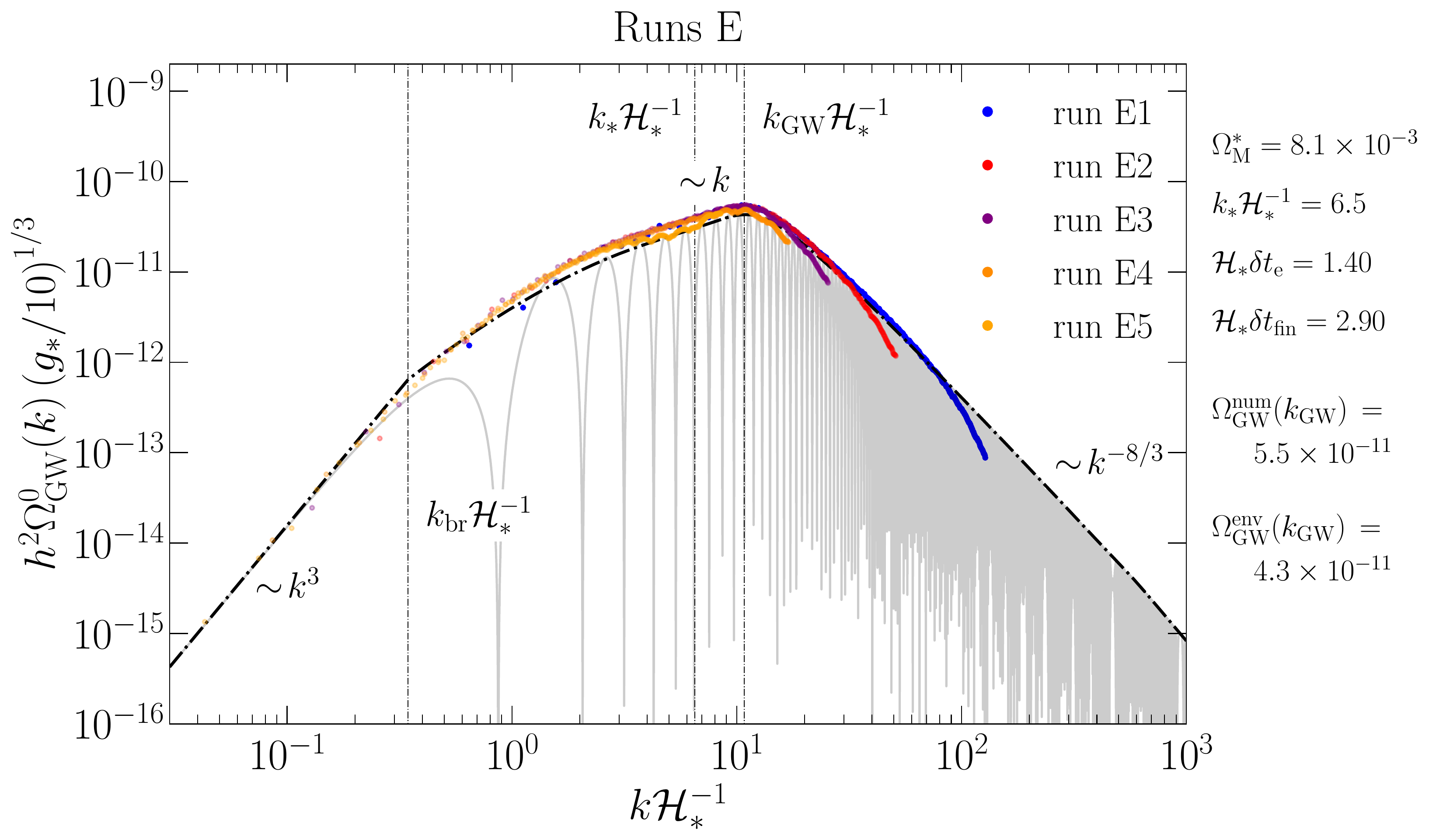}
    \caption{
    Simulated GW spectra $h^2 \OmGW^0 (k)$ of runs A--E (dots) compared to the analytical model developed in \Sec{analytical_spec} assuming constant magnetic stresses: \Eq{OmGW_constant_tfin}  at
    $\tfin$ (thin gray lines) and its envelope  \Eq{OmGW_envelope} (dot-dashed black
    lines).
    The maximal values over one oscillation of the numerical outputs at each wave number are shown in different colors for
    runs with different domain sizes, and are combined to
    show the GW spectra from subhorizon scales up to the scales
    where the inertial range is developed. Runs A1--E1 (blue dots) are computing the smallest scales resolved, up to a
    Nyquist wave number of $k_{\rm Ny}\HH_*^{-1} = 126$.
    The spectra are shown in terms of $k\HH_*^{-1}$ and compensated
    by $(g_*/10)^{-{1\over3}}$, such that they can be scaled
    to the specific value of the comoving Hubble rate at the time of generation $\HH_*$ and to different values of $g_*$. 
    }
    \label{OmGW_num_analyt2}
\end{figure*}

The GW spectra resulting from the simulations
are shown in \Fig{OmGW_num_analyt2}, together with the analytical solution obtained
at $\tfin$ [\cf \Eq{OmGW_constant_tfin}] and its envelope [\cf \Eq{OmGW_envelope}].
As observed in previous simulations (see, e.g., Refs.~\cite{RoperPol:2019wvy,RoperPol:2021uol}) and explained in \Sec{analytical_spec}, we confirm that the $k$ modes of the GW spectra initially 
grow in time as $\delta t^2$ and, after a time $\delta t \sim 1/k$, they start to oscillate
around a stationary value.
To plot the GW spectra in \Fig{OmGW_num_analyt2}, we choose the maximal amplitude of such oscillations for each wave number, since
we are interested in the envelope of the oscillations.
In previous simulations, the low wave number regime
was not captured, due to the size of the domains
\cite{RoperPol:2019wvy,RoperPol:2021uol,Brandenburg:2021tmp}.
Here, we have increased the size of the domain and we have run the simulations for long times (see \Tab{summary}, where $\tend$ denotes the end of the simulations),
such that even the smallest wave numbers of the box reach the oscillatory regime.  
Since $\tend\gg \delta\tfin\gtrsim \delta\te$, in the GW spectra of \Fig{OmGW_num_analyt2}, we observe  both the causal $k^3$ slope, expected at $k< \kbr \equiv 1/\delta\tfin$, following the model of constant
magnetic stresses presented in \Sec{sec:analytical_detail}, and  the transition toward the regime that is linear in $k$, expected at
$\max(\HH_*, 1/\delta\tfin) <k \lesssim \kGW$. 
The latter was also observed in previous numerical simulations \cite{RoperPol:2019wvy, Kahniashvili:2020jgm,
RoperPol:2021uol, Brandenburg:2021bvg, Brandenburg:2021tmp}.

In order to investigate the spectrum at the smallest wave numbers,
we have performed several runs with the same parameters $k_*$ and $\OmMmax$, but with different
sizes of the cubic domain: respectively, runs~A1 and A2, C1 and C2, D1 and D2, and
runs~E1--E5.
Runs A1, B1, C1, and D1 have a resolution of $768^3$ mesh points
and a size $\HH_* L = 6\pi$, which corresponds to $k_0\HH_*^{-1} = 1/3$
and a Nyquist wave number of $k_{\rm Ny}\HH_*^{-1} = 126$.
For runs~A, C, and D, we have performed a second set of runs, with the same initial conditions but domains doubling the size of A1, C1, and D1, so that the smallest wave number is $k_0\HH_*^{-1} = 1/6$.
We reconstruct the final GW spectrum by combining the results
of the multiple simulations.
This allows us to compute more discretized modes of the GW spectra
in different wave number ranges.
From \Fig{OmGW_num_analyt2}, one appreciates that we can accurately reproduce the break from $k^3$ to $k^1$.

Runs~E have the largest eddy turnover time; hence, the $k^3$ regime
is expected to occur at smaller wave numbers. 
We have then performed four additional runs, with the largest domain (E5) being
15 times larger than the initial one (E1), corresponding to $k_0\HH_*^{-1} = 1/30$.
Runs~E have $\HH_*\delta\tfin > 1$: the transition from the $k^3$ to the $k^1$ regimes should therefore be smoother, according to the constant stress model, and develop a logarithmic dependence in the region $1/\delta\tfin<k<\HH_*$.  
Indeed,
the GW spectrum follows the curve predicted by the
analytical model, i.e., $k^3 \ln^2(1 + \HH_*/k)$, and the transition of
this curve toward the $k^3$ regime occurs around the wave number $k_{\rm br}\HH_*^{-1} = (\HH_* \delta\tfin)^{-1} \simeq 0.3$.

\subsection{Fit of the analytical to the numerical GW spectra}
\label{sec:comparison_approx}

\FFig{OmGW_num_analyt2} shows that the analytical model based on the assumption of constant anisotropic stresses over the time interval $\delta \tfin$ accounts for most of the SGWB spectral features: the  
slopes [including the $k^3 \ln^2 (1 + \HH_*/k)$ increase characteristic of the constant source], the positions at
which the slopes change,
and the total amplitude.
In addition, it predicts accurately the early time evolution
of the spectra, starting with an initial phase of growth, proportional
to $\delta t^2$ and a subsequent oscillatory period, settling in after a time
$\delta t \sim 1/k$.
However, the analytical model does not provide a value for
$\delta \tfin$, which is related to the validity of the constant-in-time
magnetic stress approximation, and, hence, on the dynamical decay of the
turbulent magnetic field, characterized by the eddy turnover
time $\delta \te$.
Additionally, the numerical spectra have a smoother transition from $k^3$ toward the $k^3 \ln^2 (1 + \HH_*/k)$ curve than the piecewise envelope given in \Eq{OmGW_envelope}, 
leading
to larger values of the numerical GW amplitudes at the peak $\kGW$.
This is likely due to the fact that the source decays smoothly in time, instead of shutting down abruptly as we assume in the constant model. 

\begin{figure}[t!]
    \centering
    \includegraphics[width=.8\columnwidth]{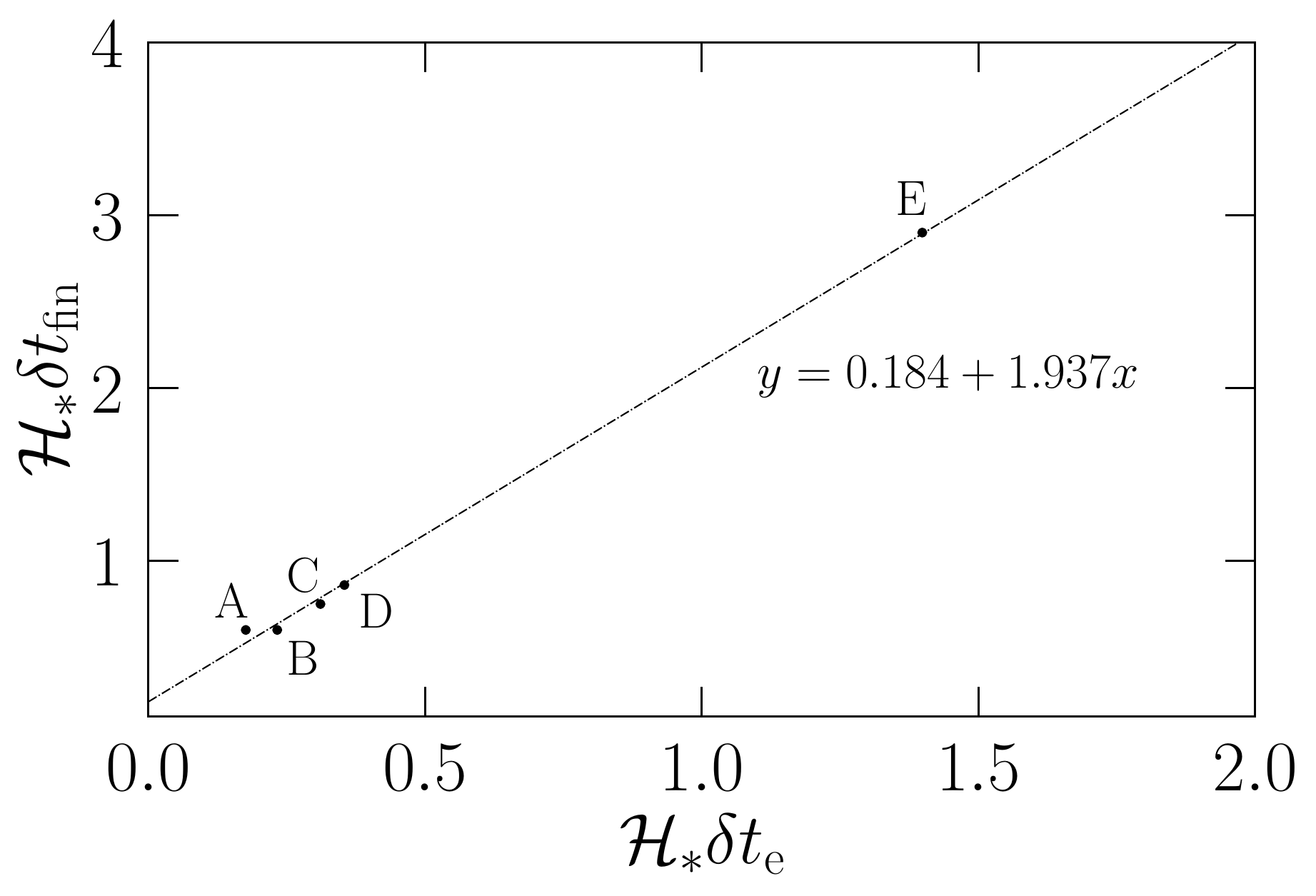}
    \includegraphics[width=.8\columnwidth]{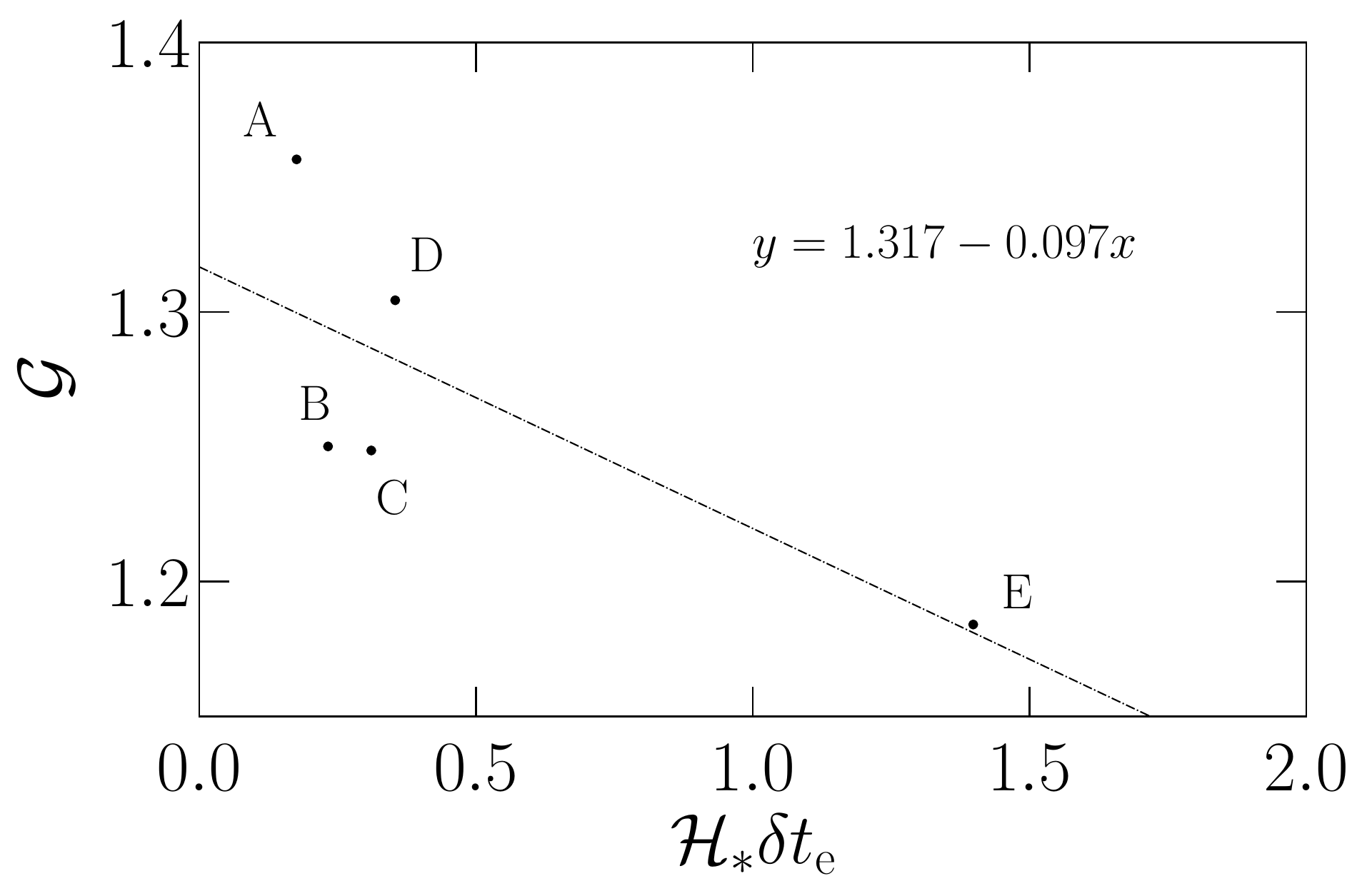}
    \includegraphics[width=\columnwidth]{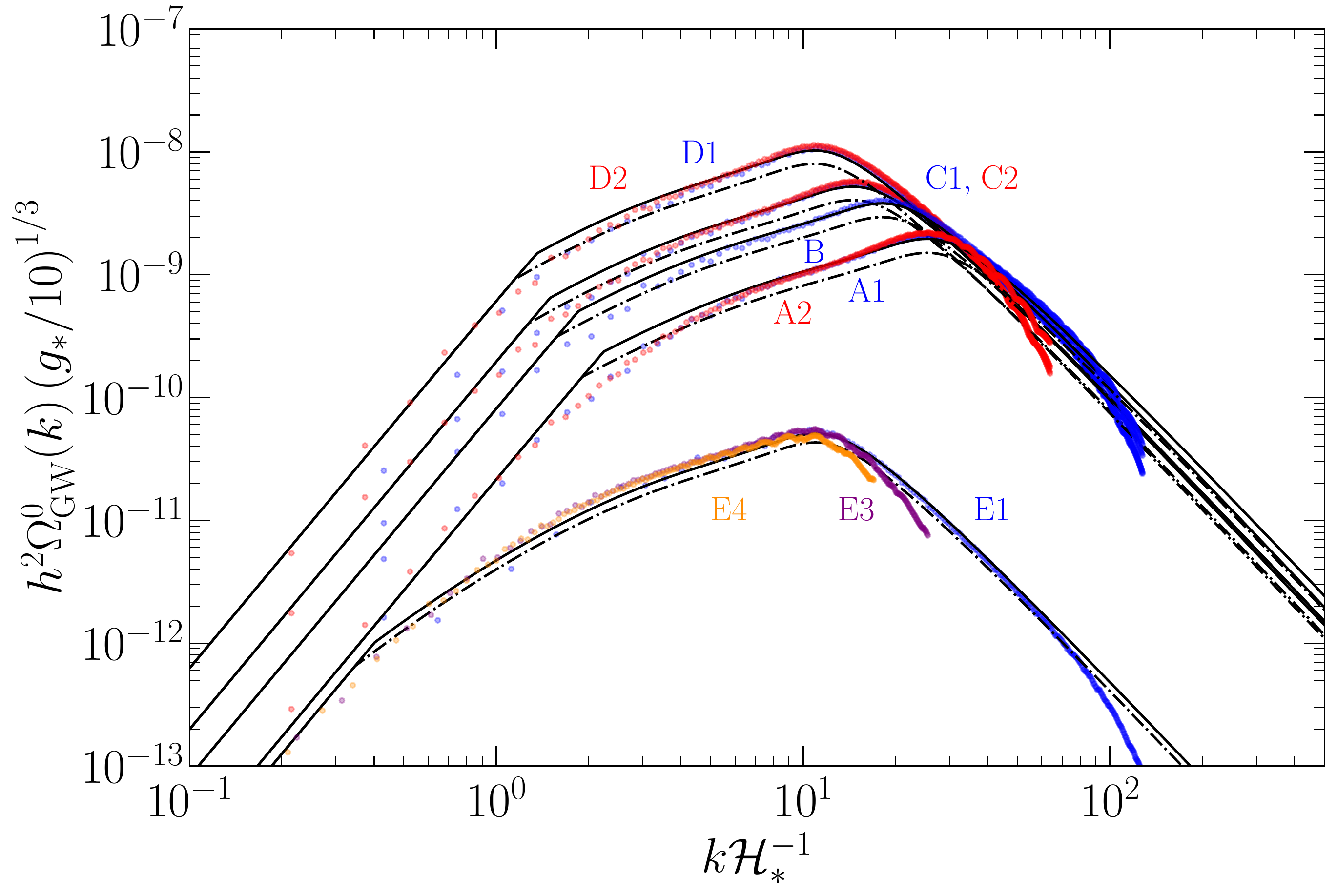}
    \caption{\emph{Upper Panel}: the points represent the eddy turnover times of the simulations
    $\delta \te$ and the corresponding values of $\delta \tfin$
    obtained by fitting the break into $k^3$ (occurring at $1/\delta \tfin$ according to the analytical model).
    The dashed line represents the fit of \Eq{delta_tfin_te}.  \emph{Middle panel}: the points represent the ratio ${\cal G}$ between the numerical
    and the analytical SGWB amplitudes at the peak.
    The dashed line represents the fit of \Eq{OmGW_num_env}.
    \emph{Bottom panel}: 
    the SGWBs computed with the analytical model given
    in \Eq{OmGW_envelope} (dot-dashed lines), and from the adjusted model given in \Eq{OmGW_empirical} (solid lines), are compared to the results of the MHD simulations (colored points). The compensated model uses the empirical
    fits shown in the upper and middle panels.
    }
    \label{dte_vs_dtf}
\end{figure}

Since the MHD turbulent decay occurs on a typical timescale of the order of the eddy turnover time, we expect the source duration parameter $\delta \tfin$ to be related to $\delta\te$. 
The specific values of $\delta\tfin$ used in the envelopes shown in \Fig{OmGW_num_analyt2} (see \Tab{summary}) have been extracted 
by fitting the analytical solution to the numerical spectra output from each simulation in the $k^3$ range.
In \Fig{dte_vs_dtf} (upper panel), we show $\delta \tfin$ inferred from the simulations vs $\delta \te$ and fit the linear relation 
\begin{equation}
    \delta \tfin = 0.184 \HH_*^{-1} + 1.937 \,\delta \te.
    \label{delta_tfin_te}
\end{equation}
Note that, in the limit $\delta \te \rightarrow 0$, this fit
yields a finite $\delta \tfin$, which is unphysical. 
Furthermore, we only have one simulated point in the region of large $\delta\te$. 
\EEq{delta_tfin_te} is therefore a tentative fit and should not be extrapolated outside the range of $\delta \te$ validated by the simulations. 

The discrepancy between the amplitude of the numerical and analytical GW 
spectra at the peak (see \Tab{summary}) is expected to decrease as 
the eddy turnover time increases, since the decay is slower and the assumption of a constant source is appropriate for a larger range of wave numbers. 
In the middle panel of
\Fig{dte_vs_dtf}, we show the ratio between 
$\OmGW^{\rm num} (\kGW)$ and  $\OmGW^{\rm env} (\kGW)$ [\cf\Eq{GW_peak}], which is decreasing, as expected, together with the following fit:
\begin{align}
    {\cal G} = \frac{\OmGW^{\rm num} (\kGW)}{\OmGW^{\rm env} (\kGW)} =
    1.317 - 0.097\, \HH_* \delta \te.
    \label{OmGW_num_env}
\end{align}

Altogether, the SGWB spectrum for given initial parameters $k_*$ and $\OmMmax$ can be obtained from the analytical model relying on constant stresses developed in \Sec{sec:analytical_detail} and, in particular, from \Eq{OmGW_envelope}, fixing $\delta \tfin$ via the empirical linear fit of \Eq{delta_tfin_te}, 
with the caveat, however, that this relation has only been validated in the range tested with the simulations. 

Additionally, one can compensate the second branch of the envelope in \Eq{OmGW_envelope}, i.e., the regime $1/\delta \tfin <k\lesssim \kGW$
proportional to $\ln^2 [1 + (\HH_*/k)]$, by the factor 
${\cal G}$
given by the empirical fit of \Eq{OmGW_num_env}.
This compensated model of the envelope of the GW spectrum
at $\tfin$ reads
\begin{align}
     \OmGW  & (k,\tfin) = \, 3 \, \biggl(\frac{k}{k_*}\biggr)^3 \, \OmMmaxsq \frac{{\cal C} (\alpha)}{{\cal A}^{2}(\alpha)} 
     \,\,p_\Pi \biggl(\frac{k}{k_*}\biggr) \nonumber \\
     \times &  \left\{ \begin{array}{ll}
        \ln^2 [1 + \HH_*\delta \tfin]
        & \text{ if } k < \kbr^{\rm comp},\\
        {\cal G}
        \ln^2[1 + (k/\HH_*)^{-1}]
        & \text{ if } k 
        \geq \kbr^{\rm comp},
        \end{array} \right.
        \label{OmGW_empirical}
\end{align}
where the specific position of the compensated break $\kbr^{\rm comp}$ is moved from
$1/\delta \tfin$ to
\begin{equation}
    \kbr^{\rm comp}\, \HH_*^{-1} = \Bigl[(1 + \HH_* \delta \tfin)^{1/\sqrt{\cal G}} - 1\Bigr]^{-1},
\end{equation}
to ensure continuity in the envelope function after compensating
one of the branches by ${\cal G}$.
The envelopes of the GW spectra are shown in \Fig{dte_vs_dtf} (bottom panel),
both with and without compensating by ${\cal G}$,
together with the output of the simulations listed in \Tab{summary}.

Whether to use the compensated model \Eq{OmGW_empirical} or directly \Eq{OmGW_envelope} depends on the particular situation.
It can be appreciated from \Fig{dte_vs_dtf} that the uncompensated model
fits the numerical simulations better in the region below the spectral peak ($k < \kGW$), but it underpredicts the amplitude at the peak; while the compensated one fits the peak but overpredicts the spectra at smaller wave numbers.
Hence, the choice between one or the other model depends on which range of
wave numbers one prioritizes to reproduce with the highest accuracy.

\section{Comparison with PTA results}
\label{sec:comparison_data}

In this section, we adopt the analytical model developed in
\Sec{analytical_spec} and validated in \Sec{sec:numerical_spec} with MHD
simulations and compare the resulting SGWB with the observations reported by
the PTA collaborations \cite{NANOGrav:2020bcs, Goncharov:2021oub,Chen:2021rqp,Antoniadis:2022pcn},
thereby inferring the range of parameters $T_*$, $\kf$, and $\OmMmax$,
which could account for the PTA results.
We remind that we 
consider a nonhelical magnetic field
and assume that the GW production starts once the magnetic field 
has a fully developed turbulent spectrum.

\subsection{PTA results}
\label{PTA_res}

The three PTA collaborations NANOGrav, PPTA, and EPTA, and the IPTA Collaboration, have independently constrained the amplitude $A_{\rm CP}$
of a red common process (CP) to several pulsars
by fitting the power spectral density $S(f)$ to a single
power law (PL) of slope $-\gamma$, as
\cite{NANOGrav:2020bcs,Goncharov:2021oub,Chen:2021rqp,Antoniadis:2022pcn}
\begin{equation}
    S(f) = \frac{A_{\rm CP}^2}{12 \pi^2} \left( \frac{f}{\fyr} \right)^{-\gamma} \fyr^{-3},
\end{equation}
or to a broken PL as
\begin{align}
    S(f) = &\,
    \frac{A_{\rm CP}^2}{12\pi^2} \left( \frac{f}{\fyr} \right)^{-\gamma}
    \nonumber \\ \times &\, 
    \left[1 + \left(\frac{f}{f_{\rm bend}}\right)^{1\over\kappa}\right]^{\kappa\gamma} \fyr^{-3},
\end{align}
with $f_{\rm bend} = 1.035\times 10^{-8}$\,Hz and $\kappa = 0.1$.
The reference frequency corresponds to 1 yr,
$\fyr \simeq 3.17 \times 10^{-8} \Hz$.

The CP reported by NANOGrav, PPTA, EPTA, and IPTA,
characterized by the amplitude $A_{\rm CP}$ and the slope
$\gamma$,
does not show enough
statistical significance toward a quadrupolar correlation over pulsars,
following the Hellings-Downs curve, to
be ascribed to a SGWB \cite{Hellings:1983fr,NANOGrav:2020bcs,Goncharov:2021oub,Chen:2021rqp,Antoniadis:2022pcn}.
Interpreting the CP as an actual GW signal, 
the characteristic strain $\hc(f)$ of the corresponding single-PL SGWB would be
\begin{equation}
\label{eq:hc}
    \hc(f) = \sqrt{12 \pi^2 S(f) f^3} = A_{\rm CP}
    \left(\frac{f}{\fyr}\right)^{{3 - \gamma\over 2}},
\end{equation}
and the SGWB spectrum $\OmGW^0(f)$, defined in \Eqs{eq:OmGWdef}{eq:evolGW}, would be
\begin{equation}
    \OmGW^0(f) =  \Omyr \left(\frac{f}{\fyr}\right)^{\beta},
    \label{SGWB_spl}
\end{equation}
with
\begin{equation}
    \Omyr = \frac{2\pi^2}{3H_0^2} \fyr^2 A_{\rm CP}^2, \quad
    \beta = 5 - \gamma.
\end{equation}
Analogously, the GW spectrum for the broken PL would be
\begin{align}
    \OmGW^0(f) = &\, \Omyr \left(\frac{f}{\fyr}\right)^{\beta} 
    \nonumber \\ \times &\,
    \left[1 + \left(\frac{f}{f_{\rm bend}}\right)^{1\over\kappa}
    \right]^{\kappa(5 - \beta)}.
    \label{SGWB_bpl}
\end{align}

\begin{figure}[t!]
    \centering
    \includegraphics[width=.49\textwidth]{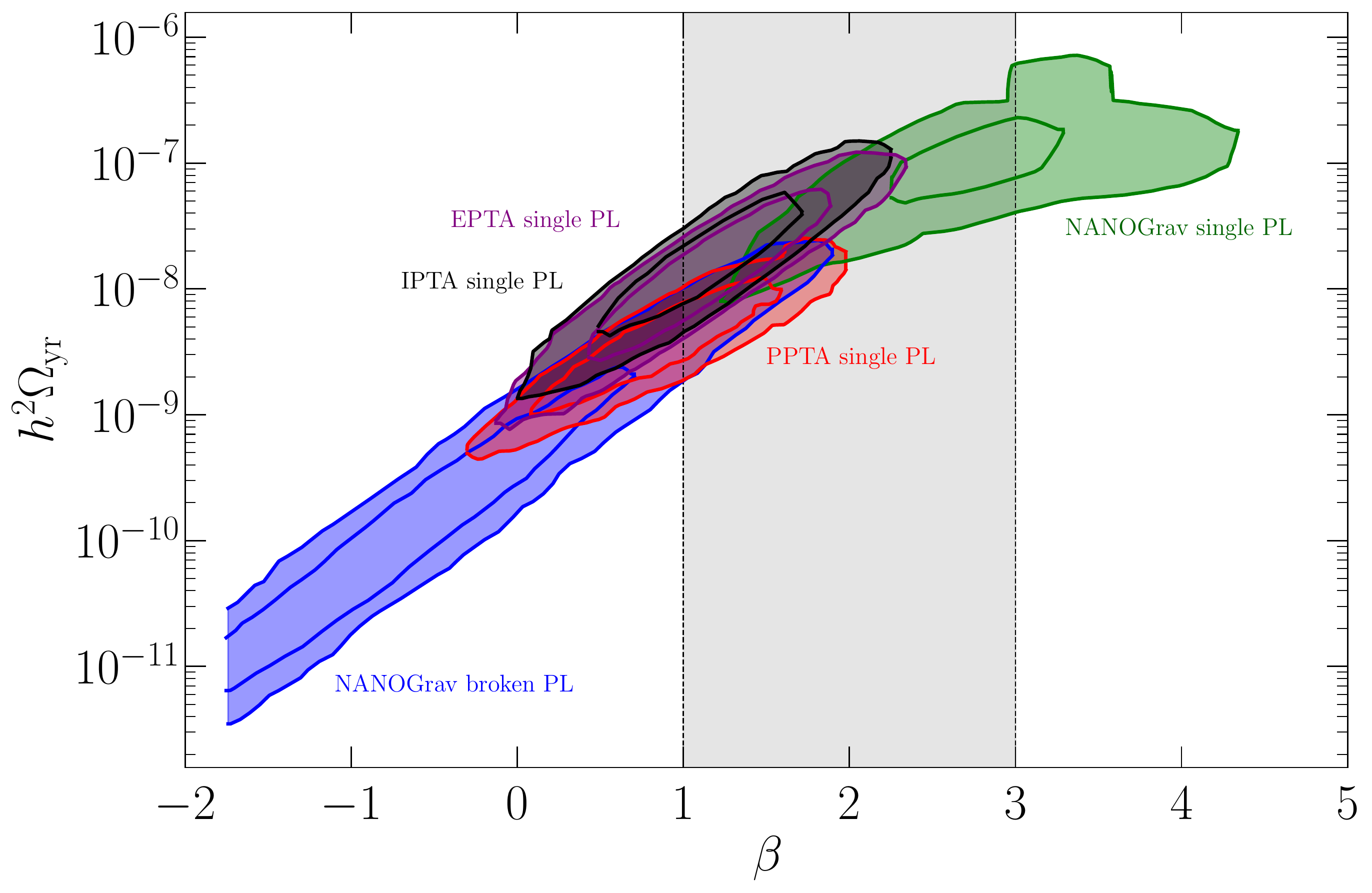}
    \includegraphics[width=.49\textwidth]{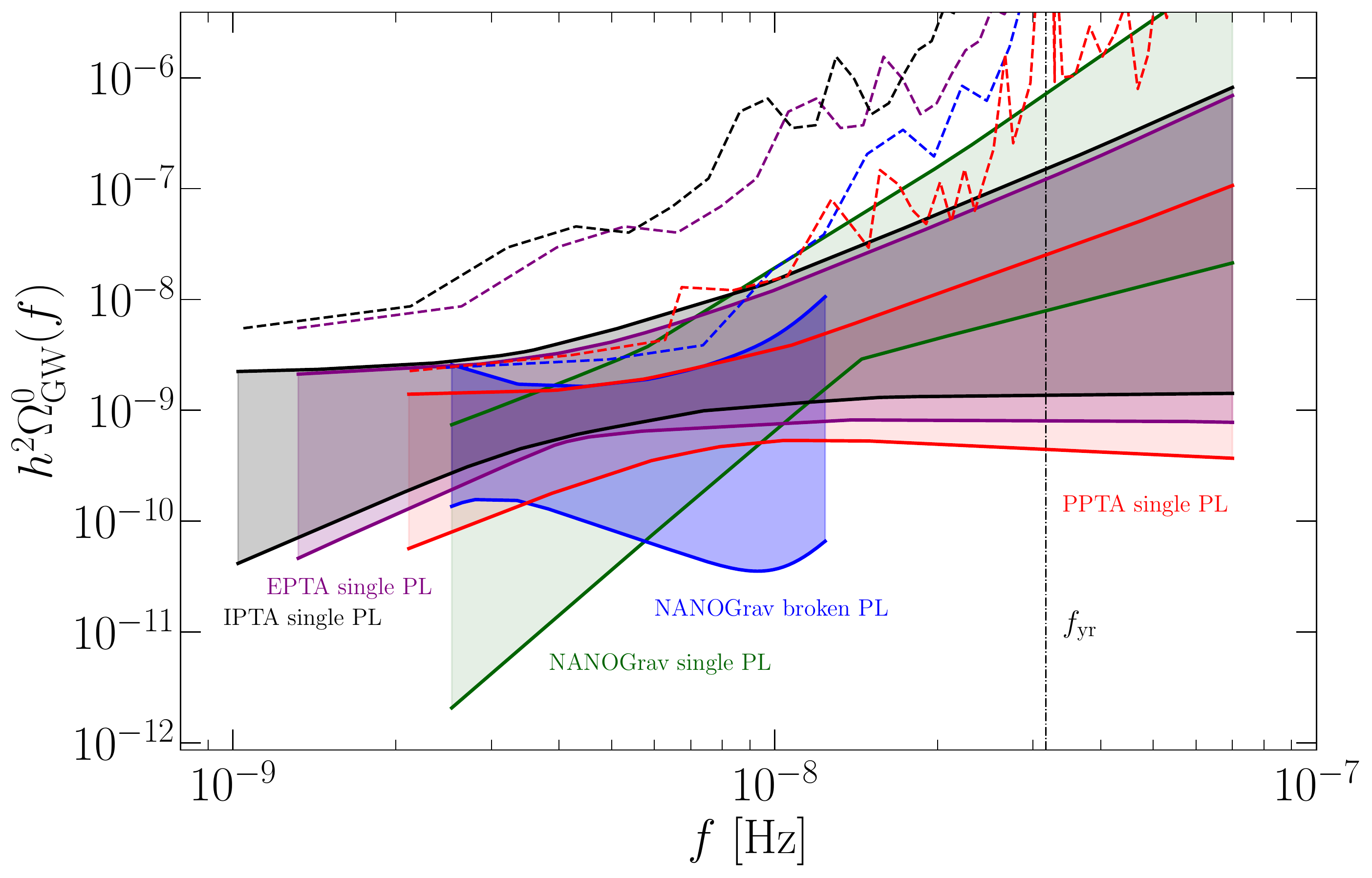}
    \caption{\emph{Upper panel}: 1$\sigma$ and
    2$\sigma$ contours of the amplitude $h^2 \Omyr$ vs slope $\beta$ [\cf \Eqs{SGWB_spl}{SGWB_bpl}] derived from the NANOGrav dataset for both the 
    broken-PL (blue) and single-PL (green) fits and from the PPTA (red), EPTA (purple), and IPTA (black) datasets for the
    single-PL fit
    \cite{NANOGrav:2020bcs,Goncharov:2021oub,Chen:2021rqp,Antoniadis:2022pcn}.
    The gray shaded area shows the slopes  $\beta \in (1, 3)$
    characteristic of the SGWB produced by primordial MHD
    turbulence below the spectral peak; \cf \Sec{sec:analytical_detail}.
    \emph{Lower panel}:
    shaded regions: range of the SGWB spectra $h^2 \OmGW^0(f)$ of \Eqs{SGWB_spl}{SGWB_bpl}, corresponding to the 2$\sigma$ contours given in the upper panel. 
    The vertical line shows the reference frequency $\fyr$.
    Dashed lines: $2\sigma$ maximum amplitude at each frequency---such that larger
    amplitudes are, in principle, excluded by the PTA observations.
    }
    \label{NANOGrav_PPTA}
\end{figure}

In \Fig{NANOGrav_PPTA} (upper panel), we reproduce the 1$\sigma$ and 2$\sigma$ contours of the amplitude $\Omyr$ as a function of slope $\beta$ reported
by NANOGrav using both the single- and broken-PL fits \cite{NANOGrav:2020bcs}, and by PPTA \cite{Goncharov:2021oub}, EPTA \cite{Chen:2021rqp}, and IPTA \cite{Antoniadis:2022pcn}
using the single-PL fit.
The PTA collaborations present their data in terms of Fourier components of the timing
spectrum of the CP. 
The frequency of the first Fourier mode corresponds to the inverse total observation time,
respectively~12.5, 15, 24, and 31 yr for NANOGrav, PPTA, EPTA, and IPTA.
From this frequency, up to $f \simeq 1.25 \times 10^{-8}\Hz$,
the NANOGrav, PPTA, EPTA, and IPTA analyses include, respectively, the first five, six, eight, and ten Fourier modes.
At higher frequencies, the Fourier modes 
have bigger uncertainty and
the presence of a PL behavior is less clear \cite{NANOGrav:2020bcs,Goncharov:2021oub,Chen:2021rqp,Antoniadis:2022pcn}.
As can be appreciated in \Fig{NANOGrav_PPTA}, the posterior SGWB amplitude and slope of the NANOGrav dataset differ, depending on whether one fits a single PL to the whole dataset or a broken PL turning to flat noise  ($\beta=5$) at high frequencies $f \gtrsim f_{\rm bend}$.
This behavior is not observed in the PPTA, EPTA, and IPTA analyses.
We therefore consider 
both the single and broken PL for the NANOGrav result, while we only keep the single PL for PPTA, EPTA, and IPTA.

The part of the MHD-produced SGWB spectrum compatible with the PTA constraints on 
the spectral slope is  
the subinertial range below
the spectral peak, where $\beta \in (1, 3)$
according to \Eq{OmGW_envelope} and the numerical results (\cf \Fig{OmGW_num_analyt2}). 
The inertial range slope $\beta = -8/3$ corresponds to $\gamma=23/3$, which is too steep compared to the slopes reported by the
PTA collaborations (\cf \Fig{NANOGrav_PPTA}).
The peak wave number, separating the subinertial and inertial parts of the spectrum, must satisfy $\kGW > k_*\geq 2\pi\HH_*$ by causality. 
Using the relation
\begin{align}
    \fGW \simeq & \,\, 1.12 \times 10^{-8} \nonumber \\
    \times & \, \frac{\kGW}{2\pi\HH_*}\, \frac{T_*}{100 \MeV} \, \biggl({g_*\over 10}\biggr)^{1\over 6}\,\Hz\,,
    \label{eq:freqH}
\end{align}
and the value of the GW peak position $\kGW \simeq 1.6\, \kf$, derived in \Sec{sec:analytical_detail},
this translates into frequencies today $\fGW \gtrsim 1.8 \times 10^{-8} \Hz$,
for temperatures around the QCD scale. 
The subinertial range is therefore expected to cover the region of highest quality PTA data (extending up to $f \simeq 1.25 \times 10^{-8}\Hz$), supporting the hypothesis that the latter are compatible with the GW signal from MHD turbulence present at the QCD scale.
At lower temperatures $T_* \lesssim 70 \MeV$, however, $\fGW$  decreases below $f \simeq 1.25 \times 10^{-8}\Hz$. 
Moreover, at $T_* \lesssim 5\MeV$, the subinertial
range exits completely the frequency range of the IPTA dataset if $\kf = 2\pi \HH_*$ (IPTA represents the lowest frequencies probed by PTA---\cf Fig.~\ref{NANOGrav_PPTA}).

For the range of initial parameters $k_*$ and $\OmMmax$ that fit the PTA observations, the break of the spectrum from the $f^3$ to the $f^1$ slope occurs in the PTA frequency band; in particular, one can roughly estimate that  
$10^{-9} \Hz \lesssim f_{\rm br} \lesssim 4\cdot 10^{-9} \Hz
    \label{eq:fbr}$ 
for temperatures of the order of $100$ MeV. 
The lowest bound in the above equation is obtained from the values of the magnetic field parameters that maximize the source duration, since $k_{\rm br,min}=1/\delta t_{\rm fin, max}$. 
Following relation \eqref{delta_tfin_te}, and given $\delta\te=\Bigl(k_*\sqrt{{3\over 2}\OmMmax}\,\Bigr)^{-1}$, one needs to insert the minimal values of both $k_*$ and $\OmMmax$. 
The former corresponds to the horizon scale $k_*=2\pi \,\HH_*$, while the latter can be roughly estimated imposing that the SGWB peak given in \Eq{GW_peak}, and evolved till today with \Eq{eq:evolGW}, is in the middle of the allowed region, say $\OmGW^0 \gtrsim 5\cdot 10^{-10}$ (\cf \Fig{NANOGrav_PPTA}). 
This leads to $\OmMmax\gtrsim 0.03$. 
From the two conditions together, one then finds $\HH_* \delta t_{\rm fin, max}\simeq 1.6$, i.e., $k_{\rm br,min} \HH_*^{-1} \simeq 0.6$, which gets translated into frequency today via \Eq{eq:freqH}.  
Conversely, the upper bound of $f_{\rm br}$ can be estimated from the maximal allowed value $\OmMmax= 0.1$ and the maximal $k_*$. 
The latter can again be estimated thanks to \Eq{GW_peak} repeating the same argument as above, leading to $k_*\lesssim 6\pi\,\HH_*$. 
From these values, one finds then $\HH_* \delta t_{\rm fin, min}\simeq 0.4$, i.e., $k_{\rm br,max} \HH_*^{-1}\simeq 2.2$.

When better quality data will be available, the presence of the break might become important to constrain the origin of the SGWB; \cf the discussion in \Secs{sec:constraints}{sec:MBHB}.
Moreover, since the maximal source duration is close to the Hubble time $\HH_* \delta t_{\rm fin, max} \simeq 1.6$, we expect the transition to occur rather sharply, i.e., without an extended logarithmic transition typical of long sources.

\subsection{Constraints on nonhelical magnetic fields using the
PTA results}
\label{sec:constraints}

In this section, we use the $2\sigma$ PTA contours of the amplitude and spectral slope of the CP (\cf \Fig{NANOGrav_PPTA}) to identify the regions in the parameter space of the primordial magnetic field $(\kf, \OmM^*)$ leading to a GW signal compatible with the PTA observations, for fixed $T_*$.
We limit the magnetic field characteristic wave number to be larger than the horizon $\kf \geq 2\pi \HH_*$ and its maximum amplitude to be below 10\%, i.e., $\OmMmax \lesssim 0.1$, according to Refs.~\cite{Shvartsman:1969mm, Grasso:1996kk, Kahniashvili:2009qi}.

For a fixed $T_*$,  varying the parameters $(\kf, \OmM^*)$, we construct the corresponding SGWBs using the analytical model of \Eq{OmGW_envelope} and
setting $\delta \tfin$ to the empirical fit
of \Eq{delta_tfin_te}, validated by the numerical simulations.
Note that we are not compensating by the factor $\mathcal{G}$ as in \Eq{OmGW_empirical} [\emph{cf.} also \Eq{OmGW_num_env}], since we are interested
in fitting the SGWB spectrum at  frequencies below the
peak: as demonstrated in \Sec{PTA_res}, only the subinertial part of the GW spectrum is expected to be in the frequency region where the PTA data could be compatible with a nonzero signal. 

For each SGWB so constructed, we compute its slope at each frequency in a subset of the frequency range of the PTA observations.
The subset is defined as follows: for the single-PL fit, we choose a range spanning from the first Fourier mode up to
$f \simeq 1.25\times 10^{-8} \Hz$, thereby excluding the highest frequencies at which the PTA results have large uncertainties (\cf \Sec{PTA_res});
for the broken-PL fit of NANOGrav, we further restrict the range to the maximal frequency $f \simeq 9 \times 10^{-9}\Hz$, excluding the part transitioning to the flat power spectral density with $\beta=5$
(\cf \Fig{NANOGrav_PPTA}).

To compute the slope of the SGWB  given in \Eq{OmGW_envelope}, 
we simplify the frequency dependence of $p_\Pi$ as
\begin{align}
    p_\Pi(f/f_*)=
    \left\{ \begin{array}{ll}
    (f/2f_*)^{0} & {\rm for}~f \leq 2f_*\,, \\
    (f/2f_*)^{-11/3} & {\rm for}~f > 2f_*\,,
    \end{array}    \right.
\end{align}
while, in general, it is computed numerically using \Eq{Pi_conv}.
The resulting SGWB slope is
\begin{align}
    &  \hspace{-2mm} \beta = \frac{\partial \ln \OmGW^0(f)}{\partial \ln f} = 
    \nonumber \\ & \hspace{-1mm} \left\{ \begin{array}{rl}
         3 \hspace{9mm} & \text{if } f < 1/(2\pi \delta  \tfin), \\
         3 - 2s (f/\HH_*) & \text{if }
        1/(2\pi \delta  \tfin) \leq f < 2f_*, \\
        \!\!\!-\twothird - 2s  (f/\HH_*) & \text{if } f \geq 2f_*,
    \end{array}  \right.
\end{align}
where the function $s$ gives the slope of the logarithmic
term appearing in \Eq{OmGW_envelope},
\begin{align}
    s(x) = &\, -\frac{1}{2}\frac{{\rm d} \ln (\ln^2[1 + 1/(2 \pi x)])}
    {{\rm d} \ln x}\nonumber \\
    = &\,\Bigl[(1 + 2\pi x) \bigl|\ln (1 + 1/(2\pi x)\bigr|\Bigr]^{-1},
\end{align}
and takes values between 0 and 1 in the low and high $f$ regimes,
respectively, yielding the slopes of the SGWB presented in
\Sec{analytical_spec}.

Via \Eqs{SGWB_spl}{SGWB_bpl}, one can calculate the range of SGWB amplitudes allowed at 2$\sigma$  by the PTA observations, for a specific slope and frequency, given as a range of $\Omyr$
(\cf \Fig{NANOGrav_PPTA}). 
For a fixed $T_*$, we consider that a point in the parameter space $(\kf, \OmM^*)$ is compatible with
the results reported by one of the PTA collaborations
if it provides a SGWB spectrum with amplitude lying
within the PTA 2$\sigma$ bounds corresponding to its slope
at, at least, one of the frequencies
in the chosen PTA frequency subset.
Note that the amplitudes reported by the PTA collaborations assume that
the GW signal follows a PL, while we expect the subinertial range
of the SGWB produced by MHD turbulence to present a spectral shape characterized
by two different regimes: one being a PL proportional to $f^3$
and the other one being approximately a PL proportional to $f^1$
(\cf \Sec{analytical_spec}).
Hence, our approach is conservative and does not rule out SGWBs that
present the break from $f^3$ to $f^1$ within the PTA range of frequencies, which has not been included in the reported analyses by
the PTA collaborations.
We also allow, in our analysis, the break from $f^1$ to $f^{-8/3}$
to occur within the PTA range.
The additional consequences of a broken-PL SGWB in cosmology,
consistent with NANOGrav observations,
have been studied in Ref.~\cite{Benetti:2021uea}.

\begin{figure}[t!]
    \centering
    \includegraphics[width=\columnwidth]{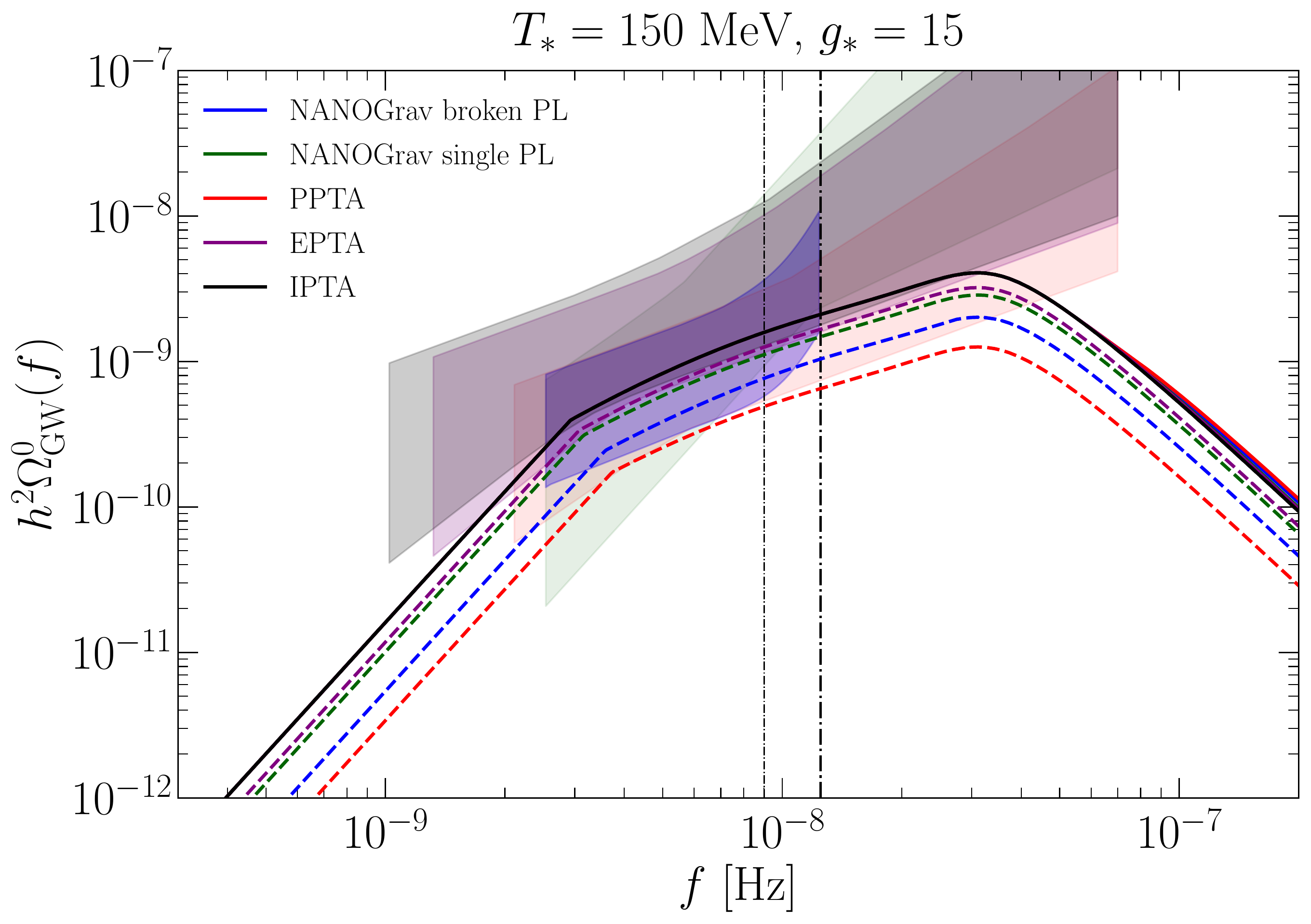}
    \includegraphics[width=\columnwidth]{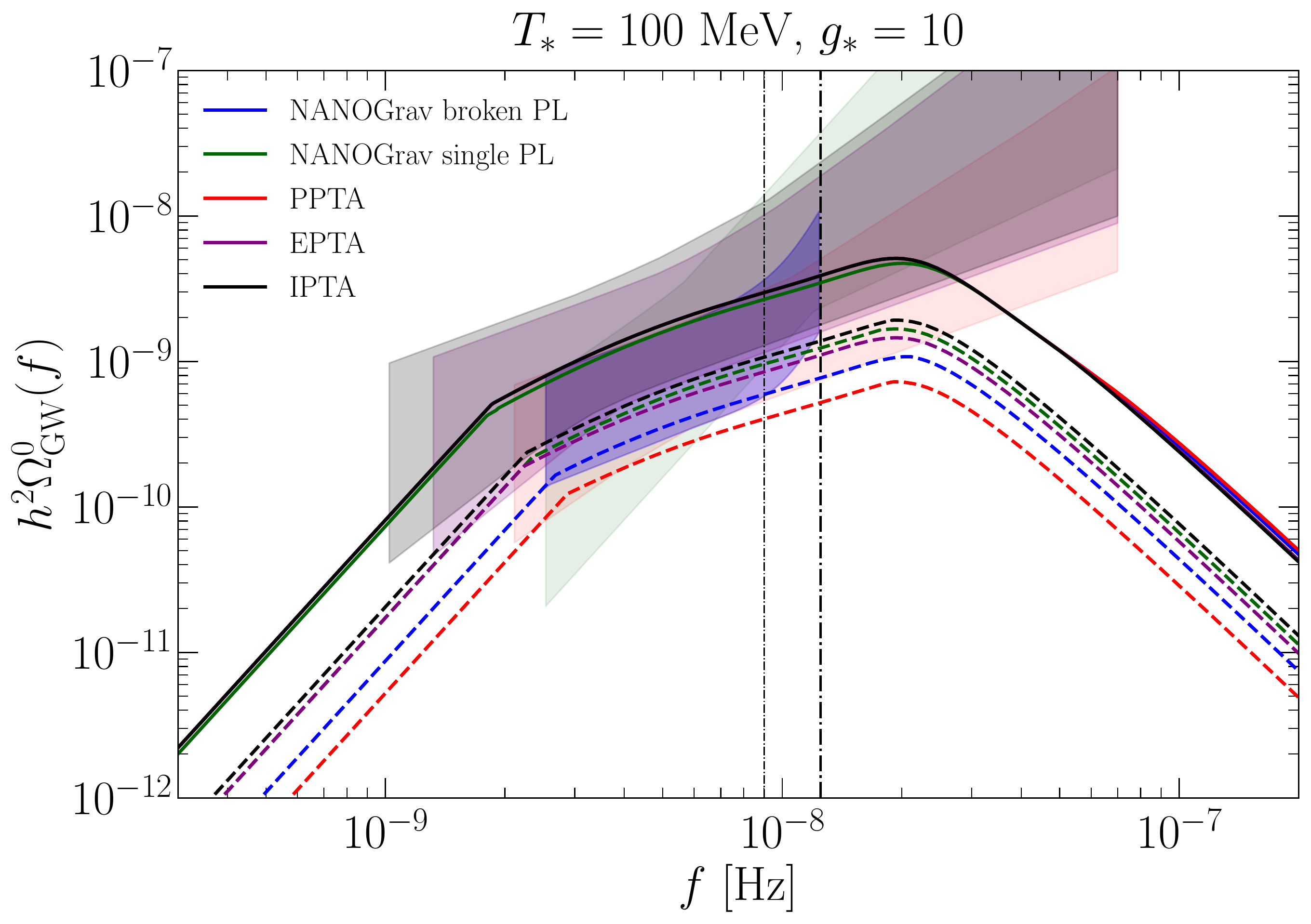}
    \caption{For $T_* = 150 \MeV$ and $g_* = 15$ in the upper panel, and  $T_* = 100 \MeV$ and $g_* = 10$ in the lower panel, we show the upper boundary (solid lines) and the lower boundary (dashed lines) of the regions compatible with the 
    PTA data at $2\sigma$. 
    To be compatible with NANOGrav with broken PL,
    each SGWB spectrum must lie in the region within the \emph{blue} solid and dashed lines; with NANOGrav with single PL, within the \emph{green} solid and dashed lines; with PPTA, within the \emph{red} solid and dashed lines; with EPTA, within the \emph{purple} solid and dashed lines; with IPTA, within the \emph{black} solid and dashed lines.
    The shaded areas correspond to the
    range of allowed values $h^2 \OmGW^0(f)$ of \Eqs{SGWB_spl}{SGWB_bpl}, restricted to the range of slopes
    of interest for a MHD-produced SGWB, i.e., $\beta \in (1, 3)$.
    The magnetic field characteristic scale is bound to $k_* \geq 2\pi \HH_*$ and the magnetic energy densities to
    $\OmMmax \leq 0.1$. 
    The vertical lines show the upper bound of the PTA frequency subset to which we restrain the analysis: $f \simeq 1.25 \times 10^{-8} \Hz$ for the single-PL cases (dot-dashed line) and $f \simeq 9 \times 10^{-9}\Hz$ for the NANOGrav broken-PL case (dashed line).   
    }
    \label{OmGW_num_analyt}
\end{figure}

\begin{figure*}[t!]
    \centering
    \includegraphics[width=.7\columnwidth]{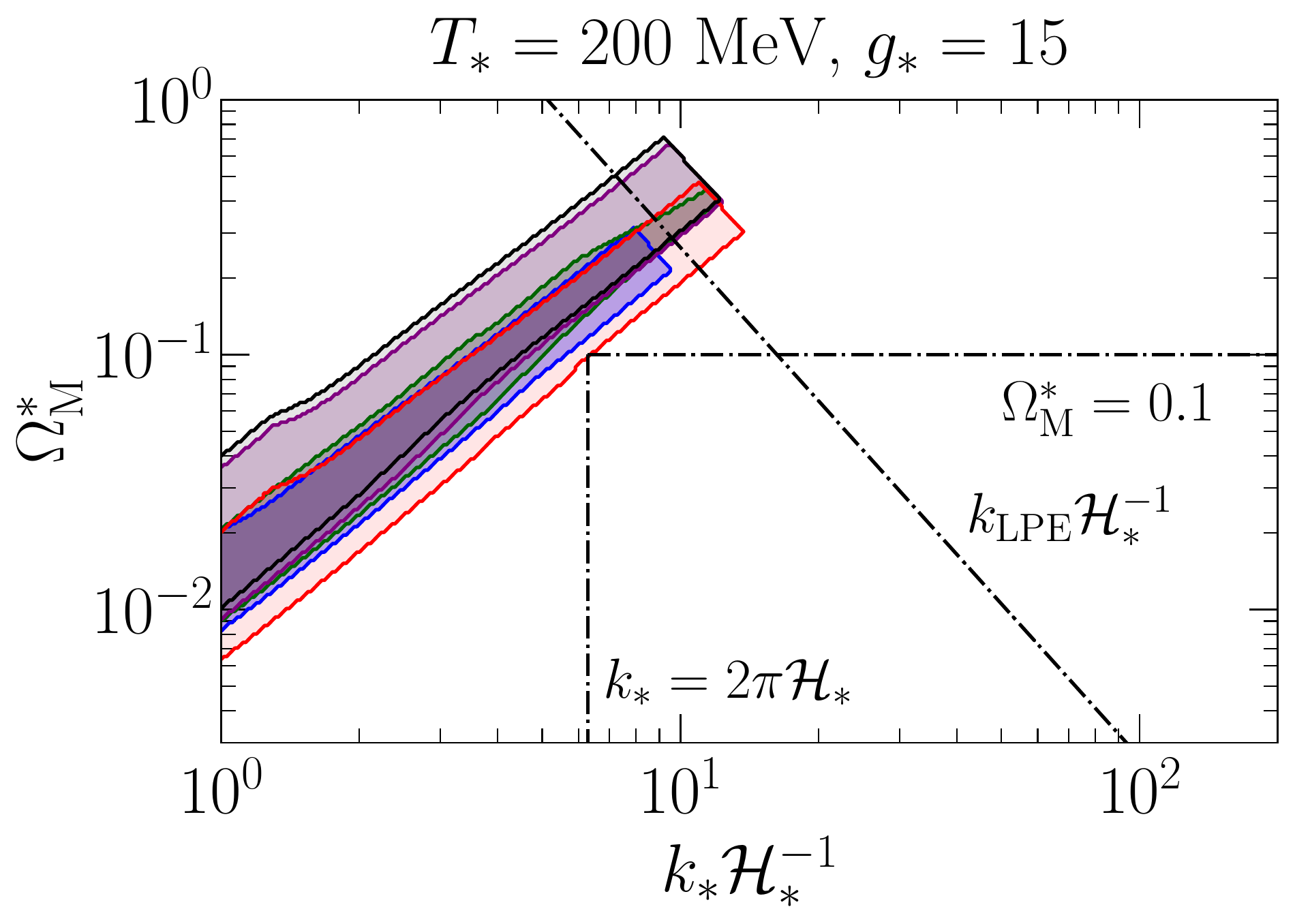}
    \includegraphics[width=.7\columnwidth]{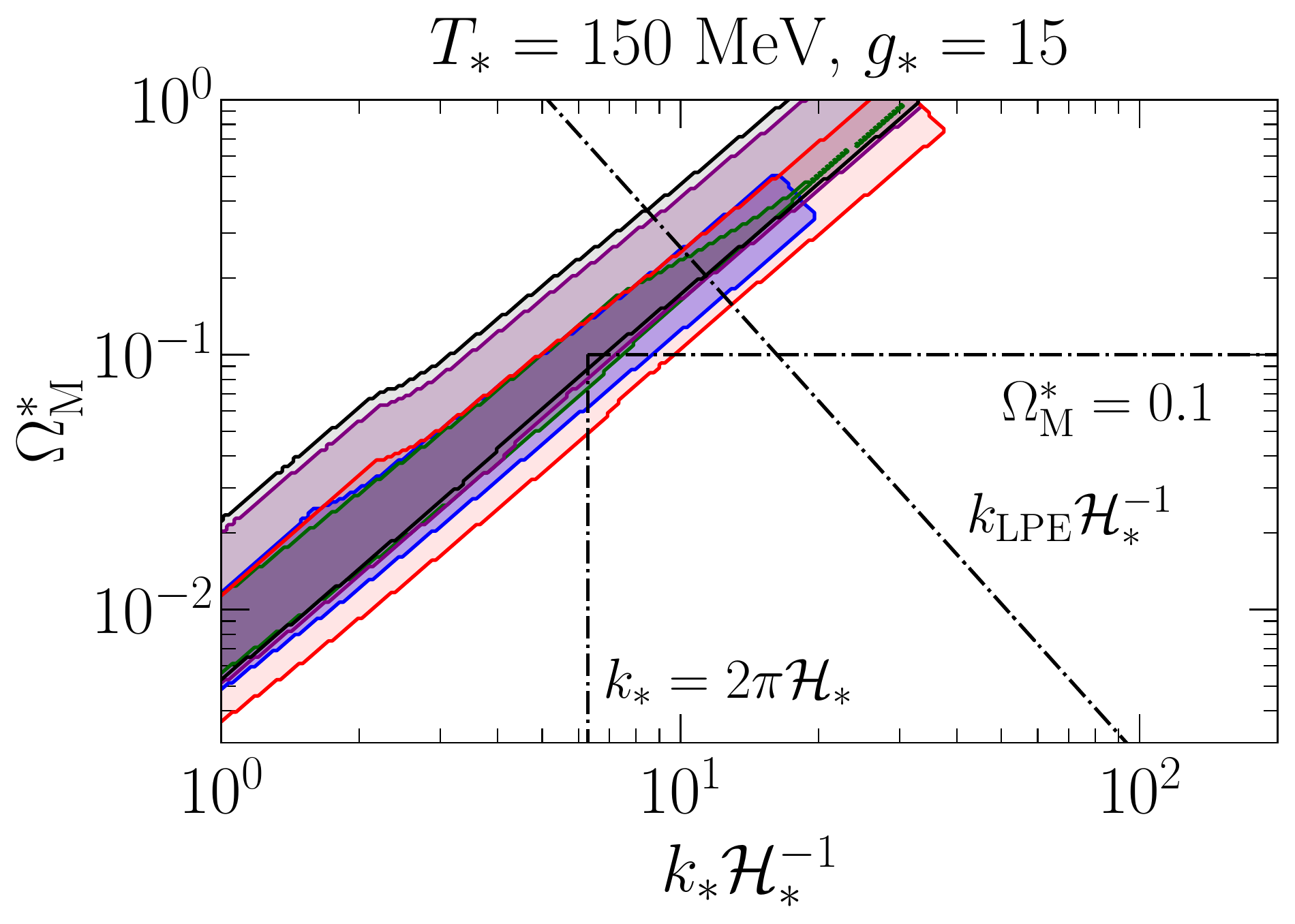}
    \includegraphics[width=.7\columnwidth]{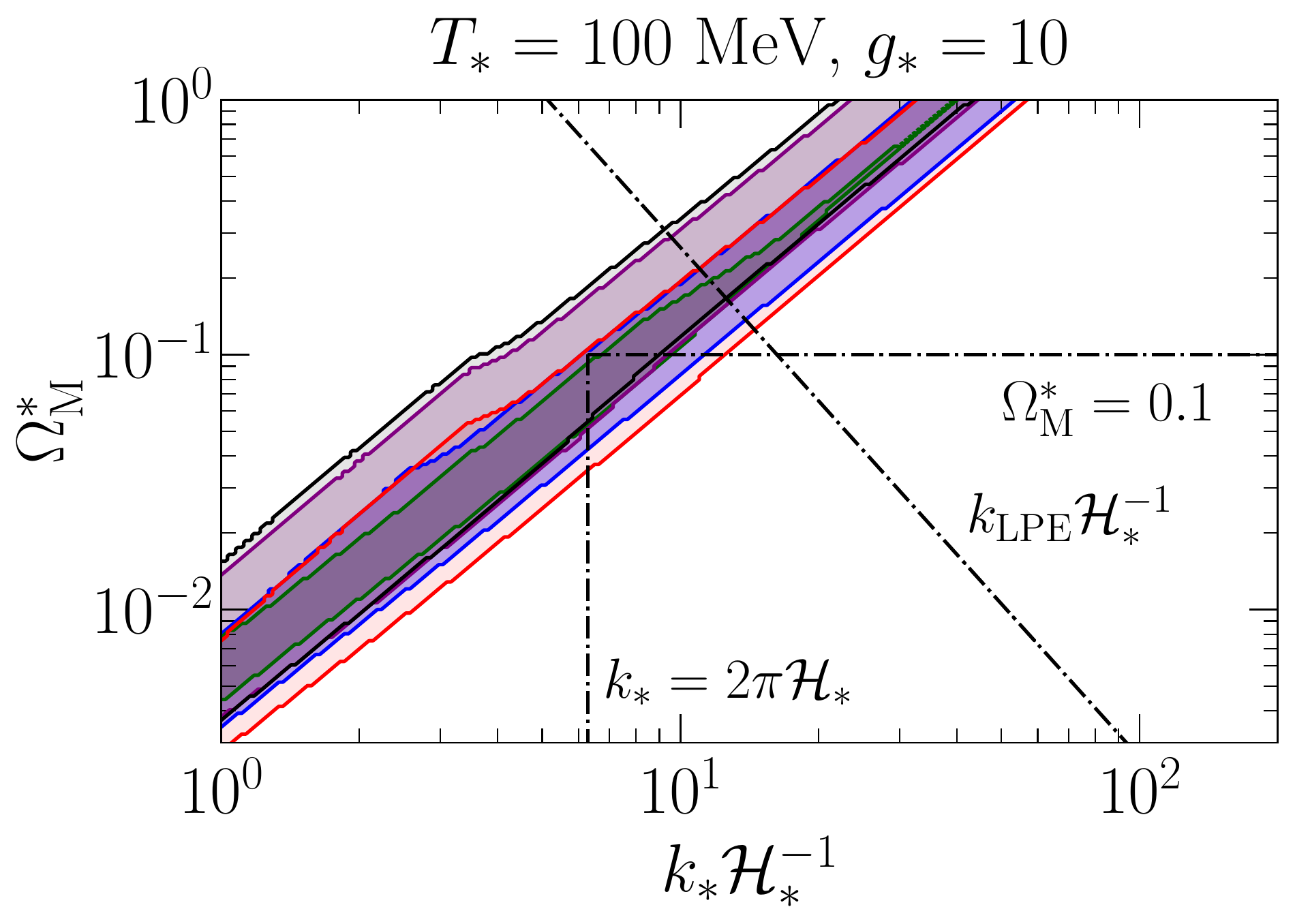}
    \includegraphics[width=.7\columnwidth]{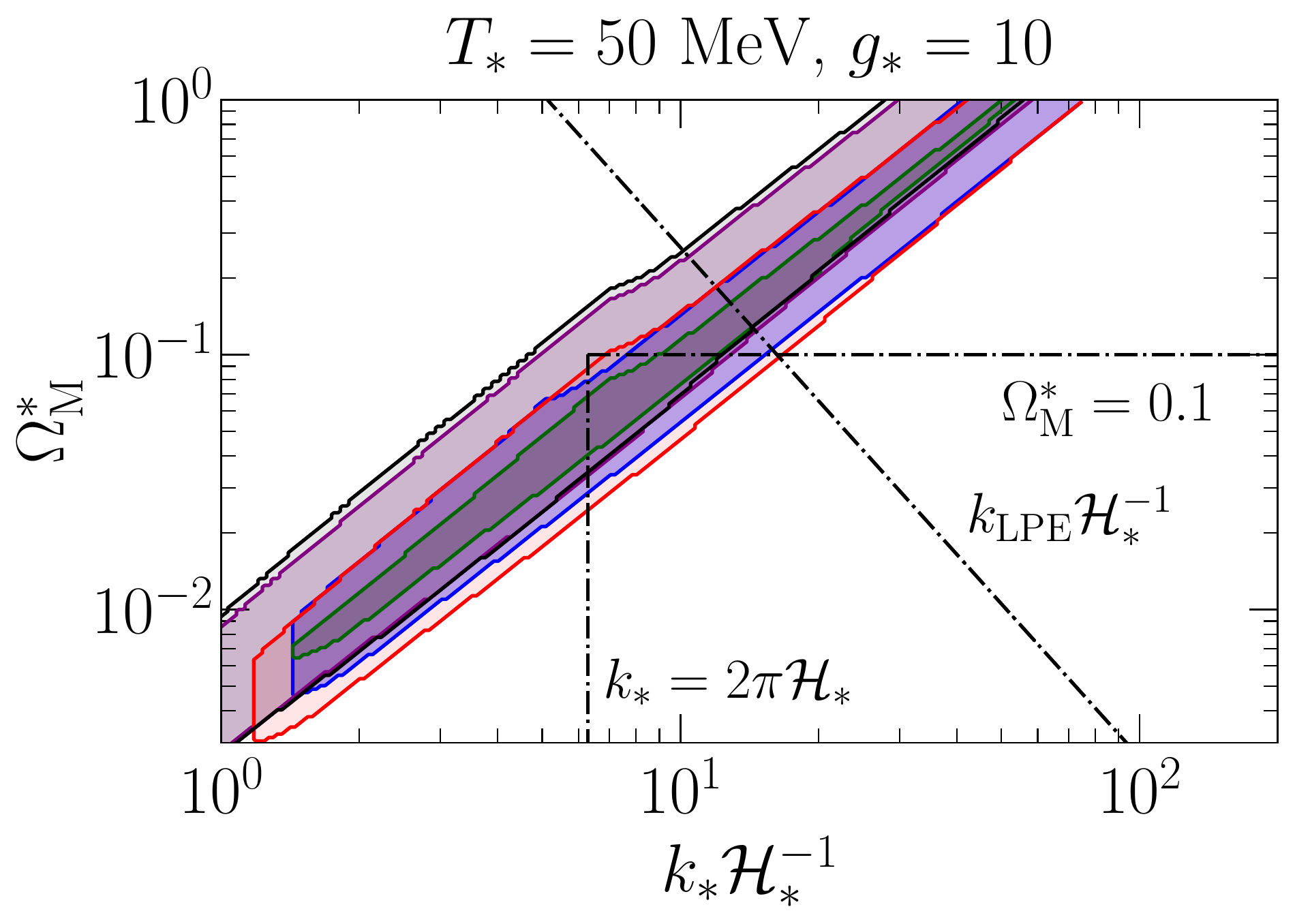}
    \includegraphics[width=.7\columnwidth]{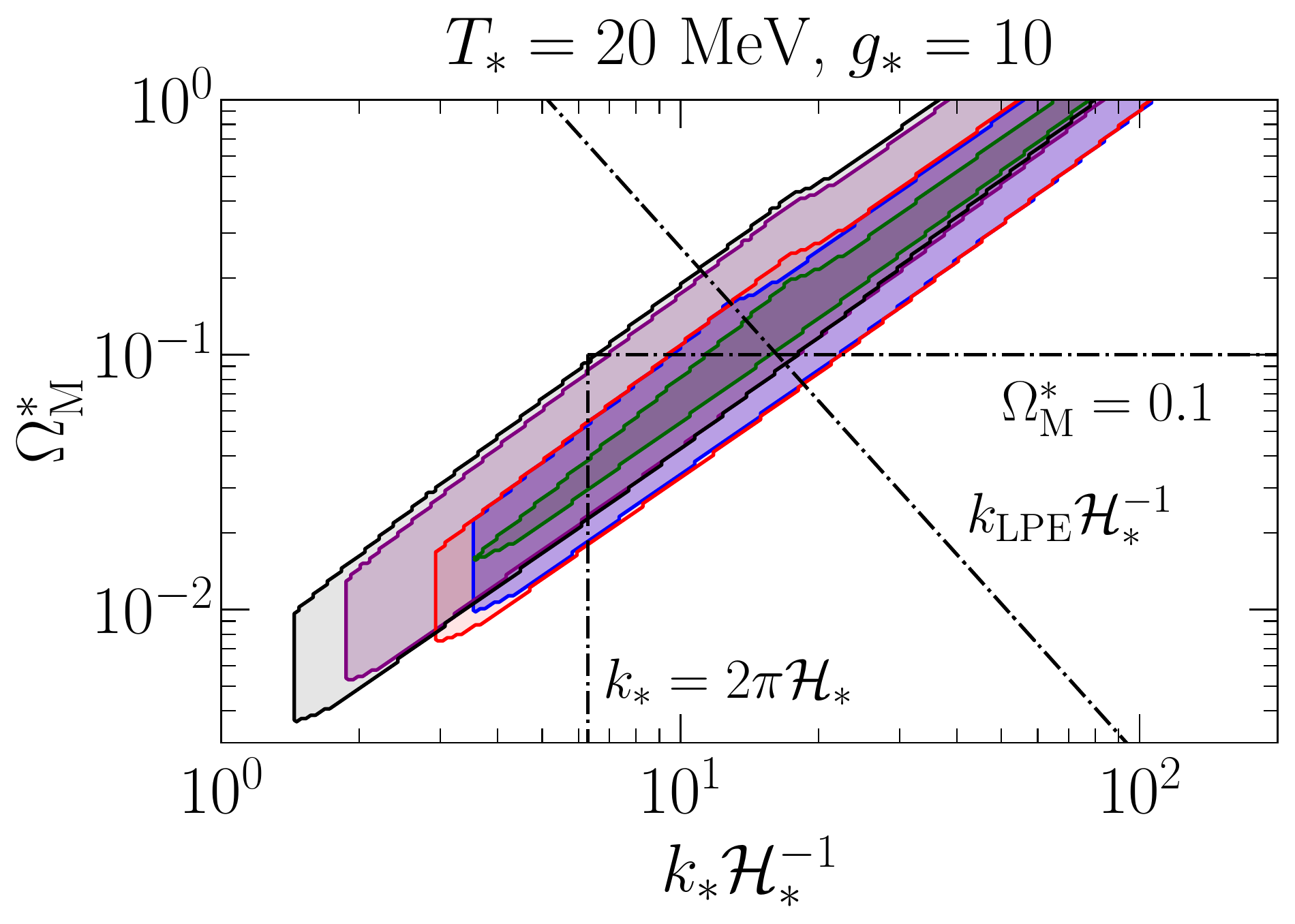}
    \includegraphics[width=.7\columnwidth]{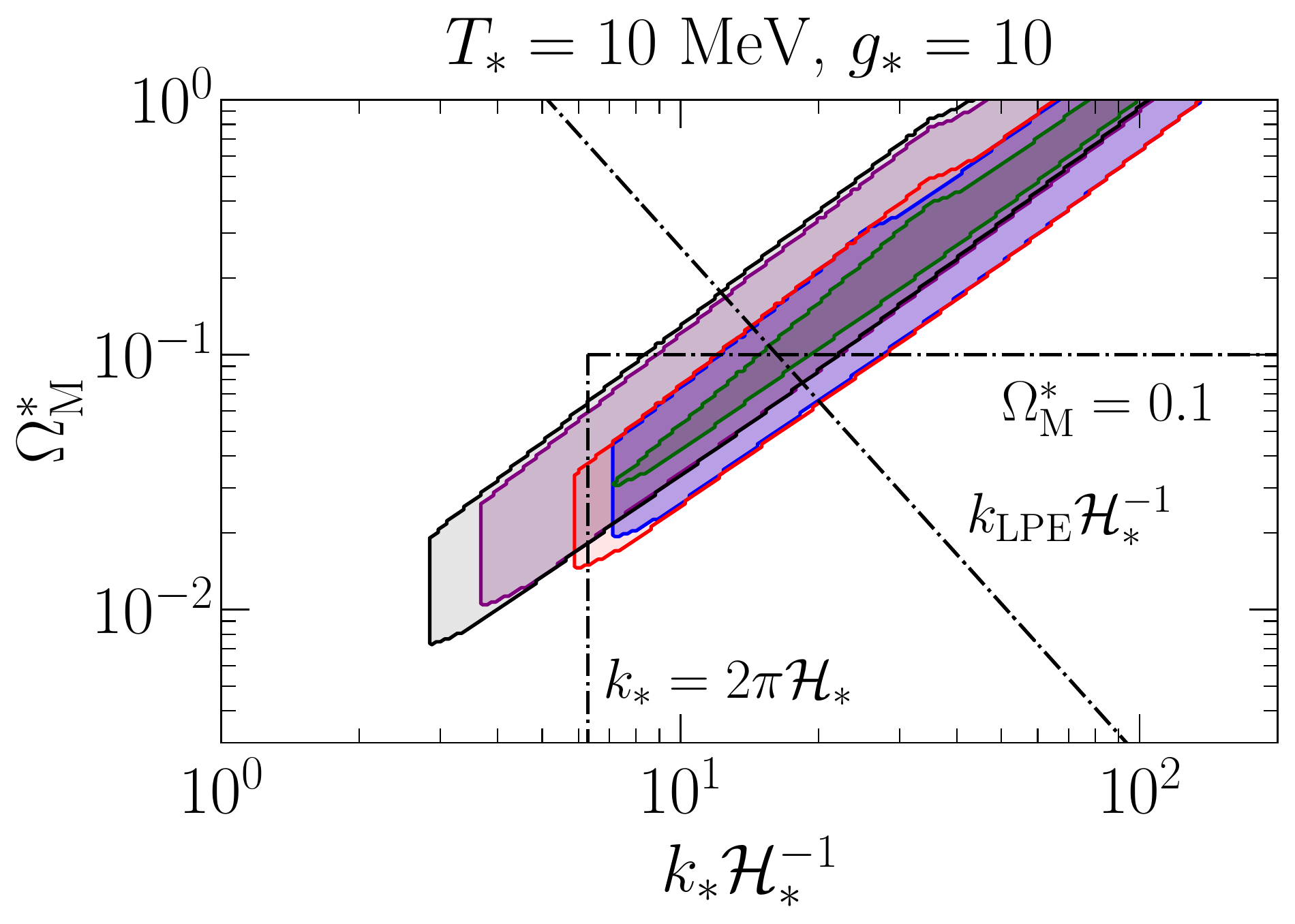}
    \includegraphics[width=.7\columnwidth]{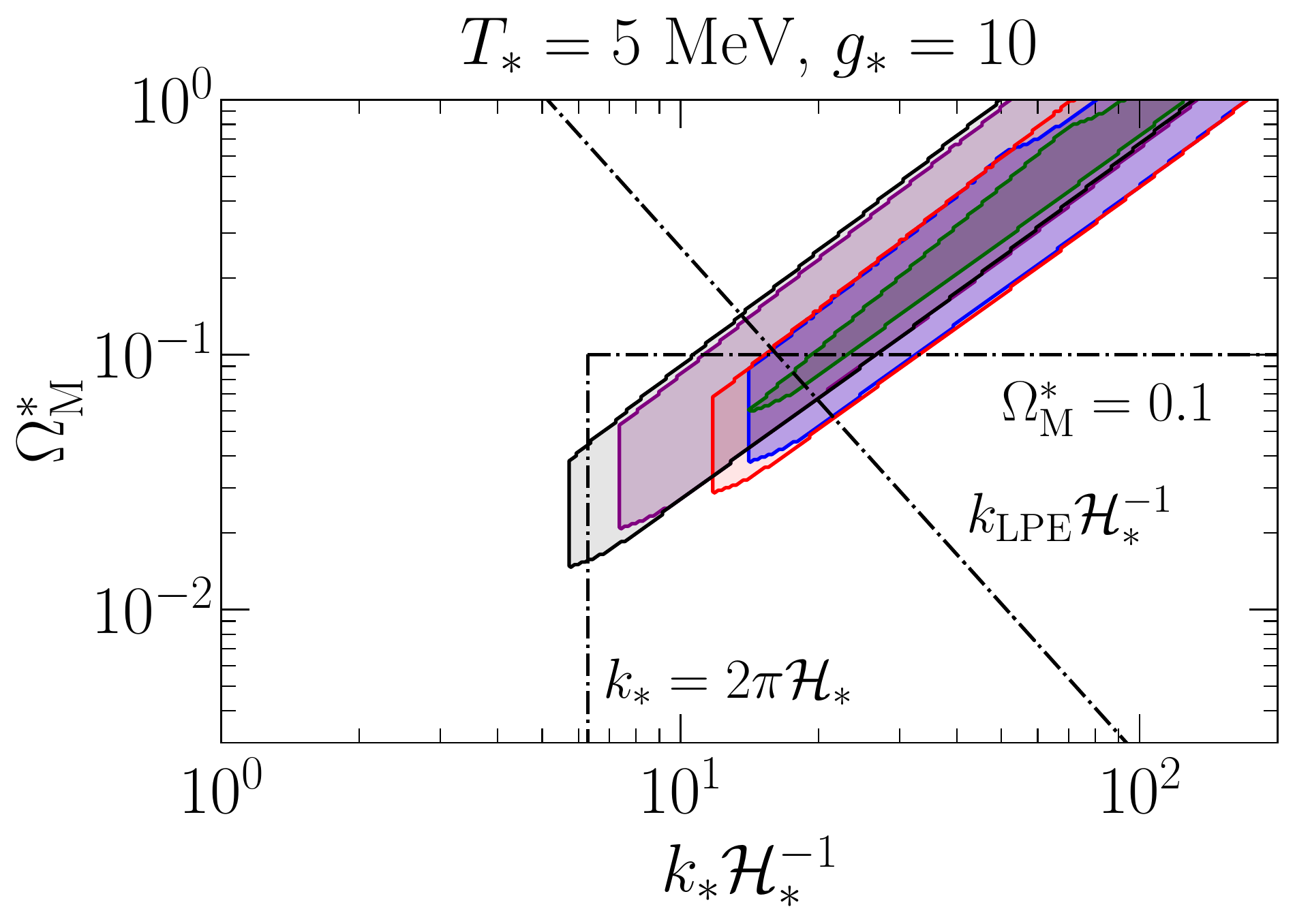}
    \includegraphics[width=.7\columnwidth]{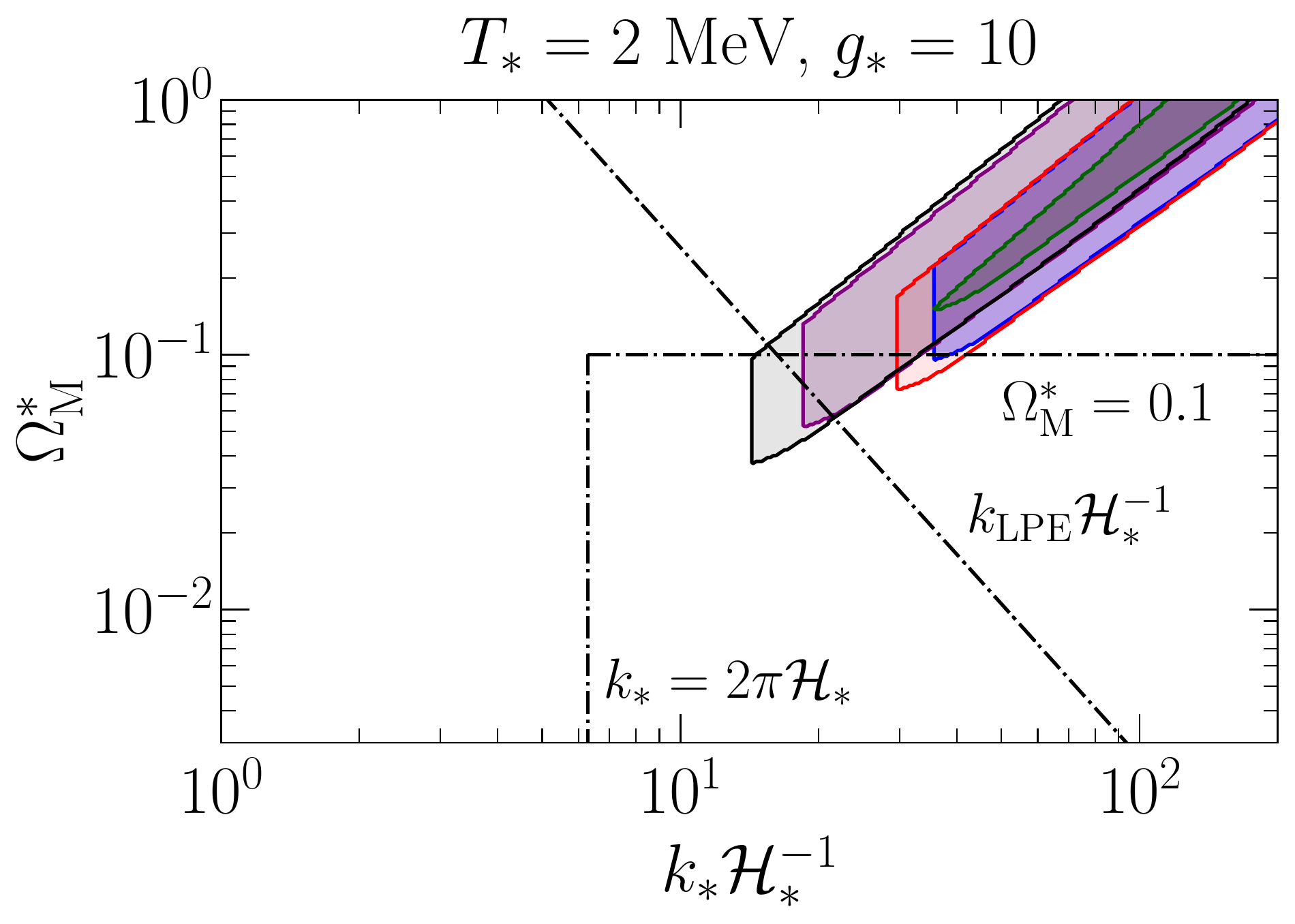}
    \caption{For different values of $T_* \in (2, 200) \MeV$, we show the allowed regions in the $(\kf, \OmMmax)$ parameter space, derived as described in the main text from the 2$\sigma$ results of NANOGrav using the
     broken-PL (blue)
    and single-PL (green) fits and from the 2$\sigma$ results of EPTA (purple), PPTA (red), and IPTA (black) using the single-PL fits.
    The vertical and horizontal dot-dashed lines show the physical limits 
    $\kf \geq 2\pi\HH_*$ and $\OmM^* \leq 0.1$, respectively: the allowed parameter region lies within the rectangle.
    The wave number of the largest processed eddies $\left.k_*\right|_{\rm LPE}$ is also shown (dot-dashed diagonal line).
    }
    \label{OmGW_nanograv}
\end{figure*}

In \Fig{OmGW_num_analyt}, we show the curves delimiting the allowed regions in which any SGWB compatible with the PTA observations must lie, obtained 
using the values of $(\kf, \OmM^*)$ derived as described above.
To be compatible with the results of a given PTA collaboration, each MHD-produced SGWB must lie within the region delimited by the dashed and solid lines corresponding to that collaboration (i.e., purple lines for EPTA, red for PPTA, and so on). 
We display the results for energy scales
close to the QCD phase transition, e.g., $T_* = 150 \MeV$ and
$g_* = 15$ in the upper panel and $T_* = 100 \MeV$ and $g_* = 10$
in the lower panel.

If $T_* = 150 \MeV$, the upper boundary of the allowed region is the same for all PTA data; hence, the solid lines superimpose. This roughly corresponds to
the point in parameter space $(\kf
=2 \pi\HH_*, \OmMmax = 0.1$) (though these values can vary slightly with frequency).
The lower boundary of the allowed region is instead different for each dataset considered: NANOGrav with both single and broken PLs, PPTA, EPTA, and IPTA.
If $T_* = 100\MeV$, the region allowed by the NANOGrav single-PL fit corresponds to $\OmM^*$ slightly smaller than 0.1 and $\kf$ slightly
larger than the horizon scale.
Note that, in general, the NANOGrav single-PL case is more constraining in terms of $(\kf, \OmM^*)$ values, since the minimum slope allowed at 2$\sigma$ is
$\beta = 1.25$ (\cf \Fig{NANOGrav_PPTA}),
while the other cases allow slopes down to $\beta= 1$.

The range of parameters $(\kf, \OmM^*)$ compatible with the data of 
the PTA collaborations at $2\sigma$ are shown in \Fig{OmGW_nanograv}
for temperature scales ranging from 2 to 200 MeV. 
At temperatures below $1 \MeV$, the PTA results cannot be
accounted for by a GW signal produced by MHD turbulence,
in the limit $\OmMmax \leq 0.1$.
For $100\MeV\leq T_*\leq 200\MeV$, the magnetic field parameters are strongly constrained: 
its characteristic wave number $k_*$ must be close to the horizon, and its amplitude must be close to the upper bound $\OmM^* \leq 0.1$.
Smaller characteristic scales and amplitudes are allowed as $T_*$ decreases.
For temperatures below $ 20 \MeV$, the point in parameter space 
$(\kf =2 \pi\HH_*, \OmMmax = 0.1$)
is no longer compatible with the data, which prefer magnetic fields with 
smaller characteristic scales but higher amplitudes, until the latter exceed again their upper bound
for temperatures smaller than 1 MeV.

In particular, setting the largest processed eddies as the
characteristic scale of the magnetic field, $\left.l_*\right|_{\rm LPE} \HH_* = \sqrt{\threehalf \OmM^*}$, the resulting SGWB is only compatible
with the PTA observations at low temperatures $T_* \in (2, 50) \MeV$
when we limit $\OmMmax \leq 0.1$.

\begin{figure}[t!]
    \centering
    \includegraphics[width=\columnwidth]{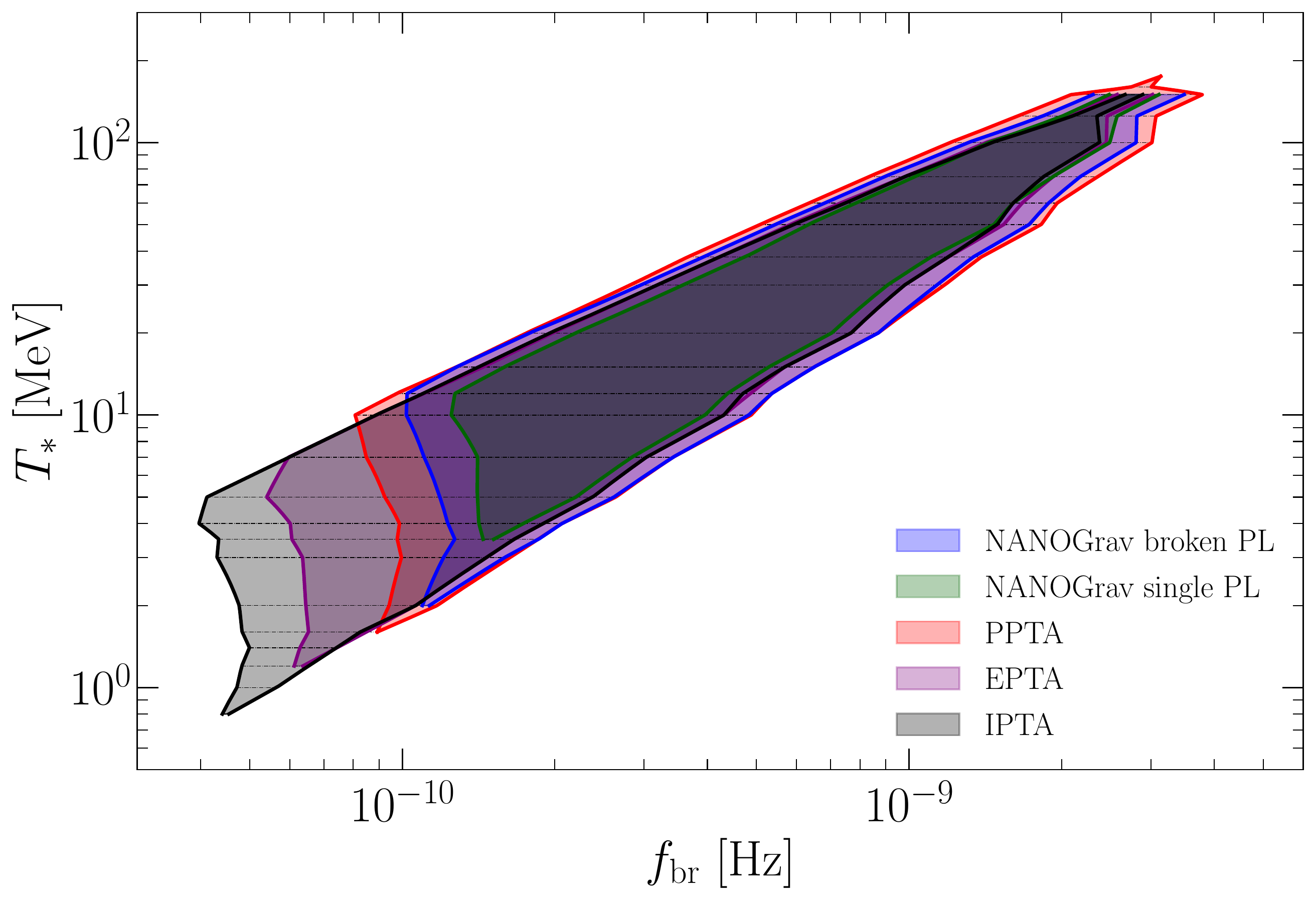}
    \caption{Range of frequencies at which the $f^3$ to $f^1$ break
    $f_{\rm br}$ occurs,
    for the parameters $(\kf, \OmM^*)$ compatible with the results of each of the PTA
    collaborations, for different $T_*$ (\cf \Fig{OmGW_nanograv}), in
    the limit $\OmMmax \leq 0.1$.
    The horizontal dot-dashed lines correspond to the
    computed values of $T_*$.}
    \label{fbr_T}
\end{figure}

From the allowed parameter regions $(\kf, \OmM^*)$ at each
$T_*$, one can predict at which frequencies the break from $f^3$ to
$f^1$ occurs.
While a rough estimate of the break frequency was given in \Sec{PTA_res} for $T_* = 100 \MeV$, we show in \Fig{fbr_T} the results of this more refined analysis.
It can be appreciated that the smaller the temperature of the phase transition, the smaller the break frequency.  
Consequently, if this break will be identified in future PTA data, it will help elucidating the SGWB origin: as for the spectral peak $\fGW$, we find that $f_{\rm br}$ is connected to the energy scale of the SGWB generating process.

\subsection{Constraints on the magnetic field amplitude and characteristic scale today}
\label{comp_Bl}

\begin{figure*}[t!]
    \centering
    \includegraphics[width=.85\textwidth]{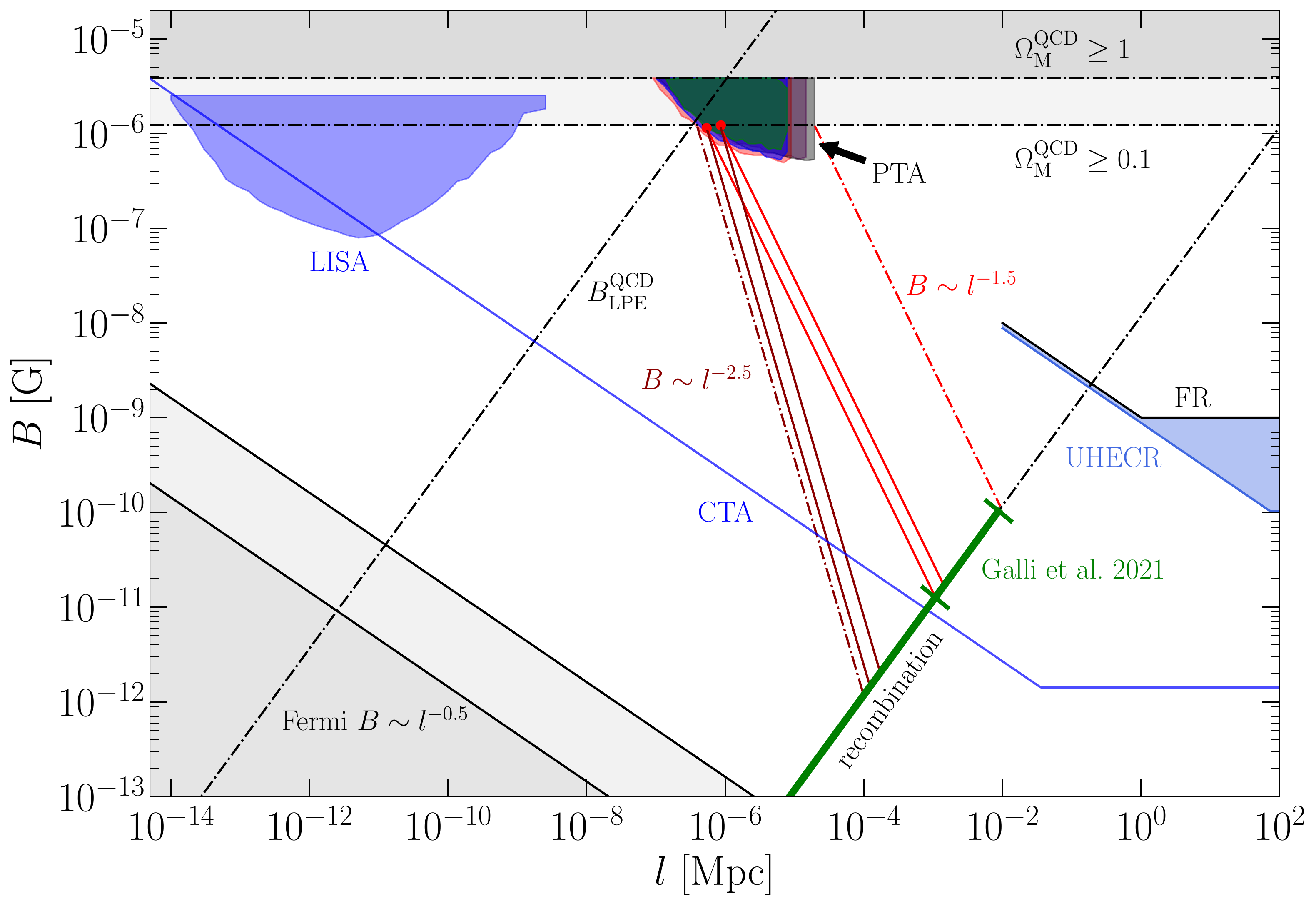}
    \caption{Region in the magnetic field parameter space, given by its comoving amplitude $B$ and characteristic scale $l$, compatible with the observations of the different PTA collaborations: in
    blue, NANOGrav with broken PL; in green, NANOGrav with single
    PL; in red, PPTA; in purple, EPTA; in black, IPTA.
    The parameter space region accessible to LISA is shown in light blue.
    The horizontal dot-dashed lines show the bounds $\OmMmax \leq 1$, and $\OmMmax \leq 0.1$ from nucleosynthesis \cite{Shvartsman:1969mm, Grasso:1996kk, Kahniashvili:2009qi}.
    The black dot-dashed diagonal lines show {\em i)} the magnetic field amplitude when the characteristic scale corresponds to the largest processed eddies at the QCD phase transition
    $\left.l_*\right|_{\rm LPE}$ [\cf \Eq{lpe_Bl}]; {\em ii)} the magnetic field amplitude reached at recombination [\cf \Eq{rec_Bl}].
    The dot-dashed red and brown lines show the 
    evolutionary paths of the extremities of the parameter space region compatible with the PTA results up to recombination,
    following compressible (red) and incompressible (brown) MHD free decay.
    The solid red and brown lines indicate the evolutionary paths of an initial field with $\kf = 2\pi \HH_*$ and $\OmMmax = 0.1$ at $T_* = 100 \MeV$ and $g_* = 10$ (right red dot) and at $T_* = 150 \MeV$ and $g_* = 15$ (left red dot).
    The green line indicates the upper limit $B \lesssim 0.1$
    nG, and the range $B_{\rm rec} \in (0.013, 0.1)$ nG, proposed to alleviate the Hubble tension,
    both derived in Ref.~\cite{Galli:2021mxk} from CMB constraints on the baryon clumping.
    The black solid diagonal lines show the Fermi Large Area Telescope (LAT) lower bound on the intergalactic magnetic field from timing of the blazar signal (darker gray area) and from the search of extended emission (lighter gray area) \cite{Fermi-LAT:2018jdy}.
    The blue line shows the expected sensitivity of CTA
    \cite{Korochkin:2020pvg}.
    At larger scales, the upper bound from Faraday rotation (FR) is
    shown \cite{Pshirkov:2015tua}, and the blue shaded region
    indicates the observations of UHECR from the Perseus-Pisces
    supercluster \cite{TelescopeArray:2021dfb,Neronov:2021xua}.
    Note that the latter constraints
    refer to present time magnetic field strength and characteristic scale,
    and they have been cut to avoid
    intersecting the evolutionary paths from the QCD phase
    transition up
    to recombination in the plot, for clarity.}
    \label{B_vs_l}
\end{figure*}

The analysis performed in \Sec{sec:constraints} allowed us to constrain the magnetic
field amplitude $\OmM^*$ and characteristic scale $\kf$ at several fixed temperature values $T_*$.
In this section, we derive the constraints
on the comoving magnetic field strength $B_*$ and characteristic length $l_*$ compatible with the PTA
results.
We then compare them with other constraints on primordial magnetic fields, 
in particular at recombination.
The results are shown in \Fig{B_vs_l}.

To begin with, we transform the constraints on $\OmM^*$ to constraints
on the comoving magnetic field root mean square amplitude $B_* = \sqrt{\bra{\BB^2}}$:
\begin{eqnarray}
    B_* &=& \sqrt{2\, \EEM^* \,\mu_0} \,\biggl(\frac{a_*}{a_0}\biggr)^2 \nonumber\\
    &\simeq & 3.87 \,  \sqrt{{\OmM^*}} \,\biggl(
    \frac{g_*}{10}\biggr)^{-{1\over6}}\mu{\rm G},
    \label{B_Gauss}
    \label{B_rms}
\end{eqnarray}
where
$\EEM^* = \OmM^* \, \EErad^* $, the factor $({a_*}/{a_0})^2$ accounts for the fact that the magnetic field is comoving, 
and we have recovered $c = 3 \times 10^8 \m/\s = 9.72 \times 10^{-15} \Mpc/\s$ and $\mu_0 = 40 \pi$ G$^2$ (J/m$^3$)$^{-1}$, otherwise set to $c = \mu_0 = 1$,
since in this section we want to express $B_*$ in Gauss (G) and
$l_*$ in parsecs. 
Consequently,  $\EErad^*=\pi^2g_*T_*^4/(30(\hbar c)^3)$ \cite{Kolb:1990vq}.
The characteristic comoving length
scale $l_*$ can be expressed in parsecs
using \Eq{HHstar}:
\begin{equation}
    l_* = \frac{2 \pi}{\kf} \simeq 5.4
     \, \frac{\HH_*}{c \kf} \frac{100 \MeV}{T_*} \biggl(\frac{g_*}{10}\biggr)^{-{1\over 6}}{\rm pc}.
    \label{l_Mpc}
\end{equation}

The region of $B_*$ and $l_*$ values allowed by the PTA
results is shown in \Fig{B_vs_l},
and it is limited by $T_* \in (1, 200) \MeV$, $l_* \in (0.4, 20)$ pc, and
$B_* \in (0.5, 1.2)\, \mu$G, when we consider
the 
nucleosynthesis constraint $\OmM^* \leq 0.1$, which gives
$B_* \leq 1.2 \, \mu$G $(g_*/10)^{-{1\over6}}$.
If we allow\footnote{%
In this subsection,
we extend our analysis up to $\OmMmax \leq 1$,
allowing a larger range of values of $T_*$.
Note, however, that values $\OmMmax \gtrsim 0.1$ require a relativistic MHD
description and therefore the SGWB derived in \Secs{analytical_spec}{sec:numerical_spec}
might be modified.} $0.1 \leq \OmMmax \leq 1$,
then the region extends to
$B_* \in (0.5, 3.8)\, \mu$G, $l_* \in (0.1, 20) \pc$,
and $T_* \in (0.2, 350) \MeV$.

\FFig{B_vs_l} also shows the region in parameter space $(B_*,\,l_*)$ that could be probed by the Laser Interferometer Space Antenna (LISA).
We have obtained it via a similar analysis to that of \Sec{sec:constraints}, using
the model developed in \Sec{sec:analytical_detail}, and considering the PL sensitivity of LISA for a threshold signal-to-noise
ratio of 10 and 4 yr of mission duration \cite{Caprini:2019pxz,Schmitz:2020syl}.
LISA could probe the SGWB from primordial magnetic fields with amplitudes in the range $B_* \in (0.08, 0.8) \, \mu$G and characteristic scales in the range
$l_* \in (2.6 \times 10^{-8},\, 8 \times 10^{-4})$ pc, with a range of temperatures
$T_* \in (50 \GeV, 2000 \TeV)$ when $\OmMmax \leq 0.1$.
If we allow $0.1 \leq \OmMmax \leq 1$, then the range of temperatures
compatible with LISA extends to $T_* \in (5 \GeV, 5000 \TeV)$
and the primordial
magnetic field parameters to $l_* \in (10^{-8}, 2.6 \times 10^{-3})\, \pc$ and $B_* \in (0.08, 2.5) \, \mu$G.

Along with the GW production, the primordial magnetic field evolves following the MHD turbulent free decay.  
For nonhelical fields, the direct cascade leads to the 
scaling
$B \propto l^{-5/2}$ for incompressible turbulence and 
$B \propto l^{-3/2}$ for compressible turbulence 
\cite{Durrer:2013pga,Banerjee:2004df}.
Furthermore, the turbulent evolution is expected to drive the magnetic characteristic scale to the one of the largest processed eddies \cite{Banerjee:2004df}, $\left.l_*\right|_{\rm LPE}\HH_* = \sqrt{{3\over 2}\OmMmax}$. 
The magnetic field amplitude at this scale can be obtained
combining \Eqs{B_Gauss}{l_Mpc}:
\begin{equation}
    \left.B_*\right|_{\rm LPE} \simeq 3.6 \, \frac{\left.l_*\right|_{\rm LPE}}{1 \pc} \,
    \frac{T_*}{100 \MeV} \,\mu{\rm G}.
    \label{lpe_Bl}
\end{equation}
This equation can be readily applied to find as well the magnetic field amplitude at recombination \cite{Banerjee:2004df,Durrer:2013pga}
\begin{align}
    \left.B_{\rm rec} \right|_{\rm LPE}
    \simeq 10^{-2} \, {\rm nG}\,
    \frac{\left.l_{\rm rec}\right|_{\rm LPE}}{1 \kpc},
    \label{rec_Bl}
\end{align}
where we have used $T_{\rm rec} = 0.32$ eV \cite{Durrer:2008eom}.
Both \Eqs{lpe_Bl}{rec_Bl} are shown in \Fig{B_vs_l} by black dot-dashed lines.

The evolutionary paths, from the QCD phase transition up to
the epoch of recombination, of the extremities of the $(B_*,\,l_*)$ region compatible with the PTA observations are shown by red (compressible) and brown (incompressible)
dot-dashed lines in \Fig{B_vs_l}.
In particular, the solid lines indicate the evolutionary paths of a primordial
magnetic field with $\kf = 2\pi\HH_*$ and $\OmM^* = 0.1$
at $T_* = 150 \MeV$ and $g_* = 15$ and at $T_* = 100 \MeV$ and $g_* = 10$.

In Ref.~\cite{Neronov:2020qrl},
it was shown that the magnetic field compatible with
the NANOGrav results \cite{NANOGrav:2020bcs} would correspond at recombination to a  magnetic field of the same order of magnitude of those analyzed in Refs.~\cite{Jedamzik:2011cu,Jedamzik:2020krr}. 
In these works, it was pointed out that a sub-nano-Gauss prerecombination magnetic field would induce additional baryon inhomogeneities, which would enhance the recombination rate, thereby changing the CMB spectrum in a way that would alleviate
the Hubble tension. 
Reference~\cite{Galli:2021mxk} derived updated constraints on the baryon clumping from data of CMB experiments, of about $b\lesssim 0.5$ at 95\% confidence level, which can be translated into an upper limit on the  prerecombination magnetic field amplitude $B_{\rm rec}\lesssim 0.1$ nG.
In addition,
they derive a range $b \in (0.16, \, 0.55)$ that is compatible
with a value $H_0 \approx 70 \km \s^{-1} \Mpc^{-1}$, relieving
the Hubble tension.
Such values of the clumping factor correspond to magnetic field strengths
$B_{\rm rec} \in (0.013, \, 0.1)$ nG, which include
phase transition and inflationary produced magnetic fields \cite{Jedamzik:2018itu, Galli:2021mxk}.
The upper limit  and, in particular, the range derived to alleviate the
Hubble tension are indicated in \Fig{B_vs_l} by
a green line and interval, respectively.

The end points of the evolutionary paths of the magnetic field amplitude and characteristic scale compatible with the PTA results, representing their values
at recombination, 
lie on the line given in \Eq{rec_Bl}, where we also superimpose the constraint $B_{\rm rec}\lesssim 0.1$ nG from Ref.~\cite{Galli:2021mxk}.
It can be appreciated that they are compatible. 
We therefore confirm that a magnetic field at the QCD scale could both account for the PTA results
and alleviate the Hubble tension, as pointed out in Refs.~\cite{Jedamzik:2020krr,Galli:2021mxk}, 
depending
on the parameters $\kf$ and $\OmM^*$ of the initial field and whether
the developed MHD turbulence of the primordial plasma is 
compressible or incompressible.

Furthermore, in \Fig{B_vs_l}, we report the lower bounds on the magnetic field amplitude from the Fermi gamma-ray telescope \cite{Fermi-LAT:2018jdy,Durrer:2013pga, Neronov:2009gh, Korochkin:2020pvg}.
It was shown recently that CTA is sensitive to primordial
magnetic fields up to 0.01 nG (\cf \Fig{B_vs_l}) in the voids of the LSS \cite{Korochkin:2020pvg}.
The signal from a primordial magnetic field produced in phase transitions can be distinguished from one produced during inflation
since the latter is expected to produce 
a coherent signal among several nearby blazars \cite{Korochkin:2021vmi}.

The magnetic field can be additionally constrained from above by  observations of ultra-high-energy cosmic rays (UHECR) sources. Recent observations of UHECR from the Perseus-Pisces supercluster \cite{TelescopeArray:2021dfb} allowed one for the first time to put an upper limit on the primordial magnetic field in the voids of the LSS \cite{Neronov:2021xua}.
Finally, the upper
bounds from Faraday rotation measurements \cite{Pshirkov:2015tua}
are shown in \Fig{B_vs_l}.

\subsection{Role of the magnetogenesis scenario on the SGWB spectrum}
\label{init}

Primordial magnetic fields can be either produced or amplified
during the QCD phase transition
(see Refs.~\cite{Durrer:2013pga,Subramanian:2015lua, Vachaspati:2020blt} for reviews
and references therein).
In particular, some magnetogenesis scenarios at the QCD scale have been proposed; see e.g., Refs.~\cite{Quashnock:1988vs,Vachaspati:1991nm,Cheng:1994yr,Sigl:1996dm,Forbes:2001vya,Tevzadze:2012kk,Miniati:2017kah}.
Previous works performing simulations to compute the SGWB produced
by MHD turbulence, both in the general context of
phase transitions \cite{RoperPol:2019wvy,Brandenburg:2021bvg,Kahniashvili:2020jgm,RoperPol:2021uol} and, more specifically, at the QCD phase transition \cite{Brandenburg:2021tmp}, 
have modeled the
magnetic field production via
a forcing term in the induction equation [\cf\Eq{dBdt}]. 
These simulations show that, in general, the efficiency of the
GW production $q = \kf\HH_*^{-1} \sqrt{h^2 \,\OmGW^0}/\OmMmax$ is larger when the magnetic field is driven than when
it is given at the initial time of the simulation \cite{RoperPol:2019wvy,RoperPol:2021uol}.
The spectral shape is also affected, mostly in the
inertial range, i.e., at frequencies larger than $\fGW$, where it presents a steeper
forward cascade toward smaller scales \cite{RoperPol:2019wvy,RoperPol:2021uol,Brandenburg:2021tmp}.
In the subinertial range, the slope can also be slightly modified,
presumably due to deviations from Gaussianity
\cite{Brandenburg:2019uzj,Brandenburg:2021bvg,RoperPol:2021uol}.

\begin{figure}[t!]
    \centering
    \includegraphics[width=\columnwidth]{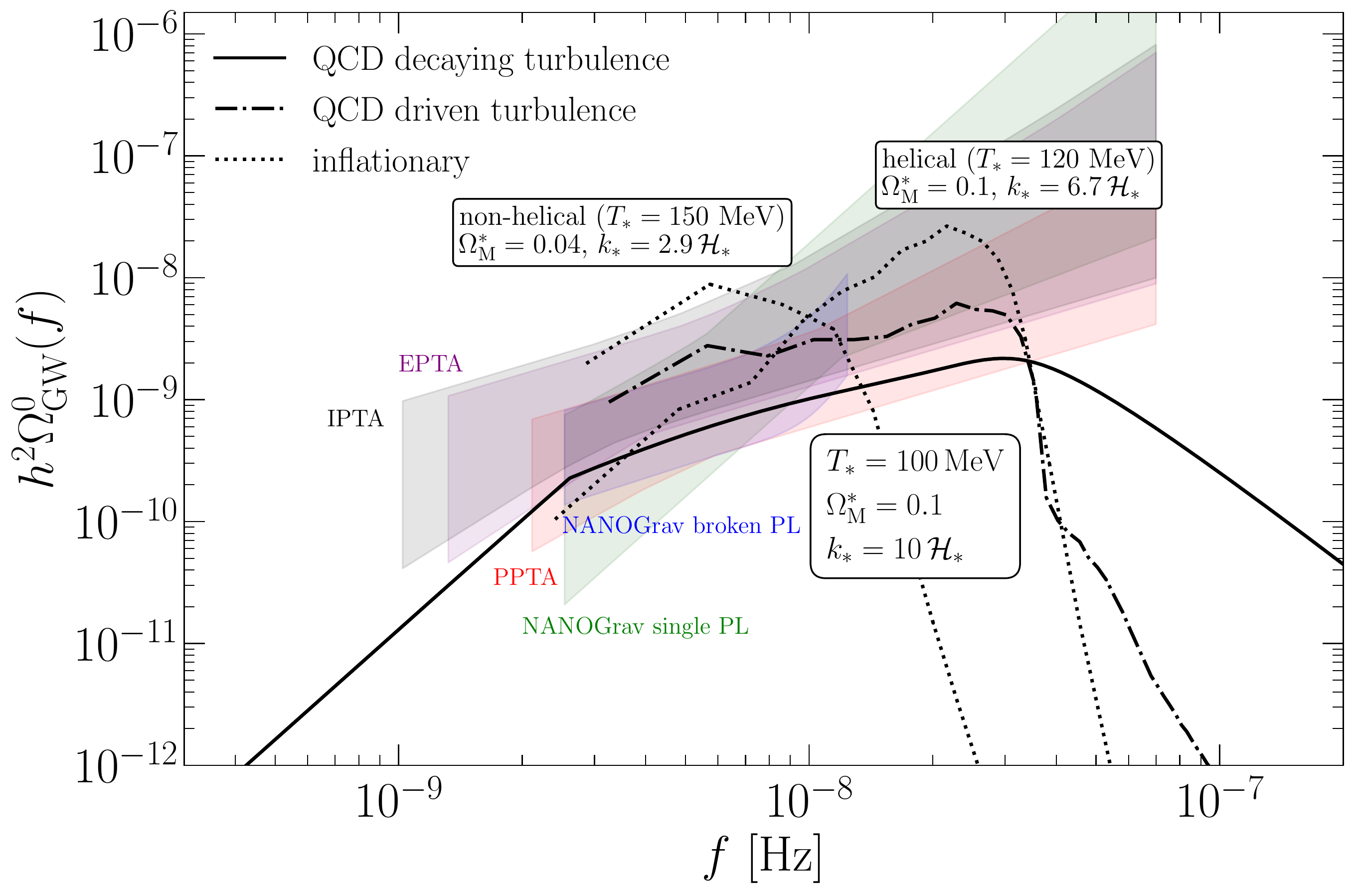}
    \caption{SGWBs generated by different sources operating around the QCD scale compared to the PTA results.
    The black solid line shows the analytical model developed in
    \Sec{analytical_spec}, validated with
    MHD turbulence simulations in \Sec{sec:numerical_spec}
    (``QCD decaying turbulence'').
    The dot-dashed line shows the SGWB obtained in
    Ref.~\cite{Brandenburg:2021tmp} by adding a forcing term
    in the induction equation to model the magnetic field generation
    (``QCD driven turbulence'').
    The dotted lines correspond to the inflationary
    magnetogenesis scenario of  Ref.~\cite{He:2021kah} with an end-of-reheating temperature around
    the scale of the QCD phase transition (``inflationary'').}
    \label{GW_efficiency_comp}
\end{figure}

However, since the magnetogenesis dynamics are still uncertain and are model dependent, previous simulations do not necessarily reproduce the actual physical
mechanism of magnetic field
production that might have operated in the early Universe.
In any case, their results
suggest that the model presented in our work,
which assumes that the magnetic field is already present
at the beginning of the simulation,
might be underpredicting the SGWB signal (or, equivalently, overestimating the magnetic field strength necessary to explain the PTA data).
This can be appreciated in \Fig{GW_efficiency_comp}, where we compare the SGWB obtained
from the analytical model of \Eqs{OmGW_envelope}{delta_tfin_te}, with the one obtained in Ref.~\cite{Brandenburg:2021tmp}
for nonhelical fields with 
$\OmMmax = 0.1$ and $\kf = 10 \, \HH_*$ at $T_* = 100 \MeV$.
We also show, for comparison, the SGWB obtained in Ref.~\cite{He:2021kah} from an inflationary magnetogenesis
scenario with an end-of-reheating temperature around the QCD scale, both for a nonhelical
magnetic field with $\OmMmax \simeq 0.04$ and $\kf \simeq 2.9 \, \HH_*$ at $T_* = 150 \MeV$
and a helical field with $\OmMmax \simeq 0.1$ and $\kf \simeq 6.7 \, \HH_*$ at $T_* = 120 \MeV$.

\section{Comparison with the SGWB from supermassive black hole binaries}
\label{sec:MBHB}

The most commonly considered model of the SGWB in the nanohertz frequency range is that of the collective GW signal from mergers of supermassive black hole binaries (SMBHB). This unrelated signal serves as a ``foreground'' for the cosmological SGWB signal detection.
It is interesting to analyze whether the two types of SGWB can be distinguished by current and future detections.

\begin{figure}[t!]
    \centering
    \includegraphics[width=\linewidth]{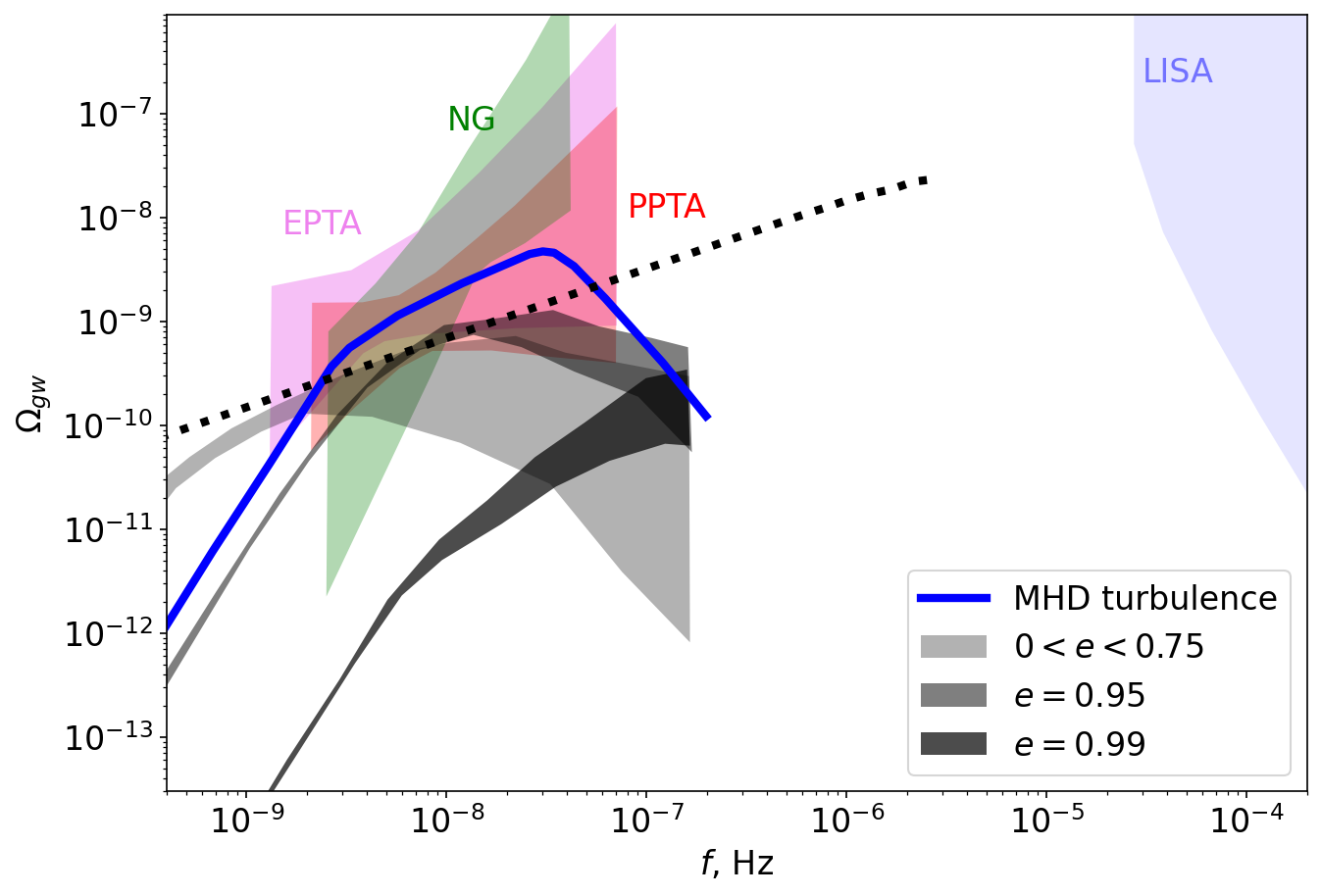}
    \caption{Possible SGWB from supermassive black hole mergers
    and by MHD turbulence, using the analytical model developed
    in \Sec{analytical_spec},
    compared to the PTA results.
    The black dotted line is proportional to $f^{2/3}$.}
    \label{fig:smbh}
\end{figure}

Straightforward analytical estimates \cite{Phinney:2001di,Sampson:2015ada} show that the cumulative spectrum of the GW emission from a population of SMBHB losing energy exclusively via gravitational radiation is expected to follow a PL  with the slope $(3-\gamma)/2=-2/3$ or, equivalently, $\beta = 2/3$ [\cf \Eqs{eq:hc}{SGWB_spl}].
This naive model is shown by the black dotted line in \Fig{fig:smbh}.
The large error bars of the PTA measurements do not allow one to distinguish between this slope and the expected slope of the SGWB produced by primordial MHD turbulence (shown by the blue line in \Fig{fig:smbh}).

This simple analytical model $\OmGW^0(f) \sim f^{2/3}$ does not take into account a number of effects that influence the shape of the SGWB from supermassive black hole mergers. 
One of these effects is related to the ``last parsec'' problem \cite{Sampson:2015ada}, the fact that the timescale of the gravitational energy loss on GW emission is longer than the Hubble time for binaries with subparsec  binary separations. Orbital periods of such binaries are about 10 yr and the GW emission from these systems falls into the
frequency range of the PTA results.
SMBHB can occur on the time span of the age of the Universe only if there exists a nongravitational energy loss that resolves the last parsec problem. In any case, this alternative energy loss channel removes energy from the GW signal and suppresses the GW spectral power. This results in deviations from the PL scaling $\beta = 2/3$. 

Dynamical friction produced by scattering of stars may be a viable solution to the last parsec problem if the eccentricity of the binary black hole systems is taken into account.  Examples of modeling of this effect  \cite{Kelley:2017lek} are shown by gray-shaded bands in \Fig{fig:smbh}. The suppression of the GW power occurs in the frequency range of PTAs for highly eccentric systems, with  $e\ge 0.95$. 

Still another effect may produce a second break in the spectrum at higher frequency, as seen in \Fig{fig:smbh}. This break occurs due to the discreetness of the spatial distribution of sources contributing to the SGWB
\cite{2008MNRAS.390..192S,Kelley:2017lek}.

Overall, the combination of the two breaks may result in a
SGWB spectrum from SMBHB similar to that produced from MHD turbulence.  This is clear from a comparison of the model spectrum discussed above (blue line in \Fig{fig:smbh}) with the state-of-art models for the supermassive black hole SGWB spectra  (gray bands in \Fig{fig:smbh}), calculated based on the cosmological hydrodynamical model Illustris \cite{2015MNRAS.452..575S}. 

It still should be possible to distinguish between the SMBHB and cosmological models using the statistics of individual binary system  detections at higher frequencies. Even though the diffuse background flux is suppressed at high frequencies because of the discreetness of the source distribution, individual sources (not considered anymore as part of the diffuse flux) become detectable. Their spectra typically extend well into the frequency range of LISA and their cumulative flux still follows the analytical $f^{2/3}$ scaling, with a moderate suppression in the LISA frequency range due to the fact that the GW emission from higher mass systems does not reach LISA sensitivity band. If the supermassive black hole SGWB is at the level of the current PTA measurements, LISA should be able to detect numerous individual merging systems and independently constrain the normalization of the supermassive black hole merger part of the background \cite{2008MNRAS.390..192S}.

\section{Conclusions}
\label{conclusions}

In this work, we have analyzed the GW signal produced by the anisotropic stresses of a primordial nonhelical magnetic field. 
We suppose that some process related to a primordial phase transition---in particular, here we focus on the QCD phase transition---generates the initial magnetic field. 
Since both the kinetic viscosity and the resistivity are very low in the early Universe, the magnetic field induces MHD turbulence in the primordial plasma. 
For simplicity, we do not model the magnetic field generation nor the buildup of the turbulent cascade, but we set as initial condition for the GW production a magnetic field with fully developed turbulent spectrum.
This is an important caveat of our analysis. 
We have chosen this approach to better keep under control the physics of the GW production and consequently gain insight on the resulting GW spectral shape, starting from simple initial conditions.
The chosen initial conditions are conservative: we
expect
the MHD turbulent magnetic spectrum to
develop for any magnetogenesis mechanism and
the amplitude of the SGWB to increase if an
initial period of magnetic field
generation is included in the analysis.
We plan to increase the level of complexity by analyzing concrete magnetic field production mechanisms in future works.

The first important result of our analysis is that the GW signal can be easily reproduced by assuming that the magnetic stresses sourcing the GWs are constant in time over a time interval $\delta\tfin$. 
The reason is that, for most of the spectral modes of the GW signal, the typical time of the GW production is shorter than the typical time of the magnetic field evolution since, by causality, the Alfv\'en speed is smaller than the speed of light. 
We provide in \Eq{OmGW_envelope} a simple formula for the resulting SGWB spectrum, in which the spectral slopes and the scaling with the source parameters (i.e., energy scale of the source and magnetic field's amplitude and characteristic scale) are apparent. 

This formula can be used in general as a template for the SGWB spectrum from fully developed MHD turbulence. 
We have in fact validated it with a series of MHD simulations initiated with a fully developed magnetic field spectrum and no initial bulk velocity. These have been performed using the {\sc Pencil Code} \cite{PencilCode:2020eyn} and consist in several runs that cover a wide range of modes, from the large super-Hubble scales up to the high wave numbers of the magnetic field inertial range. 
The GW spectra outputs from the simulations are well reproduced by the template obtained under the assumption of constant anisotropic stress. 
In particular, they both feature a break at a characteristic wave number corresponding to the inverse duration of the GW source, where the causal $k^3$ increase transitions to a linear increase, more or less smoothly depending on whether the source lasts more or less than one Hubble time. 
We indeed use the break position in the simulations to fix the source duration parameter $\delta\tfin$, which is a free parameter of the analytical model, in terms of the eddy turnover time $\delta \te$. 

We have then applied our results to the case of the QCD phase transition. As pointed out in a previous work \cite{Neronov:2020qrl}, the GW signal from MHD turbulence occurring close to the QCD energy scale in the early Universe can account for the CP reported recently by the observations of the PTA collaborations: NANOGrav, PPTA, EPTA, and IPTA.
Here we have used the simulation-validated SGWB template \Eq{OmGW_envelope} and compared it to the PTA results. 

Several points deserve to be highlighted concerning this particular possible explanation of the PTA CP. 
First of all, the region of the MHD-produced SGWB spectrum that is compatible with the PTA constraints on the CP spectral index is the subinertial region, and for temperature scales of the order of the QCD phase transition, the subinertial region naturally falls in the frequency range where the PTA data present less uncertainty. 

Second, the break in the SGWB spectrum is also expected to fall in the same best quality data frequency region, for temperatures around 100 MeV. 
The position of the break is correlated to the energy scale of the process that generated the magnetic field and, in turn, the SGWB. 
Therefore, measuring the position of the break in the future PTA data offers the interesting opportunity to pin down its origin, especially if the PTA observations can be combined with LISA to help disentangle this  SGWB of primordial origin from the one due to SMBHB. 

Third, the energy scale of the magnetogenesis mechanism, and therefore of the GW production, is quite constrained already by the PTA data: it must be in the range $1 \MeV<  T_*< 200\MeV$; otherwise, this scenario fails to explain the PTA results in the limit $\OmMmax \leq 0.1$. 
At the same time, the initial amplitude of the magnetic field must be at least 1\% of the radiation energy density, and its characteristic scale must be within 10\% of the horizon scale.
It is therefore not unreasonable to expect that future PTA data will be able to falsify the hypothesis of the SGWB signal from MHD turbulence. 

At last, the ranges of magnetic field amplitudes and characteristic scales that can account for the PTA CP through the GW signal they generate could also affect the evolution of the baryon density fluctuations at recombination, effectively enhancing the recombination process and lowering the sound horizon at recombination \cite{Jedamzik:2020krr,Galli:2021mxk,Jedamzik:2020zmd}.
The presence of a magnetic field at recombination with present-time strength of about $0.01 \lesssim B_{\rm rec}\lesssim 0.1$ nG was recently proposed as a possible way to alleviate the Hubble tension \cite{Galli:2021mxk,Jedamzik:2020krr}.
Such a field could be detected in the voids of Large Scale Structure with a future CTA gamma-ray observatory \cite{Korochkin:2020pvg}.
We find here that the SGWB which such a magnetic field would produce offers a further observational channel to test this hypothesis. 

\begin{center}
    {\bf \small DATA AVAILABILITY}
\end{center}

The source code used for the simulations of this study,
the {\sc Pencil Code}, is freely available \cite{PencilCode:2020eyn}.
The simulation datasets are also publicly available \cite{DATA}.
The calculations, the simulation data, and the routines generating the
plots are publicly available on
GitHub\footnote{\url{https://github.com/AlbertoRoper/GW_turbulence/tree/master/PRD_2201_05630}.}
\cite{GH}.\\

\begin{center}
    {\bf \small ACKNOWLEDGEMENTS}
\end{center}

We are grateful to Ruth Durrer and Tina Kahniashvili
for their useful feedback and comments.
Support through the French National
Research Agency (ANR) project MMUniverse (ANR-19-CE31-0020)
is gratefully acknowledged.
A.R.P.~also acknowledges
support from the Shota Rustaveli National Science Foundation (SRNSF) of Georgia (Grant No.~FR/18-1462).
We acknowledge the allocation of computing resources provided by the
{\em Grand \'Equipement National de Calcul 
Intensif} (GENCI) to the project ``Opening new windows on Early Universe 
with multi-messenger astronomy'' (A0090412058).

\bibliography{references}

\end{document}